\documentclass[english]{iopart}

\usepackage{graphicx}
\usepackage{url}
\usepackage{iopams} 

\expandafter\let\csname equation*\endcsname\relax
  \expandafter\let\csname endequation*\endcsname\relax

\usepackage{amssymb}
\usepackage[latin1]{inputenc}
\usepackage{amsmath}
\usepackage{epsfig}
\newcounter{fig}
\begin{document}

\title[Factorization of  Ising 
correlations  and their lambda extensions]{Factorization of Ising correlations
  $ C(M,N)$ for $ \nu= \, -k$ and
  $ \, M+N$ odd, $\, M \, \le N$,  $\, T < T_c$ and their lambda extensions
}

\vskip .1cm 

\author{S. Boukraa$^1$,  C. Cosgrove$^2$,  
J.-M. Maillard$^3$, B. M. McCoy$^4$,}
\address{1 LSA, IAESB,
 Universit\'e de Blida 1,  Algeria}
\address{2 The University of Sydney, Faculty of Sciences, Carslaw
  Building, Sydney, Australia}
\address{3 LPTMC, Sorbonne Universit\'e,  Tour 23
 5\`eme \'etage, case 121, \\
 4 Place Jussieu, 75252 Paris Cedex 05, France} 
\address{4 Institute for Theoretical Physics,
State University of New York, 
Stony Brook, NY 11794-3840, USA}
\ead{maillard@lptmc.jussieu.fr, jean-marie.maillard@sorbonne-universite.fr, 
christopher.cosgrove@sydney.edu.au,
bkrsalah@yahoo.com, mccoy@max2.physics.sunysb.edu}

\vskip .1cm


\begin{abstract}

We study the factorizations of Ising low-temperature correlations $\, C(M,N)\, $ for
$\, \nu=\, -k\, $ and $\, M+N\, $ odd, $\, M \, \le N$,  for both the cases $\, M\, \neq\,  0$ 
where there are two factors, and $\, M=\, 0\, $ where there are four factors. 
We find that the two factors for $\, M\, \neq\,  0$  satisfy the same
non-linear differential equation and, similarly, for $\, M=\, 0\,$ the
four factors each satisfy  Okamoto sigma-form of Painlev\'e VI  equations with the same
Okamoto parameters. Using a Landen transformation we show, for $\, M\, \neq\,  0$, that
the previous non-linear
differential equation can actually be reduced to  an Okamoto sigma-form of Painlev\'e VI  equation.
For both the two and four factor case, we find that
there is a one parameter family of boundary conditions on the
Okamoto sigma-form of Painlev\'e VI   equations which
generalizes the factorization of the correlations $\, C(M,N)\, $ to an
additive decomposition of the corresponding sigma's 
 solutions of the  Okamoto sigma-form of Painlev\'e VI   equation
which we call lambda extensions.
At a special value of the parameter,  the lambda-extensions of the
factors of $\, C(M,N)$ reduce to homogeneous polynomials in the
complete elliptic functions of the
first and second kind. We also generalize some Tracy-Widom (Painlev\'e V) relations
between the sum and difference of sigma's to this  Painlev\'e VI framework.

\end{abstract}

\vskip .1cm


\noindent {\bf AMS Classification scheme numbers}: 34M55, 
47E05, 81Qxx, 32G34, 34Lxx, 34Mxx, 14Kxx 

\vskip .2cm


\section{ Introduction}
\label{Introduc}

In a previous paper~\cite{bmm} we considered the two-point correlation $ \, C(M,N)$
of spins at sites $\, (0,0)$ and  $\, (M,N)$
of the anisotropic Ising model defined
by the interaction energy
\begin{eqnarray}
  \hspace{-0.7in} \quad  \quad  \quad \quad  \quad \quad 
         {\mathcal E \,}
\, \, = \, \,\, -\sum_{j,k}\{E_v\sigma_{j,k}\sigma_{j+1,k}
\, +E_h\sigma_{j,k}\sigma_{j,k+1}\}, 
\end{eqnarray}
where $ \, \sigma_{j,k}\, = \,\, \pm  \, 1 \, $ is the spin at row $\, j$
and column $\, k$,  and where the sum is over all lattice sites. 
Defining
\begin{eqnarray}
\hspace{-0.9in} 
  k\,\,= \,\, \,(\sinh 2E_v/k_BT \, \sinh 2E_h/k_BT)^{-1}
  \quad \quad \hbox{and} \quad \quad \quad
  \nu \, = \, \, \frac{\sinh 2E_h/k_BT}{\sinh 2E_v/k_BT}, 
\end{eqnarray}
we found~\cite{bmm} that in the special
case\footnote[5]{The condition
  $\, \nu \,  \,  = \, \, \, -k \, $ (as well as the isotropic case $\, \nu \, = \ 1$) is special
  because it is such that the complete elliptic integrals of the third kind 
  reduce to complete elliptic integrals of the second kind (see equation (30) in~\cite{bmm}).}
\begin{eqnarray}
  \quad  \quad  \quad \quad  \quad \quad  \quad
\nu \,  \,  = \, \, \, -k, 
\end{eqnarray}
the correlation\footnote[1]{Which is the same as the Toeplitz
  determinants~\cite{Comedy} of Forrester-Witte~\cite{fw,fw2} as given in~\cite{gil}.}
$ \, C(M,N)$ satisfies an Okamoto sigma-form of the 
Painlev{\'e} VI equation.

For $\, \, T\, <\, T_c$,  $\, M \, \le \,  N\, $ and $\, \,  \nu\, =\, -k \, $
with $\,\,  t=\, k^2\, $ and
\begin{eqnarray}
  \label{defsigma}
\hspace{-.08in}
\sigma\,\, =\,\, \,
t \cdot \, (t-1) \cdot \, \frac{d \ln C(M,N)}{dt} \,\, \,-\frac{t}{4}, 
\end{eqnarray}
we have~\cite{bmm}:
\begin{eqnarray}
&&\hspace{-.38in}
  t^2 \cdot \, (t-1)^2 \cdot \, \sigma''^2 \, \,  \,  \, 
  +4 \cdot \,  \sigma' \cdot \, (t\, \sigma'\,  -\sigma) \cdot \,
  \Bigl((t -1) \cdot \, \sigma'\,  -\sigma \Bigr)
\nonumber\\
&&\hspace{-.08in}
-M^2 \cdot \, (t\, \sigma'-\sigma)^2 \,\, \, \,  -N^2 \cdot \, \sigma'^2
\nonumber\\
\label{eqnm}
&&\hspace{-.08in}
+\Bigl( M^2 +N^2 \,
-\frac{1}{2} \cdot \, (1 \, +(-1)^{M+N}) \Bigr)
\cdot \,  \sigma' \cdot \, (t\, \sigma'\, -\sigma)
\, \,  \,  = \,  \, \, \, 0. 
\end{eqnarray}
When  $ \, M+N \, $ is odd, $\, M \, \le \,  N$,  the previous Okamoto sigma-form of the 
Painlev{\'e} VI equation (\ref{eqnm}) becomes:
\begin{eqnarray}
\label{eqnmodd}
&&\hspace{-.68in}
  t^2 \cdot \, (t-1)^2 \cdot \, \sigma''^2 \, \,  \,  \, 
  +4 \cdot \,  \sigma' \cdot \, (t\, \sigma'\,  -\sigma) \cdot \,
  \Bigl((t-1) \cdot \, \sigma'\,  -\sigma \Bigr)
\nonumber\\
&&\hspace{-.48in}
-M^2 \cdot \, (t\, \sigma' -\sigma)^2 \,\, \, \,  -N^2 \cdot \, \sigma'^2
\,\,   +(M^2 +N^2)
\cdot \,  \sigma' \cdot \, (t\, \sigma'\, -\sigma)
\, \,  \,  = \,  \, \, \, 0. 
\end{eqnarray}

We noted~\cite{bmm}, when $ \, M+N \, $ is odd, that the {\em low-temperature} 
correlation $ \, C(M,N) \, $ factors into two factors. 
In the even more special
case of $\, M= \, 0 \, $ and $ \, N \, $ odd,
the previous sigma-form of Painlev\'e VI
non-linear ODE (\ref{eqnmodd}) reads
\begin{eqnarray}
&&\hspace{-.38in}
  t^2 \cdot \, (t-1)^2 \cdot \, \sigma''^2 \, \,  \,  \, 
  +4 \cdot \,  \sigma' \cdot \, (t\, \sigma'\,  -\sigma) \cdot \,
  \Bigl((t-1)  \cdot \, \sigma'\,  -\sigma \Bigr)
\nonumber\\
\label{eqnmoddMzero}
&&\hspace{-.08in}
 \,\, \,  -N^2 \cdot \, \sigma'^2
\, \, \,  +N^2 \cdot \,  \sigma' \cdot \, (t\, \sigma'\, -\sigma)
\, \,  \,  = \,  \, \, \, 0,
\end{eqnarray}
and  the  {\em low-temperature} 
correlation $ \, C(M,N)$ factors into {\em four}
factors\footnote[2]{Homogeneous polynomial in the complete elliptic integrals
  of the first and second kind.}, each of which were shown to satisfy  sigma-form
Painlev\'e VI non-linear differential
equations with the same Okamoto parameters\footnote[8]{But with different boundary conditions.}.

In~\cite{bmm} we gave the four Okamoto parameters for
Okamoto sigma-form of Painlev{\'e} VI 
equations which are satisfied by each of the four factors.
In this paper we continue the study of~\cite{bmm} and 
find the second-order non-linear differential  equation for the sigma's of the two factors 
of $\, C(M,N)$ with $\, M+N \, $ odd and $\, M\, \neq\,  0$. 
We study the necessary boundary conditions for
both the two and four factor decompositions and show that the
factors of $\, C(M,N)$ can be generalized to a {\em one-parameter} family of
solutions of the Okamoto sigma-form of Painlev{\'e} VI equation analytic
at $ \, t= \, k^2= \, 0$. 
In the remainder of this introduction we outline the methods 
and results of this study.

\subsection{ Outline of the methods and the results}
\label{Introduc}

We begin by recalling that in~\cite{bmm} we showed, for $\, M+N$ odd, $\, M \, \le N$, 
 that the representation of $ \, C(M,N)$ for $ \, k= \, -\nu \, \, $ as a Toeplitz determinant 
is symmetric. In~\cite{wilf}
it is shown, by elementary row column operations, that any $ \, N \times \,  N$
symmetric Toeplitz determinant $ \, {\rm det}(a_{i-j})$ with $ \, a_j= \, a_{-j}$
has a factorization for $ \, N$ even ($N= \, 2m$) into two
$ \, m\times \,  m$ determinants:  
\begin{eqnarray}
  &&\hspace{-.46in}
 {\rm det}(a_{i-j})_{i,j=1,\cdots, 2m}
 \nonumber\\
  \label{factoreven}
&&\hspace{-.46in}
  \quad  = \, \,  \, {\rm det}(a_{i-j}-a_{i+j-1})_{i,j=1,\cdots, m}  \cdot \, 
{\rm det}(a_{i-j}+a_{i+j-1})_{i,j=1,\cdots, m}.
\end{eqnarray}
This can be extended to $\, N$ odd $(N= \, 2\,  m \, +1)$ as a
factorization into an
$ \, m \times \,  m$ determinant and a $ \, (m+1)\times \, (m+1) \, $ determinant: 
\begin{eqnarray}
  &&\hspace{-.46in}
 {\rm det}(a_{i-j})_{i,j=1,\cdots, 2 m \, +1}
\nonumber\\
\label{factorodd}  
  &&\hspace{-.46in}
\quad  =  \, \,  \,
\frac{1}{2} \cdot \, {\rm det}(a_{i-j}-a_{i+j})_{i,j=1,\cdots, m} \cdot \, 
{\rm det}(a_{i-j}+a_{i+j-2})_{i,j=1,\cdots, m+1}.
\end{eqnarray}
Thus the existence of factorizations of the $ \, C(M,N)$'s into two factors is not surprising.

To obtain explicit expressions for the factors  we use the method discussed
in~\cite{bmm} expressing $ \, C(M,N)$ as {  \em homogeneous} polynomials
in terms of the complete elliptic integrals of the first and second kind
\begin{eqnarray}
&& \hspace{-.36in} \quad \quad 
  {\tilde K}(k)  \, \, = \,  \,   \, \frac{2}{\pi} \cdot \, K(k) \,  \,= \, \,\, \, 
   {}_2F_1\Bigl([\frac{1}{2},\frac{1}{2}], \, [1], \,k^2\Bigr),
   \quad
  \nonumber\\
\label{elliptic}  
  &&\hspace{-.36in} \quad \quad 
  {\tilde E}(k) \, \, = \,  \,  \,  \frac{2}{\pi} \cdot \, E(k) \, \,  = \,  \,\, \, 
  {}_2F_1\Bigl([\frac{1}{2},-\frac{1}{2}], \, [1],  \,k^2\Bigr), 
\end{eqnarray}
by first solving the
{\em quadratic difference}\footnote[1]{See for instance equations (41) and (42) in~\cite{Importance}. Note that
these quadratic difference equations~\cite{mccoy1,mccoy2,mccoy3,perk} are actually valid for the {\em anisotropic}
Ising model.  Do note that the  {\em lambda-extensions}~\cite{Holonomy} of the $\, C(M, \, N)\, $ correlation functions 
{\em also verify these quadratic difference equations}.}
{\em  equations}~\cite{mccoy1,mccoy2,mccoy3,perk},  the $\, C(M, \, N)$'s being then factored. 
We will write the factorizations of $ \, C(M,N)$   as
\begin{eqnarray}
 \label{twofactors}
&& \hspace{-.46in} \quad 
 (1 \, -t)^{-1/4} \cdot \, C(M,N;t)  \,\,\, = \, \,\,\,\,  g_{+}(M,N;t)\cdot \, g_{-}(M,N;t), 
\end{eqnarray}
with 
\begin{eqnarray}
  \quad \quad \quad  \quad \quad   \quad \quad 
t \, = \, \, k^2, 
\end{eqnarray}
where the two factors $ \, g_{\pm}$ are homogeneous polynomials of the complete
elliptic integrals of the first and second kind and have the expansion about $ \, t= \, 0$
\begin{eqnarray}
  \label{expansion}
 \hspace{-.46in}\,\,
  g_{\pm}(M,N;t) \,  \, = \,  \,  \,  \, \,
   1 \,  \,\, \,  \pm \,  t^{(N+1)/2}   \cdot \, f_1(t) \, \,  \, +t^{N+2}  \cdot \, f_2(t), 
\end{eqnarray}
where both $ \, f_1(t)$ and $ \, f_2(t)$ are analytic at $\, t= \, 0$. 
Thus from (\ref{twofactors}) we have:
\begin{eqnarray}
\label{Cf1f2}
\hspace{-.46in}\,\,
C(M,N;t)\, \, = \, \, \, \,
(1 \, -t)^{1/4} \cdot \, \Bigl(-t^{N+1} \cdot \, f_1^2
\, \,  +\Bigl(1 \, +t^{N+2} \cdot \, f_2\Bigr)^2 \Bigr).
\end{eqnarray}
Examples of the factorizations (\ref{twofactors}), and of the expansions
(\ref{expansion}), are given in \ref{appA}.

We consider  the following logarithmic derivatives of
the previous two factors: 
\begin{eqnarray}
 \label{loggpm}  
\hspace{-.46in} \quad  \quad  \quad 
 \sigma_{\pm}(M,N;t) \, \,\,  = \, \, \,\,
  t \cdot \,(t-1) \cdot \, \frac{d\ln g_{\pm}(M,N;t)}{dt}.
\end{eqnarray}
The sigma functions have {\em additive} decompositions which follow from
the mulplicative decompositions  (\ref{twofactors})
\begin{eqnarray}
\label{additive}
&&\hspace{-.56in} \quad \quad  \quad \quad 
 \, \sigma(M,N;t)\,\, = \, \,\,  \, \sigma_{+}(M,N;t)\,\, \, +\sigma_{-}(M,N;t), 
\end{eqnarray}
where $\, \sigma(M,N;t) \, $ is defined by (\ref{defsigma}) and
\begin{eqnarray}
  \label{sigmapmrhorho}
\hspace{-.66in} \quad 
  \sigma_{\pm}(M,N;t)\, \,  = \, \, \, \, \,  
   \pm  \, t^{(N+1)/2} \cdot \, \rho_1(M,N;t) \, \, \, +t^{N+1} \cdot \, \rho_2(M,N;t),
\end{eqnarray}
where $\, \rho_1 \, $ and  $\, \rho_2 \, $ (related to $\, f_1\, $ and $\, f_2\, $
by equation (\ref{Cf1f2})) are power series\footnote[2]{
From (\ref{additive}) it is clear that $ \,   2 \cdot \, t^{N+1} \cdot \, \rho_2(M,N;t) \, $
 is the sigma  function for $ \, C(M,N;t)$.} of $\, t$,
analytic at $\, t=\, 0$. Examples are given in \ref{appB}. 

In~\cite{bmm} we found in Appendix D 2.1, in the special case $ \, k= \, -\nu$,
that the sigma-form of Painlev\'e VI for the sigma function of $ \, C(M,N) \, $ 
{\em admits extensions\footnote[5]{For an introduction of the concept of
    {\em lambda extension} of correlation functions
    see for instance equations (9), (10) in~\cite{Holonomy}.}
  to  a one parameter family of solutions} which are analytic at
$ \, t= \, 0$. This {\em one-parameter} family of solutions analytic at $ \, t= \, 0 \, $
extends to the additive decomposition (\ref{additive}) as 
\begin{eqnarray}
\label{lambdaadditive}
 \hspace{-.46in} \quad  \quad  \, 
 \, \sigma(M,N;t;\lambda)\, \, = \,\, \, \,
  \sigma_{+}(M,N;t;\lambda_{+})\,\, \,  +\sigma_{-}(M,N;t;\lambda_{-}), 
\end{eqnarray}
where
\begin{eqnarray}
  \label{lambdasigmapm}
 \hspace{-.96in} \quad  \quad  \quad  \quad  \quad 
\sigma_{\pm}(M,N;t; \, \lambda_{\pm})  
  \, \,  =\, \, \,\, \,  
  \sum_{n=1}^{\infty} \, \Bigl(\lambda_{\pm} \cdot \, t^{(N+1)/2}\Bigr)^n \cdot \, B_n(M,N;t), 
\end{eqnarray}
where the $\, B_n(M,N;t)$'s are
power series\footnote[1]{The  $\, B_n(M,N;t)$'s are D-finite series and, in fact, polynomials
  in the complete elliptic integrals of the first and second kind $\,  {\tilde K}$
  and  $\, {\tilde E}$ (see (\ref{htilde1}),  (\ref{htilde1bis}) below).} 
analytic at $\, t \, = \, 0$, 
and where we must choose (see (\ref{lmbdapmlambda}) below) 
$\,  \lambda_{+} \,\, = \,\,\, -\lambda_{-}\,=\,\,\,\lambda \,\,$
in order to match with the {\em lambda extension}~\cite{Holonomy}
solutions of (\ref{eqnmodd}).

In~\cite{bmm} the second order non-linear differential equations (\ref{eqnmodd}), (\ref{eqnmoddMzero})
were found to be of the ``master Painlev\'e equation'' form (see the so-called
SD-I equation (4.9) with $\, c_1=\, 0$, $\, c_4=\, 0$, $\, c_3=\, -c_2$,
in Cosgrove and Scoufis~\cite{cosgrove})
\begin{eqnarray}
&&\hspace{-.38in} \,\,\,
  x^2 \cdot \, (x -1)^2 \cdot \, y''^{\, 2} \,\,\,\,
  +4 \cdot \,  y'  \cdot \,  (x \, y' \, -y) \cdot \, ( (x -1) \, y' \, -y) 
   \nonumber\\
  &&\hspace{-.08in} \,\,\,
+c_5 \cdot \, (x \, y' \, -y)^2 \,\, 
+c_6 \cdot \, y' \cdot \, (xy'-y) \, \, \, +c_7 \cdot \, (y')^2
\nonumber\\
\label{cosgrove}
&&\hspace{-.08in} \,\,\,
+c_8\cdot \, (x \,y' \, -y) \, \, \, \, +c_9 \cdot \, y'\,  \, \, +c_{10}
\, \, \,= \,\, \, \,  0, 
\end{eqnarray} 
which has the {\em Painlev{\'e} property} of {\em fixed critical points}~\cite{Handbook,Ince}.
The  non-linear differential equation (\ref{cosgrove}) preserves its form under
the linear shift:
\begin{eqnarray}
  \quad   \quad  \quad 
y \quad \,  \longrightarrow  \quad  \quad  y \, \,\, +A \,\, \, +B \cdot \, x. 
\end{eqnarray}
This shift may be used to eliminate $ \,c_5 \, $ and $ \, c_6\, $ which reduces
(\ref{cosgrove}) to the canonical form of the  sigma-form of Painlev\'e VI  equation
obtained by Okamoto~\cite{okamoto} with $ \, c_5= \, c_6= \, 0 \, $
which is birationally equivalent to the original Gambier form of
Painlev{\'e} VI: 
\begin{eqnarray}\hspace{-.08in}
  &&\hspace{-.78in}
 \frac{d^2y}{dt^2}
   \,  \,= \, \, \,\,
   \frac{1}{2} \cdot \, \left(\frac{1}{y} \,+\frac{1}{y-1}
+\frac{1}{y-t}\right) \cdot \, \left(\frac{dy}{dt}\right)^2 \,\,
   -\left(\frac{1}{t} \, +\frac{1}{t-1} \, +\frac{1}{y-t}\right) \cdot \, \frac{dy}{dt}
\nonumber\\
  \label{gambier}
  &&\hspace{-.6in}
     +\frac{y \cdot \, (y-1) \cdot \, (y-t)}{t^2 \cdot \, (t-1)^2} \cdot  \,
\left(\alpha \, \, +\beta \cdot\,  \frac{t}{y^2} \,\,  +\gamma \cdot \, \frac{t-1}{(y-1)^2} \, \,\, 
+\delta \cdot\,  \frac{t \cdot \, (t-1)}{(y-t)^2}\right).
\end{eqnarray} 
  
In section \ref{nonlin} we will obtain non-linear differential equations for $\, \sigma_{\pm}(M,N;t)$
by using the method of~\cite{bmm} of expanding the factors as power series in $\, k$ (or
$\, t= \, k^2$), and then using Jay Pantone's program {\it guessfunc}~\cite{pantone}
to produce a {\em non-linear differential equation}
{\em quadratic in the second derivative}.  These non-linear differential equations
for $\, \sigma_{\pm}(M,N;t)$ are {\em not of the  ``master Painlev\'e equation'' form}
(\ref{cosgrove}). We will show that they
can  be (non-trivially) reduced\footnote[1]{If these non-linear differential equations
 for $\, \sigma_{\pm}(M,N;t)$ were of the  ``master Painlev\'e equation'' form,
 their reduction to the form (\ref{cosgrove}), or to
 the Okamoto sigma-form of Painlev\'e VI would probably correspond to
 algebraic transformations~\cite{Kitaev,Manin,CubicQuartic,Okamoto2,Ramani}
 called  ``folding transformations'' using the term coined by the Tokyo school~\cite{TOS}.
 Here we need some (slight) generalization of the concept of ``folding transformations''.
} to the form (\ref{cosgrove}), or to
the Okamoto sigma-form of Painlev\'e VI, by
introducing a {\em Landen transformation}~\cite{Heegner}.
In section \ref{SelecteD} we find the selected values of
$\, \lambda \,$  for which $ \, g_{\pm}(M,N;t;\lambda) \, $  
reduce to homogeneous polynomials in the complete
elliptic integrals of the first and second kind $ \, {\tilde K}(k)$ and $ \, {\tilde E}(k)$.
In section \ref{TracyWidom}, recalling the Tracy and Widom paper~\cite{tw} we introduce,
besides the sum (\ref{additive}), the {\em difference} of the two $\, \sigma_{\pm}$,
and find the second order non-linear ODE satisfied by this difference. 

In section \ref{bound}  we recall {\em the Okamoto sigma form of 
Painlev\'e VI equations}~\cite{bmm} (having the {\em same} Okamoto 
parameters) satisfied by the {\em four}  sigma's
corresponding to the four factors of $\, C(0,N;t)$
with $\,N$ odd. This allows us to write $\, \sigma(0,N;t)$ 
{\em as the sum of four sigma's}. 
We  find the boundary conditions needed to generalize this additivity relation
to one-parameter lambda-extensions of these  sigma's.
We also find the selected values
of the  lambda parameters such that the four factors of $\, C(0,N;t) \, $
are (homogeneous) polynomial expressions
 of the complete elliptic integrals.  
 In section \ref{TracyWidom2} we also introduce the {\em  difference} of
 two sigma's among the four. We find that
 the second order non-linear ODE,
satisfied by this difference, is  compatible with the second
order non-linear ODE
satisfied by the sum of these two sigma's. We also show that the situation,
where the four factors of $\,C(0,N;t) \, $ are
actually homogeneous polynomial expressions of the complete elliptic integrals
$ \, {\tilde K}(k)$ and $ \, {\tilde E}(k)$,
associated with the previous selected values
of the  lambda parameters, corresponds, in fact, to the existence
of a polynomial relation,
$\, {\cal P}_N(\sigma, \, \sigma', \, t) \, = \, \, 0$,
compatible with the second order non-linear ODE. 
We finally present, in section \ref{detFW}, a discussion of the Forrester-Witte
determinants of~\cite{fw} and~\cite{gil} and of the boundary conditions
on the Okamoto non-linear differential equation needed to specify
these determinants for the factors of $\, C(M,N)$ when $ \, (M+N) \, $ is odd.
 
\vskip .2cm

\section{Non-linear equation for the two 
  factors of $ \, C(M,N) \, $ with $ \, M+N \, $ odd, $\, M \, \le \, N$}
\label{nonlin}

The earliest study of factorizations of Painlev{\'e} $\, \tau$-functions~\cite{PainlTracy} was
made by Tracy and Widom~\cite{tw} in the context of random matrix theory and
Painlev{\'e} V representation of {\em Fredholm determinants}.

Here we begin with the factorizations (\ref{twofactors})  of $ \, C(M,N)$'s
with $ \, M+N \, $ odd, $\, M \, \le \, N$,
for miscellaneous values of $ \, M$ and $ \, N$, and,
by use of the methods described in~\cite{bmm} and of the
program {\em guessfunc} of Jay Pantone~\cite{pantone}, we find that
both $ \, \sigma_{+}(M,N;t)\,$ and $ \, \sigma_{-}(M,N;t)\,$ in (\ref{additive}) satisfy the
{\em same second-order non-linear differential equation}
\begin{eqnarray}
  &&\hspace{-0.92in}\,\, 
 32 \,\, t^3 \cdot \, (t\, -1)^2 \cdot \, \sigma''^2
 \,\, \, + 4 \,\, t^2 \cdot \, (t-1) \cdot \,
\Bigl( 8 \cdot \, \sigma \, \, - 8 \cdot \,(t+1) \cdot \, \sigma'  \, \,
  +M^2 -N^2 \Bigr) \cdot \, \sigma'' 
\nonumber\\
&&\hspace{-0.9in} \, \,\, \,\, 
- \Bigl(8 \, \sigma \, \, -16 \cdot \, t \, \sigma' \,  +M^2 \, t \, -N^2 \, +1 \,  -t\Bigr)
\cdot
\Bigl(8 \cdot \, t \cdot \, (t-1) \cdot \, \sigma'^2 \,  \,
   -16 \, t \cdot \, \sigma  \cdot \, \sigma'
\nonumber\\
\label{nonlineareq}  
&&\hspace{-0.9in}
\quad   \quad   \quad \quad 
\, +8 \cdot \, \sigma^2 \, \, \, + (M^2-N^2) \cdot  \, \sigma \Bigr)
     \, \, \,  = \, \, \,\, \,  0, 
\end{eqnarray}
where the prime indicates a derivative with respect to $ \, t$, and where $\, \sigma$ reads:
\begin{eqnarray}
  \quad \quad \quad 
\sigma  \,\,  = \, \, \, t \cdot \,(t-1) \cdot \,\frac{d\ln g}{dt}.  
\end{eqnarray}
The two solutions of (\ref{nonlineareq}),  $ \,  \, \sigma_{+}(M,N;t)\,$ and $ \, \sigma_{-}(M,N;t)$,
have different boundary conditions. 
Note that $\, \sigma_{\pm} \, = \, 0 \, \, $ {\em is a selected solution} of (\ref{nonlineareq}).  

Similar to~\cite{bmm}, these  non-linear differential equations
are obtained for
particular values of $\, M$ and $\, N$, when restricted to order three derivatives and, then, 
finding a first integral to obtain a non-linear differential  equation
quadratic in the second derivative. For  small values of  $\, M$ and $\, N$
one may get several (compatible) non-linear differential equations, however with larger
values of  $\, M$ and $\, N$ one gets a cleaner situation with a unique
and {\em stable} form
corresponding to the previous pattern (\ref{nonlineareq}). Note that this form (\ref{nonlineareq})
is actually valid for the very small values of  $\, M$ and $\, N$ when other compatible
non-linear differential equations also occur.

The second order non-linear differential equation (\ref{nonlineareq}) is {\em not} of the SD-I 
``master Painlev\'e equation''  form (\ref{cosgrove}) given in~\cite{cosgrove}.  On the contrary  (\ref{nonlineareq})
is, at first sight,  of the general form studied by Bureau~\cite{Bureau1,Bureau2}
\begin{eqnarray}
\quad \quad \quad
  &&\hspace{-0.3in}
y''^2 \, \, \, = \, \, \, \,
       E(x, \, y, \, y') \cdot \, y'' \, \, + F(x, \, y, \, y'),
\end{eqnarray}
having the interesting feature that {\em movable essential singularities}
and  {\em movable natural boundaries} are known to be possible. Consequently, it is not
garanteed that the non-linear differential equation (\ref{nonlineareq})
can be simply reduced to a sigma-form of Painlev\'e VI or the
SD-I``master Painlev\'e equation''  form\footnote[1]{And if this is the case, one expects quite unpleasant
  B\"acklund correspondences, like (5.19) in~\cite{cosgrove}, to take place.}.

In the present case (\ref{nonlineareq}) this reduction can actually be carried out
by making the (Landen~\cite{Heegner}) substitution
\begin{eqnarray}
  \label{tx}
  \quad   \quad 
k^2 \, =\,\, t \,=\,\, \,  \left(\frac{1 \, -\sqrt{1-x}}{1 \, +\sqrt{1-x}}\right)^2, 
\end{eqnarray}
which is the (compositional) inverse of
\begin{eqnarray}
  \label{inverse}
\quad   \quad  
&&\hspace{-.8in}
x\,\,= \,\,\, \frac{4 k}{(1 \, +k)^2},
\quad   \quad   \quad    \hbox{where:}  \, \, \,\quad
 x \, = \, k_L^2
 \,\,\,   \quad   \hbox{with:}  \quad  \quad  
  k_L \,\, = \, \, \, {{2 \, \sqrt {k}} \over {1 \, +k}},  
\end{eqnarray} 
together with $\,\, \sigma(t) \,  = \, \,  {\tilde \sigma}(x) \,\, $ given by:
\begin{eqnarray}
\label{salah}
  &&\hspace{-.8in} \quad 
 \sigma(t)\, \,  = \, \, \, {\tilde \sigma}(x)
 \nonumber\\
  &&\hspace{-.6in} \quad \quad 
  = \, \,\,\,
  \frac{2}{x} \cdot \, \frac{1 \, \, -\sqrt{1-x}}{1 \, \, +\sqrt{1-x}} \cdot \, 
\Bigl(  h(x) \, \,  \, \, 
   -\frac{M^2 -3N^2 +1}{16}  \, \, \,  \,  +\frac{M^2-N^2+1}{16}   \cdot \,  x
   \nonumber\\
\label{final}
 &&\hspace{-.6in} \quad \quad \quad  \quad \quad \quad \quad \quad \quad \,\,
+ \, \frac{M^2-N^2}{16} \cdot \, x \cdot \,
 \left(\frac{1\,-\sqrt{1-x}}{1\,+\sqrt{1-x}}\right) \Bigr).
\end{eqnarray}
With the previous change of variable (\ref{inverse}) and function transformation (\ref{salah}), 
$ \, h(x) \, $ {\em satisfies the Okamoto non-linear differential equation}
\begin{eqnarray}
  &&\hspace{-.4in} \quad 
     x^2 \cdot \, (x-1)^2 \cdot \, h''^2 \, \, \,
     +4 \, \, h' \cdot \, (x \, h'\, -h) \cdot \, ((x -1) \cdot \, h' \,-h)
     \nonumber\\
\label{okamoto}  
  &&\hspace{-.3in} \quad \quad  \quad 
     +c_7 \cdot \, h'^2 \, \, \, +c_8 \cdot \, (x\, h'\, -h) \, \, \, +c_9 \cdot \, h' \,\, \,  +c_{10}
     \, \,  \, = \,\,\,  \, 0.
\end{eqnarray}
where the prime now indicates a derivative with respect to $ \, x$ and where the $\, c_n$'s read
\begin{eqnarray}
 &&\hspace{-.4in}
 c_7\, =\,\,  -(n_1^2\, +n_2^2\, +n_3^2\, +n_4^2), 
\hspace{.8in} c_8 \, = \, \, -4\, n_1n_2n_3n_4, 
\nonumber\\
&&\hspace{-.4in}
c_9\, =\,\,
  -(n_1^2n_2^2\, +n_1^2n_3^2\, +n_1^2n_4^2\, +n_2^2n_3^2\, +n_2^2n_4^2\, +n_3^2n_4^2\, 
   -2\, n_1n_2n_3n_4), 
\nonumber\\
\label{okaparameters}
 &&\hspace{-.4in}
 c_{10}\, =\,\,
  -(n_1^2n_2^2n_3^2\, +n_1^2n_2^2n_4^2\, +n_1^2n_3^3n_4^2+\, n_2^2n_3^2n_4^2), 
\end{eqnarray}
the four Okamoto parameters being (unique up to permutations and
sign changes of any pair of $ \, n_k$):
\begin{eqnarray}
  \hspace{-1in}
  n_1 =\, \frac{M+N+1}{4}, ~~n_2 =\,   \frac{M+N-1}{4},~~n_3 =\,    \frac{N -M+1}{4},
~~n_4  =\,    \frac{N -M -1}{4}.
\label{parameters}
\end{eqnarray}
The previous Okamoto non-linear differential equation (\ref{okamoto}) can be rewritten:
\begin{eqnarray}
 &&\hspace{-.98in}     \, \, \, \,
   x^2 \cdot \, (x-1)^2 \cdot \, h''^2 \, \, \,\,
   +4 \, \, h' \cdot \, (x \, h'\, -h) \cdot \, ((x -1) \cdot \, h' \,-h) \, \,  \, \, 
    \nonumber\\
\label{okamotorewrit}    
 &&\hspace{-.98in} \quad  \quad  
 - {{ (M^2+N^2+1)} \over { 4}} \cdot \, h'^2 \, \, \, \, - {{ (M^2+N^2+1)^2} \over {64}}  \cdot \, h'
    \\
 &&\hspace{-.98in} \quad   \quad  
    - {{(M+N+1) \cdot \, (M+N-1) \cdot \, (M-N+1) \cdot \, (M-N-1) } \over { 64 }}   \cdot \, (x\, h'\, -h)
\nonumber\\
&&\hspace{-.98in} 
     \,\, \,  -{{(M^6 -M^4\,N^2 -M^2\,N^4 +N^6 \, -M^4 +10\,M^2\,N^2 -N^4 \,-M^2  -N^2 \, +1) } \over {1024 }}
       \, \,  \, = \,\,\,  \, 0.
       \nonumber
\end{eqnarray}
Such a reduction of the non-linear differential equation (\ref{nonlineareq})
to the Okamoto sigma-form of Painlev\'e VI (\ref{okamotorewrit}),
is illustrated in \ref{Reduc} on their respective solutions in $\, t$ and $\, x$, associated with
the two factors of the low-temperature correlation function $\, C(2, \, 3)$.

\vskip .2cm

\subsection{A few remarks on the previous substitutions (\ref{tx}) and  (\ref{salah})}
\label{fewr}

Even if the sum (\ref{additive}) (see also  (\ref{twofactors}),  (\ref{loggpm}), (\ref{additive2})) 
\begin{eqnarray}
\label{additive2}
\hspace{-.46in} \quad  \quad  \quad 
\sigma(M,N;t)  \,\, = \, \,\,  \, \sigma_{+}(M,N;t)\,\, \, +\sigma_{-}(M,N;t), 
\end{eqnarray}
satisfies the  Okamoto sigma form of Painlev\'e VI equation (\ref{eqnmodd}), and
the two $\, \sigma_{\pm}(M,N;t)$, in the right-hand-side of (\ref{additive2}),
verify another non-linear differential equation (\ref{nonlineareq}),
this is {\em far from sufficient} to show that (\ref{nonlineareq}) has the {\em Painlev\'e property},
namely having {\em fixed critical points}~\cite{Handbook,Ince}. To prove
that a  non-linear differential equation like
(\ref{nonlineareq}) actually has fixed critical points remains a quite technical proof. 
We have actually achieved such a  demonstration, but it is too cumbersome to be given here.
Clearly the simplest way to show that (\ref{nonlineareq}) has the Painlev\'e property amounts to
reducing  (\ref{nonlineareq}) to a sigma form of Painlev\'e VI equation, 
finding the change of variables (\ref{tx}) and (\ref{salah}) to perform
this reduction. The non-linear differential equation (\ref{nonlineareq}) is clearly different of an 
Okamoto form (\ref{okamoto}) {\em because of the presence  of a term in} $\, \sigma''$ next to
the term in $\, \sigma''^{\, 2}$: in contrast the Okamoto forms (\ref{okamoto})  have
{\em no  term} in $\, h'' \, $ next to the term in $\,  h''^{\, 2}$.

Finding the  well-suited transformations  (\ref{tx}) and (\ref{salah})
is, however, far from being obvious.  Recalling transformation (\ref{salah})
let us first note  that transformations of the general form 
\begin{eqnarray}
\label{salahgener}
  \hspace{-.98in} \quad \quad \quad \quad \quad \quad \quad \quad  \quad \quad 
  \sigma(t) \, \, = \, \,  \,\,  \,
  \alpha(t) \cdot  h(t)\,\,\, + \beta(t), 
\end{eqnarray}
where $\, \alpha(t)$ and $\, \beta(t)$ are some functions to be found, {\em are not sufficient}
to reduce (\ref{nonlineareq}) into an Okamoto form (\ref{okamoto}), 
and {\em  not even  sufficient to get rid of the  term in} $\, h'' \, $ next to the term in $\,  h''^{\, 2}$.

The (Landen~\cite{Heegner}) change of variable  (\ref{tx}) {\em is in fact crucial to achieve that goal}.
Once one has discovered this key  change of variable  (\ref{tx}) one can, for instance,
seek for  transformations of the form
\begin{eqnarray}
\label{pattern}
&&\hspace{-.98in}   \quad \quad \quad \quad \quad \quad \quad \quad \quad 
  \sigma(t) \,\,  = \, \, \,   \alpha(x) \cdot \,
  \Big( h(x)\,\,  +\beta(x) \Big), 
\end{eqnarray}
where $\, \alpha(x) \, $ and $\, \beta(x) \, $ are arbitrary functions, 
such that one gets {\em no term} in $\, h'' \, $ next to the term in $\,  h''^{\, 2}$,  which is a
first {\em necessary condition} to be an Okamoto sigma form of Painlev\'e VI.
One first finds  that $\, \alpha(x) \, $ must  necessarily be a solution of the following
linear differential equation
\begin{eqnarray}
\label{alphalineq}
&&\hspace{-.98in}  \quad 
\Bigl(\sqrt{1-x} \, \,  +1 -x \Bigr) \cdot \, \alpha(x)
     \, \, \,
+(x \, -1) \cdot \, \Bigl(2 \cdot \, \sqrt{1-x} \, \, +2 \, -x\Bigr)\cdot \, {{ d \alpha(x)} \over {dx}}
     \, \, = \, \, \, \, 0, 
\end{eqnarray}
which has the following solution  
\begin{eqnarray}
\label{alphasol}
  &&\hspace{-.98in} \, \, 
     \alpha(x)  \,  \, = \, \,\, 
    \frac{ \rho}{(1  \, +\sqrt{1-x})^2} 
     \, \,  = \, \, \, 
    \,  \frac{\rho}{x} \cdot \, \frac{1 \, -\sqrt{1-x}}{1  \, +\sqrt{1-x}}
    \, \,  = \, \,\,  {{ \rho} \over {x^2}} \cdot \, \Bigl(2\, -x \,\, -2 \cdot \,   \sqrt{1-x}\Bigr),
\end{eqnarray}
where $\, \rho$ is an arbitrary constant. One finds 
another second order linear ODE which also has (\ref{alphasol}) as a solution,
and another  linear ODE, in $\, \alpha(x)$ and $\, \beta(x)$, of the form
(the $ \,a_n$'s  are simple algebraic expressions of $\, x$ with $\, \sqrt{1\, -x}$):
\begin{eqnarray}
\label{alphabeta}
  &&\hspace{-.98in} \, \quad
     a_0(x) \,  \, +\Bigl(a_1(x) \cdot \, \alpha(x)\, +a_2(x)  \cdot \,  {{ d \alpha(x)} \over {dx}}
   \,  +a_3(x)  \cdot \,  {{ d^2 \alpha(x)} \over {dx^2}} \Bigr) \cdot \, \beta(x)
\nonumber \\
  &&\hspace{-.98in} \,\quad   \,  \, 
     +\Bigl(a_4(x) \cdot \, \alpha(x)\, +a_5(x)  \cdot \,  {{ d \alpha(x)} \over {dx}}  \Bigr)
     \cdot \, {{ d \beta(x)} \over {dx}}
\, \, +a_6(x) \cdot \, \alpha(x)  \cdot \, {{ d^2 \beta(x)} \over {dx^2}}
 \,\, = \, \, \, 0.
\end{eqnarray}
Taking into account (\ref{alphasol}), equation (\ref{alphabeta}) reduces to
\begin{eqnarray}
\label{alphabetareduced}
&&\hspace{-.98in} \quad \quad \quad \quad \quad \quad
16 \cdot \,    \rho   \cdot \, (1\, -x)^{3/2}  \cdot \,  {{ d^2 \beta(x)} \over {dx^2}}
 \,\,\, + (N^2-M^2)  \,\, = \, \, \, 0,   
\end{eqnarray}
yielding  the following expression for $\, \beta(x)$
\begin{eqnarray}
\label{alphabetareduced}
&&\hspace{-.98in} \quad \quad  \, \, \quad
\beta(x) \, \,\, = \, \, \, \,
-{{ M^2 -N ^2} \over {4 \cdot \,  \rho}} \cdot \, \sqrt{1\, -x}
\,  \, \, \,\, \,  + \alpha_0 \cdot \, x \, \, + \beta_0
   \\
&&\hspace{-.98in} \quad \ \quad  \quad \quad
\,\, = \, \, \, \,
{{M^2 -N^2} \over{8 \cdot \, \rho }} \cdot \, x \cdot \,  \frac{1  \, -\sqrt{1-x}}{1  \, +\sqrt{1-x}} 
\, \, \,\,   \, \, \,\,  + \Bigl( \alpha_0 +{{ 1} \over {4 \cdot \, \rho}} \Bigr) \cdot \, x
\, \,  \, \, \, + \Bigl(\beta_0 - {{ 1} \over {4 \cdot \, \rho}}\Bigr),
\nonumber
\end{eqnarray}
where $\, \alpha_0\,$ and $\, \beta_0\,$ are arbitrary constants.
This  yields to the following form\footnote[1]{To be compared with (\ref{salah}).}  
\begin{eqnarray}
\label{salahalphabeta}
&&\hspace{-.99in} 
\sigma(t)  = \,   \frac{2 }{x} \cdot \, \frac{1 \, -\sqrt{1-x}}{1  \, +\sqrt{1-x}} \cdot \,
\Big( \gamma \cdot \, h(x)\, 
 +\alpha \,  +\beta \cdot  x   \, 
+{{M^2 -N^2} \over{16 }} \cdot \, x \cdot \, \frac{1  \, -\sqrt{1-x}}{1  \, +\sqrt{1-x}} \Big), 
\end{eqnarray}
where $\, \gamma \, = \,\rho/2\, $ and where  $\, \alpha \, $ and $\, \beta \, $ are arbitrary constants.
This form is such that one reduces to 
an Okamoto form (\ref{okamoto}) up to the usual
$\, h(x) \, \rightarrow \, \, \gamma \cdot \,  h(x) \,   +\alpha \,+ \, \beta \cdot \, x \, \, $
transformations. To sum-up (\ref{salah}) can be deduced from (\ref{tx}).

The main question is how to discover the  key (Landen) change of variable (\ref{tx})?
Assuming that the non-linear differential equation (\ref{nonlineareq})
has the Painlev\'e property, one can probably assume, because of the explicit form
of (\ref{nonlineareq}), that its critical points are the three points $\, 0, \, 1, \, \infty$.
Consequently, a change of variable to reduce (\ref{nonlineareq}) to an  Okamoto form (\ref{okamoto}),
must map the three critical points $\, 0, \, 1, \, \infty$
of (\ref{nonlineareq}) onto the three critical points  $\, 0, \, 1, \, \infty \, $
of an Okamoto form (\ref{okamoto}). Unfortunately this condition, reminiscent of {\em Belyi maps}~\cite{Belyi},
is {\em not} sufficient enough to actually discover the well-suited change of variable  (\ref{tx}).
At this step it is worth recalling that Painlev\'e VI functions
{\em can be seen as deformations of elliptic functions}~\cite{Manin}
(see  also \ref{appManin}). A {\em lattice of periods}
is canonically attached to elliptic functions. 
If a change of variable maps a non-linear differential equation of the Painlev\'e type
onto another non-linear differential equation of the Painlev\'e type, namely (\ref{nonlineareq})
onto an Okamoto form (\ref{okamoto}), {\em it must map the two lattice of periods}  of the 
underlying elliptic functions. In other words the  change of variable  {\em  must correspond to a quite
selected and rigid set of transformations}: it has to be a {\em modular correspondence}~\cite{Modular}.
These are {\em algebraic transformations} given by the so-called
{\em modular equations}\footnote[1]{We must consider {\em modular curves} associated with modular forms. One excludes
Shimura curves associated with automorphic forms~\cite{Heun}.}~\cite{Modular,Maier}. The Landen (or inverse Landen) 
transformation is the simplest example of {\em  modular correspondence}.
In \ref{appManin} we recall Manin's viewpoint showing explicitely that Painlev\'e VI functions
can be seen as  {\em  deformations of elliptic functions}, and also underlying the
{\em Landen transformation as a symmetry  in the family of  Painlev\'e VI equations}. The crucial role
of other modular correspondences
for Painlev\'e VI equations is also underlined in \ref{deformManin}. 
 
\vskip .2cm

\subsection{Completing the characterization of the factors $ \, g_{\pm}$.}
\label{completing}

Both of the sigma functions $ \, \sigma_{\pm}$ satisfy the same  Painlev{\'e}-type
non-linear differential equation (\ref{nonlineareq}). To complete the characterization
of the factors $ \, g_{\pm}$ we need to obtain the boundary conditions on the equations
for $  \, \sigma_{\pm}$ which allow (homogeneous) polynomial in $ \, {\tilde K}(k)$ and $ \, {\tilde E}(k)$
factors to occur.
By direct substitution in  (\ref{salah}) we see that
\begin{eqnarray}
  &&\hspace{-.4in} \quad \quad \quad 
     h_0(x) \, \, = \, \, \,
     \frac{M^2 -3N^2 +1}{16}  \, \, \,\, \,  \,   -\frac{M^2-N^2+1}{16}  \cdot \, x
 \nonumber\\
\label{h0exact}
&&\hspace{0.1in}  \quad \quad \quad  \,  \, 
-\, \frac{M^2-N^2}{16} \cdot \, x \cdot \, \left(\frac{1\,\,  -\sqrt{1-x}}{1 \, \, +\sqrt{1-x}}\right), 
\end{eqnarray}
is {\em an exact solution of} (\ref{okamoto}) with Okamoto parameters
(\ref{parameters}). This exact {\em  algebraic} solution of  (\ref{okamoto})  precisely corresponds
to the exact solution $\, \sigma \, = \, 0 \, $ of (\ref{nonlineareq}).  
This algebraic function is in fact of the form
\begin{eqnarray}
  &&\hspace{-.4in} \quad \quad \quad \quad 
     h_0(x)\, \, = \, \, \, x \cdot \, (x\, -1) \cdot {{d \ln({\cal H}_0(x))} \over {dx}}, 
\end{eqnarray}
where $\, {\cal H}_0(x)\, $ is {\em  an algebraic function}:
\begin{eqnarray}
  &&\hspace{-.98in} 
{\cal H}_0(x)  \, \, = \, \, \, 
\\
 &&\hspace{-.99in} \, \, \, \,  
\Bigl(1   -\sqrt{1-x}\Bigr)^{-(M^2-3\,N^2+1)/16} \cdot 
\Bigl(1  +\sqrt{1-x}\Bigr)^{(3\,M^2-N^2-1)/16} \cdot  (1 -x)^{-(M^2+N^2)/16}.
\nonumber 
\end{eqnarray}
Thus we may write:
\begin{eqnarray}
  \quad   \quad 
h(x) \,  \, \, = \,   \, \,  H(x) \, \, \,\,+ h_0(x). 
\end{eqnarray}
We need only the power series solutions of (\ref{okamoto}) which are analytic at
$ \, x= \, 0$:
\begin{eqnarray}
  \label{series}
\quad   \quad 
h(x) \,\,  = \, \, \,  \sum_{n=0} \,  \,a_n \, \, x^n.
\end{eqnarray} 
We found, in Appendix D of~\cite{bmm}, that there are, in general, 
four classes of these analytic solutions which are related by 
changing the signs of any pair of $ \, n_k$. For the present purpose, we 
need the class 4 solutions given by (D.7), (D.11) and (D.15)
\begin{eqnarray}
&&\hspace{-1.01in} \quad
a^{(4)}_0\, \, = \,\,\,
{{-n_1n_2-n_3n_4 \, -(n_1+n_2) \cdot \, (n_3+n_4) } \over { 2}},
\nonumber \\
&&\hspace{-1.01in} \quad
a^{(4)}_1 \, =\,\, \,
\frac{(n_1+n_2) \cdot \, n_3n_4 \, \, \, +(n_3+n_4) \cdot \, n_1n_2}{n_1 +n_2 + n_3 +n_4},
\\
&&\hspace{-1.11in} \quad \quad
a^{(4)}_2 = \,
\frac{(n_1+n_2) \cdot \, (n_1+n_3) \cdot \,(n_1+n_4) \cdot \,(n_2+n_3) \cdot \,
(n_2+n_4) \cdot \,(n_3+n_4)}
{(n_1+n_2+n_3+n_4)^2 \cdot \, (n_1+n_2+n_3+n_4+1)(n_1+n_2+n_3+n_4-1)},
\nonumber
\end{eqnarray}
which, with the Okamoto parameters (\ref{parameters}), read:
\begin{eqnarray}
\label{a0}
&&\hspace{-.6in}\quad 
a^{(4)}_0\, =\, \, \frac{1}{16} \cdot \, (M^2-3N^2+1), \quad \quad
a^{(4)}_1\, =\,\, -\frac{1}{16} \cdot \, (M^2-N^2+1),
\nonumber \\
\label{a2}
&& \quad \quad
a^{(4)}_2\, = \,\, \, \frac{1}{64} \cdot \, (M^2-N^2).
\end{eqnarray}
These agree with the expansion at $\, x=\, 0 \, $ of (\ref{h0exact}).
Because
\begin{eqnarray}
  \hspace{-.3in}
  \quad \quad \quad 
n_1\, +n_2\, +n_3\, +n_4 \, \, = \, \,\, N, 
\end{eqnarray}
we see, from (D.23) of~\cite{bmm}, that $ \, a^{(4)}_{N+1}$, the coefficient
of $ \, x^{N+1}$, is an arbitrary constant.
To proceed further we extend the recursive analysis
of~\cite{bmm} (see appendix D in~\cite{bmm})
beyond the term $ \, x^{N+1}$. We find that the coefficients
of $ \, x^n$ for $ \, (N+1) \, \leq \, n \, \leq \, (2 \, N \, +1)\, $ depend only on $ \, c_{N+1}$,
the coefficient of $\, x^{N+1}\, $ but that starting with $\, c_{2(N+1)}\, $ the
coefficients depend on $ \, c_{N+1}^2\, $ as well as $ \, c_{N+1}$. 
Continuing in this fashion we obtain the form (\ref{lambdasigmapm})
\begin{eqnarray}
  \label{lambdasigmapmbis}
 \hspace{-.96in} \quad  \quad  \quad    \quad  \quad  \quad 
\sigma_{\pm}(M,N;t; \, \lambda_{\pm})  
  \, \,  =\, \, \,\, \,  
  \sum_{n=1}^{\infty} \, \Bigl(\lambda_{\pm} \cdot \, t^{(N+1)/2}\Bigr)^n \cdot \, B_n(M,N;t), 
\end{eqnarray}
where the $\, B_n(M,N;t)$'s are
power series
of $\, t$,  analytic at $\, t \, = \, 0$, such that:
\begin{eqnarray}
\label{hn}
  \hspace{-.96in} \quad  \quad  \quad \quad  \quad
  \Bigl( {{ N\, +1} \over {2}}  \Bigr)^{n-1} \cdot \,  B_n(M,N;t)  \,  \, \,  =\, \, \,\, \,
  1 \, \, \, \,  + \, o(t).
\end{eqnarray}
We must choose 
\begin{eqnarray}
  \label{lmbdapmlambda}
  \quad  \quad \quad  \quad \quad  \quad 
\lambda_{+}\,\, = \,\,\,-\lambda_{-}\,=\,\,\,\lambda, 
\end{eqnarray}
in order to match with the {\em lambda extension} solutions of (\ref{eqnmodd}):
\begin{eqnarray}
  \label{sigmapm}
\hspace{-.66in} \quad \quad  \quad   \quad 
  \sigma(M,N;t; \, \lambda)\, \,  = \, \, \, \, \, 
    \, 2 \cdot \, \sum_{n=1}^{\infty} \,  \Bigl(\lambda^2 \cdot \, t^{N+1}\Bigr)^n \cdot \, B_{2\, n}(M,N;t).
\end{eqnarray}
However, these lambda extensions of $\, \sigma_{\pm}(M,N;t; \lambda_{\pm})$
{\em do not in general have a representation as homogenerous polynomials}  in 
$\, {\tilde K}$ and $ \, {\tilde E}$ for the corresponding $\, g_{\pm}(M,N;t; \lambda_{\pm})$.
We note in particular that $\, B_1$ reads 
\begin{eqnarray}
\label{htilde1} \quad  \quad 
  B_1\,\, =\,\, \,
  {}_2F_1\Bigl([\frac{N+M}{2},\frac{N-M}{2}], \, [N+1], \, t\Bigr), 
\end{eqnarray}
which may be conjectured from the expansions in  \ref{appB}, and proven
by the recursive procedure outlined in section 3.
A step further one can find that:
\begin{eqnarray}
  \label{htilde1bis}
&& \hspace{-.98in}     \quad  \quad  \, 
   {{ N\, +1} \over {2}}  \cdot \,  B_2
   \nonumber \\
  && \hspace{-.96in}   \quad  \quad  \quad
     \,\, = \,\, \,  {{ N^2 \, -M^2 } \over {4  \cdot \, (N \, +1) }} \cdot \, t  \cdot \,  (t\, -1) \cdot \, 
   {}_2F_1\Bigl([\frac{N+M+2}{2},\frac{N-M+2}{2}], \, [N+2], \, t\Bigr)^2 \, \,
  \nonumber \\
 && \hspace{-.96in}   \quad  \quad  \quad  \quad 
    +(N \, +1)  \cdot \, {}_2F_1\Bigl([\frac{N+M}{2},\frac{N-M}{2}], \, [N +1], \, t\Bigr)^2
  \\
 && \hspace{-.96in}   \quad  \quad  \quad  \quad  \quad  \quad \, \,
    + N \, \cdot \,  (t\, -1)  \cdot \, {}_2F_1\Bigl([\frac{N+M}{2},\frac{N-M}{2}], \, [N +1], \, t\Bigr)
  \nonumber \\
  && \hspace{-.96in}   \quad  \quad  \quad  \quad  \quad  \quad  \quad  \quad \, \,
     \times \, {}_2F_1\Bigl([\frac{N+M+2}{2},\frac{N-M+2}{2}], \, [N+2], \, t\Bigr).
    \nonumber
\end{eqnarray}

\vskip .1cm 

\subsection{Selected values of $\, \lambda$  for which $ \, g_{\pm}(M,N;t;\lambda) \, $
  reduce to polynomials in  $ \, {\tilde K}(k)$ and $ \, {\tilde E}(k)$.}
\label{SelecteD}

To complete the illustration of the factorization of the Toeplitz 
determinant for $ \, C(M,N; \, t)$, we need to determine the value of $ \, \lambda \, $
for which the (differentially algebraic) lambda extension $ \, g_{\pm}(M,N;t;\lambda_{\pm})$  reduces to a determinant of
{\em finite-dimensional matrices}\footnote[1]{In contrast with Fredholm determinants. For generic values
of $ \, \lambda$ the lambda extensions of $ \, C(M,N;\, t; \, \lambda) \, $ are Fredholm determinants.}.
In table 1, we list the coefficients of the terms $\, \pm \, k^{N+1}$ and
$ \, k^{2 \, (N+1)}$ in $ \, \sigma_{\pm}(M,N) \, $ for some low values of $ \, M$ and $ \, N$. 
\begin{table}[ht!]
\center
\caption{Coefficients of $\pm k^{N+1}$ and $k^{2(N+1)}$ in $\sigma_{\pm}(M,N)$} 
\begin{tabular}{|l|c|c|}\hline
$M,N$& $\pm k^{N+1}$ coefficient&$k^{2(N+1)}$ coefficient\\ \hline
1,~2&$\frac{3}{2^5}=\frac{3}{2}\left(\frac{1}{2^4}\right)$
&$\frac{3}{2^9}=\frac{3}{2}\left(\frac{1}{2^4}\right)^2$\\
1,~4&$\frac{3\cdot 5}{2^9}=\frac{5}{2}\left(\frac{3}{2^8}\right)$
&$\frac{3^2\cdot 5}{2^{17}}=\frac{5}{2}\left(\frac{3}{2^8}\right)^2$\\
3,~4&$\frac{5\cdot 7}{2^9}=\frac{5}{2}\left(\frac{7}{2^8}\right)$
&$\frac{5\cdot 7^2}{2^{17}}=\frac{5}{2}\left(\frac{7}{2^8}\right)^2$\\
1,~6&$\frac{5 \cdot 7}{2^{12}}=\frac{7}{2}\left(\frac{5}{2^{11}}\right)$
&$\frac{5^2\cdot 7}{2^{23}}=\frac{7}{2}\left(\frac{5}{2^{11}}\right)^2$\\
3,~6&$\frac{7\cdot 9}{2^{12}}=\frac{7}{2}\left(\frac{9}{2^{11}}\right)$
&$\frac{7\cdot 9^2}{2^{23}}=\frac{7}{2}\left(\frac{9}{2^{11}}\right)^2$\\
5,~6&$\frac{3\cdot 7 \cdot 11}{2^{12}}
=\frac{7}{2}\left(\frac{3\cdot 11}{2^{11}}\right)$
&$\frac{3^2 \cdot 7 \cdot 11^2}{2^{23}}
=\frac{7}{2}\left(\frac{3 \cdot 11}{2^{11}}\right)^2$\\
\hline \hline
2,~3&$\frac{5}{2^6}=2\left(\frac{5}{2^7}\right)$
&$\frac{5^2}{2^{13}}=2\left(\frac{5}{2^7}\right)^2$\\
2,~5&$\frac{3\cdot 7}{2^{10}}=3\left(\frac{7}{2^{10}}\right)$
&$\frac{3 \cdot 7^2}{2^{20}}=3\left(\frac{7}{2^{10}}\right)^2$\\
4,~5&$\frac{3^2\cdot 7}{2^{10}}=3\left(\frac{3\cdot 7}{2^{10}}\right)$
&$\frac{3^3\cdot 7^2}{2^{20}}=3\left(\frac{3\cdot 7}{2^{10}}\right)^2$\\
2,~7&$\frac{5\cdot 9}{2^{13}}=4\left(\frac{5 \cdot 9}{2^{15}}\right)$
&$\frac{3^4\dot 5^2}{2^{28}}=4\left(\frac{5\cdot 9}{2^{15}}\right)^2$\\
4,~7&$\frac{9\cdot 11}{2^{13}}=4\left(\frac{11 \cdot 9}{2^{15}}\right)$
&$\frac{11^2 \cdot 9^2}{2^{28}}=4\left(\frac{11\cdot 9}{2^{15}}\right)^2$\\
\hline
\end{tabular}
\end{table}
From this table we see that the coefficients of $ \, \pm  \, k^{N+1}$,
 and the coefficients of $\,k^{2(N+1)}\,$ (or $ \, \pm \,  t^{(N+1)/2}\, $ and  $\,t^{N+1}\,$
 in (\ref{lambdasigmapm})), have respectively the form
\begin{eqnarray}
\label{coeff1}
\quad  \quad 
\frac{N+1}{2} \cdot \,  \alpha_{M,N}, 
  \quad   \quad  \quad  \quad 
\frac{N+1}{2} \cdot \,  \alpha^2_{M,N}.
\end{eqnarray}
where\footnote[5]{A demonstration of this result which requires the introduction of Schlesinger's transformations  will not be given here.}
\begin{eqnarray}
\label{alphaMN}
\hspace{-1.01in}
\quad \quad \quad 
\alpha_{M,N} \, \, \,= \, \, \,\,\,
\frac{(N+M)! \cdot \, (N-M)!}{2^{2N}
 \cdot \, (N+1)! \cdot \, ((N+M-1)/2)! \cdot \, ((N-M-1)/2)!}.
\end{eqnarray}
The selected values of $\,\,  \lambda \, \, = \, \, \lambda_{+} \, $ read:
\begin{eqnarray}
\label{coeff1}
\quad  \quad 
  \lambda_{+} \, \, = \, \,   \lambda \, \, = \, \, \, \frac{N+1}{2} \cdot \,  \alpha_{M,N}. 
\end{eqnarray}
  
\vskip .1cm

\section{Tracy-Widom viewpoint}
\label{TracyWidom}

Recalling the Tracy-Widom paper~\cite{tw} we introduce, besides
the sum (\ref{additive}), the {\em difference}: 
\begin{eqnarray}
\label{difference}
&&\hspace{-.56in} \quad \quad  \quad \quad 
   \, \delta(M,N;t)\,\, = \, \,\,  \,
      \sigma_{+}(M,N;t)\,\, \, -\sigma_{-}(M,N;t). 
\end{eqnarray}
In this section we simply denote the difference $\, \delta(M,N;t)$ by $\, \delta$, and the sum
$\, \sigma(M,N;t) \, = \, \,  \sigma_{+}(M,N;t)\,\, \, +\sigma_{-}(M,N;t) \, $ by $\, \sigma$.
One has the following non-trivial relation\footnote[2]{One can easily verify this relation for the
  two factors of $\, C(2, \, 3)$ (see (\ref{g23t}) below).} between the sum
(\ref{additive}) and this difference  (\ref{difference})
\begin{eqnarray}
\label{differencerelation}
&&\hspace{-.96in} \quad \quad \quad \quad \quad \quad \quad
   \, \delta^2 \,\, \, + \, 
   t \cdot \, (t-1) \cdot \, \sigma' \,   \,\, \, -t \cdot \, \sigma
   \,  \,\, = \, \,\,  \, 0.
\end{eqnarray}
This relation, independent of $\, M$ and $\, N$, can easily be
obtained\footnote[9]{In fact this relation was obtained as the first of
the two equations of a B\"acklund transformation, but for the simplicity
of the presentation we will not give such transformations and
other Schlesinger transformations.}
by guessing from the series expansions of $\, \delta$ and $\, \sigma$
for various values of $\, M$ and $\, N$. 

Relation (\ref{differencerelation}) also  yields
\begin{eqnarray}
\label{yields}
 &&\hspace{-.96in}  \quad  \quad  \quad  \quad  \quad  \quad  \quad 
 \sigma'' \,\,\,  = \, \,\,  \,
{{\delta^2 } \over { t^2 \cdot \, (t\, -1)}} \, \, \,\,
    - 2 \cdot \,  {{\delta \cdot \, \delta'} \over { t \cdot \, (t\, -1)}},
\end{eqnarray}
or:
\begin{eqnarray}
\label{yields2}  \quad  \quad  \quad  \quad  \quad  \quad  \quad 
&&\hspace{-.96in}
\sigma \,\,  = \, \,\,
-(t-1) \cdot \, \int^{t} \, {{\delta^2} \over {(t-1)^2 \cdot \, t }} \cdot \, dt. 
\end{eqnarray}
Relation (\ref{differencerelation}) is the generalization\footnote[8]{We have
carried out the limiting contraction of relation (\ref{differencerelation})
for  Painlev\'e VI to relation (\ref{TracyWidomPV}) for Painlev\'e V, but
we will not give these calculations here.} to
Painlev\'e VI of the Tracy and Widom
relation (82)  in~\cite{tw} associated with Painlev\'e V, which reads:
\begin{eqnarray}
\label{TracyWidomPV}
&&\hspace{-.96in} \quad \quad \quad  \quad \quad \quad \quad \quad \quad
   \, \delta^2 \,\, \, \, + 
   t  \cdot \, \sigma'  \,\, \,\, - \sigma
    \,\, \,= \, \,\, \, \, 0. 
\end{eqnarray}
Using (\ref{differencerelation}) and (\ref{yields}) one can eliminate
$\, \sigma'$ and  $\, \sigma''$ in the Okamoto
relation (\ref{eqnmodd}), and  deduce\footnote[1]{Again one can easily
  verify this relation for the
  two factors of $\, C(2, \, 3)$ (see (\ref{g23t}) below).}: 
\begin{eqnarray}
\label{deduces}
&&\hspace{-.98in}
\sigma  \,= \, \,
- {{ t^2 \cdot \, (t \, -1)^2 } \over { t \, +1}} \cdot \,  {{\delta'' } \over {\delta }}
\, +  2 \cdot \, {{ \delta^2} \over {t \, +1 }}
\,  + {{ (t\, -1) \cdot  \, \Bigl((M^2 \, -1) \cdot \, t \,
 -(N^2\, -1)\Bigr)} \over {4 \cdot \, (t \, +1) }}.
\end{eqnarray}
Furthermore, using Pantone's program one can first find that the {\em difference}
(\ref{difference}) actually satisfies an {\em order-three}
non-linear differential equation:
\begin{eqnarray}
\label{orderthree}
&&\hspace{-.98in}  \quad \quad 
4 \cdot \, t^3 \cdot \,  (t-1)^2 \cdot \, (t+1) \cdot \, \delta  \cdot \, \delta'''
\nonumber \\
&&\hspace{-.96in}  \quad  \quad \quad 
   +4 \cdot \, t^2 \cdot \, (t-1) \cdot \,
   \Bigl(2 \cdot \, (t^2+t-1) \cdot \, \delta \, \,
   -t \cdot \, (t^2-1) \cdot \, \delta' \Bigr) \cdot \, \delta''
  \nonumber \\
  &&\hspace{-.96in}   \quad  \quad  \quad \quad  \, \,
  \, \,  -16 \cdot \, t \cdot \, (t+1) \cdot \, \delta^3 \cdot \, \delta'
     \\
   &&\hspace{-.96in}  \quad  \quad \quad \quad \quad   \, \,
     \, \, 
     +4 \cdot \, (3\,t \, +1) \cdot \, \delta^4 \, \, \,
     -(M^2+N^2-2)\cdot \, t \cdot \, (t-1)\cdot \, \delta^2 \, \, = \, \, \, 0.
      \nonumber 
\end{eqnarray}
Injecting the expression (\ref{deduces}) of $\, \sigma\,$
in terms of $\, \delta\,$ and $\, \delta''\,$ 
in the Okamoto relation (\ref{eqnmodd}), one find a non-linear ODE on $\, \delta$
of {\em order four}. One can use the order three non-linear ODE (\ref{orderthree})
to express $\, \delta'''$ in terms of  $\, \delta$, $\, \delta'$, $\, \delta''$,
but also the fourth derivative  $\, \delta^{(4)}$ in terms of
$\, \delta$, $\, \delta'$, $\, \delta''$.  Injecting these expressions
of  $\, \delta^{(4)}$ and  $\, \delta'''$ 
in the previous {\em order four} non-linear ODE, one finally finds\footnote[5]{Note that this
order-two non-linear ODE (\ref{ordertwoquadra}) could have been 
obtained directly using Pantone's program,
but this requires many more coefficients of the power series of $\, \delta$
to be found (1600 coefficients in $\, k$).} the
{\em order-two} non-linear ODE (quite similar to (\ref{nonlineareq}))
\begin{eqnarray}
\label{ordertwoquadra}
&&\hspace{-.98in}  
16 \cdot \, t^5  \cdot \, (t\, -1)^2 \cdot \, \delta''^{\, 2}   \, \, \, 
+4 \cdot \, t^2  \cdot \, (t\, -1)^2 \cdot \, 
\Bigl( 4 \cdot \, \delta^2 \, \, \, -(M^2 +N^2 -2) \cdot \, t\Bigr) \cdot \,\delta \cdot \, \delta''
\nonumber \\
&&\hspace{-.98in}   \quad 
-16 \cdot \, t \cdot \, (t\, +1)^2 \cdot \,
\Bigl(t \cdot \,  \delta' \, - \delta\Bigr) \cdot \, \delta^2 \cdot \, \delta' 
 \, \, \,   \, \, -16 \cdot \, \delta^6 \, \, \, \, \, 
+8 \cdot \,t \cdot \,(M^2+N^2-2) \cdot \, \delta^4
\nonumber \\
&&\hspace{-.98in} \quad 
\, \, +t \cdot \, \Bigl((M^2 -1) \cdot \, t \, \, -(N^2\, -1)\Bigr) \cdot \,
\Bigl((N^2 \, -1) \cdot \, t \, \, -(M^2 -1) \Bigr) \cdot \, \delta^2
\, \, = \, \, \, 0, 
\end{eqnarray}
which is {\em not} of
the SD-I ``master of Painlev\'e form'' (\ref{cosgrove}).

Let us denote  the LHS of the order-three non-linear ODE (\ref{orderthree}) by $\, {\cal R}_3$, 
and  the LHS of the order-two non-linear ODE (\ref{ordertwoquadra}) by $\, {\cal R}_2$.
We have the following relation:
\begin{eqnarray}
\label{relation}
&&\hspace{-.98in}   \quad \quad \, \, \, 
   \Bigl(8 \cdot \, t^3 \cdot \,\delta'' \,  \, \,
+ \Bigl( 4 \cdot \, \delta^2 \, -(M^2 +N^2 -2) \cdot \, t\Bigr) \cdot \, \delta \cdot \,\Bigr)
   \cdot \, {\cal R}_3
\nonumber \\
&&\hspace{-.98in}  \quad \, \, \, \, \, \,  \quad
   \, \, = \, \, \, \,
t \cdot \, (t \, +1) \cdot \, \delta  \cdot \, {{ d {\cal R}_2 } \over { dt}}
\, \, \, 
-\Bigl(2 \cdot \, t  \cdot \,(t+1)  \cdot \,  \delta' \, \,
+(3\,t \,+1) \cdot \, \delta \Bigr) \cdot \, {\cal R}_2.
\end{eqnarray}

Similar to what has been performed in section \ref{nonlin}, one would like to find the
change of variable, and function transformation, enabling the
reduction of  the order-two non-linear ODE (\ref{ordertwoquadra})
to an Okamoto sigma-form of Painlev\'e VI. Again one notes (see (\ref{pattern})) that a transformation
of the form $\, \delta(t) \, = \, \alpha(t) \cdot \, h(t) \, + \, \beta(t) \, $ is {\em not} sufficient to
get rid of the $\, h'' \, $ term next to the $\, h''^{\, 2} \, $ term. One does need
to find a change of variable like the Landen transformation (\ref{tx}). 

Another (simpler) route amounts to saying that the
Tracy-Widom-like transformation (\ref{differencerelation})
will change the second order non-linear ODE (\ref{ordertwoquadra})
into a third-order  non-linear ODE in $\, \sigma$,
$\, {\cal S}_3 \, = \, 0$, 
that will eventually reduce to (\ref{eqnmodd}) because of the compatibility
of all these equations. 
Let us write (\ref{eqnmodd}) as $\, {\cal S}_2 \, = \, 0$,
we actually have the following compatibility relation\footnote[1]{In fact, stricto sensu,
stating $\, {\cal S}_3 \, = \, 0$ does not imply $\, {\cal S}_2 \, = \, 0$ by 
relation (\ref{followingx}). We have, here with (\ref{followingx}), just
a compatibility relation not a reduction
of  $\, {\cal S}_3 \, = \, 0$ to $\, {\cal S}_2 \, = \, 0$.}:
\begin{eqnarray}
  \label{followingx}
 &&\hspace{-.98in}
  \sigma''^{\, 2}  \cdot \,  {\cal S}_3 \,\, \, = \,\,\,
  \, \, t^5 \cdot \, \Bigl((t-1) \cdot \, \sigma'  \,
  -\sigma\Bigr)^2 \cdot \Bigl({{ d {\cal S}_2} \over {dt}}\Bigr)^2
    \, \, +\sigma'' \cdot \, t^5 \cdot \, (t-1)^2 \cdot \, {\cal S}_2^2
  \nonumber \\
 &&\hspace{-.98in}
  \, \, +\sigma'' \cdot \, t^5  \cdot \, (t+1) \cdot \,
    \Bigl((t-1) \cdot \, \sigma'  \, -\sigma\Bigr)^3 \cdot \, \Bigl(M^2-N^2 \, \, - 4 \cdot \,
    \Bigl((t+1) \cdot \,  \sigma'  \, -\sigma\Bigr) \Bigr) \cdot \, {{ d {\cal S}_2} \over {dt}}
    \nonumber \\
 &&\hspace{-.98in}
  \, \,  -\sigma'' \cdot \, t^5  \cdot \, \Bigl((t -1) \cdot \, \sigma'  \, -\sigma\Bigr) \cdot \,
    \Bigl(  2 \cdot \, (t-1)    \cdot \,   {{ d {\cal S}_2} \over {dt}} \, \,
    +\sigma'' \cdot \,   (t-1)    \cdot \,  \Bigl((t -1) \cdot \, \sigma'  \, -\sigma\Bigr)
 \nonumber \\
 &&\hspace{-.98in} \quad \quad  \quad 
    \times \, 
    \Bigl( 8 \cdot \, t \cdot \sigma \,  \,
    -8 \cdot \, (t^2 \, -1) \cdot \, \sigma' \, \,  +(t - 1) \cdot \, (M^2 \, -N^2) \Bigr) 
  \Bigr)
  \cdot \, {\cal S}_2.
\end{eqnarray}

\vskip .1cm

\section{Boundary conditions for the four 
factors of $ \, C(0,N)$ with $ \, N$ odd}
\label{bound}

In~\cite{bmm}, we discovered that $ \, C(0,N)$ with $ \, N$ odd and $ \, k= \, -\nu$, 
in the low-temperature regime, 
{\em factors into four factors} instead of two. The four factors for $ \, C(0,N)\, $
were presented as
\begin{eqnarray}
  \label{ffactors}
\hspace{-.3in}
  C(0,N) \, \, \, = \, \,  \,\, 
  {\rm constant}  \cdot  \, (1-t)^{1/2} \cdot  \, t^{(1-N^2)/4 }\cdot \,  f_1f_2f_3f_4, 
\end{eqnarray}
where the factors $ \, f_j$ all vanish at $ \,t= \, 0 \,$ in such a way to cancel
the factor $ \, t^{(1-N^2)/4}$.
Here again we change the factors $\, f_i$ in (\ref{ffactors}) in such a way 
to extract a factor of $ \, (1-t)^{1/4} \, $
which is the limiting
behavior of $ \, C(0,N)$ as $ \, N \, \rightarrow \, \infty$,
and we impose the condition that
{\em the four new factors satisfy the same non-linear differential equation}.
The previous factorization (\ref{ffactors}) in
{\em four} factors\footnote[1]{Examples of $\,  g_1(0,N)$'s
  for $ \, C(0,5)$ and $ \, C(0,7)$ are given in \ref{appC}.}  now reads:
\begin{eqnarray}
  \label{fourfactors}
\hspace{-.5in}
  (1\, -t)^{-1/4} \cdot \, C(0,N) \, \,   \, = \, \,  \, \,
  g_1(0,N) \cdot \,   g_2(0,N) \cdot  \,  g_3(0,N) \cdot \,  g_4(0,N).
\end{eqnarray}
If one defines
\begin{eqnarray}
  \label{sigmadef}
  \quad \quad \quad 
\sigma_j \, \,  = \, \, \, t  \cdot \, (t-1)  \cdot \, \frac{d\ln g_j(t)}{dt}, 
\end{eqnarray}
the previous factorization (\ref{fourfactors})  in four factors  becomes
an {\em additivity property} of the corresponding $\, \sigma_i$'s:
\begin{eqnarray}
\label{sigmaaddfour}
\hspace{-.8in} \quad \quad \quad
  \sigma(0, \, N) \, \, \, = \, \, \, \,\, 
   \sigma_1(0, \, N) \, \,  \,  +\sigma_2(0, \, N) \, \, \, 
      +\sigma_3(0, \, N) \, \, \,   +\sigma_4(0, \, N).  
\end{eqnarray}
In~\cite{bmm}, we showed that the sigma's, associated with the
four  factors $\, f_j$ in (\ref{ffactors}),
satisfy  Okamoto sigma-form of Painlev\'e VI equations (\ref{okamoto})
with  the {\em same} Okamoto parameters $ \, n_i$
(unique up to permutations and sign changes of any pair) 
\begin{eqnarray}
  \label{4factorparams}
\hspace{-.3in} \quad 
  n_1= \, \frac{N+1}{4}, \quad \quad
  n_2= \, \frac{N-1}{4}, \quad \quad
  n_3= \, -\frac{1}{2}, \quad \quad n_4= \, 0, 
\end{eqnarray}
which specializes to
\begin{eqnarray}
  &&\hspace{-.69in}
t^2 \cdot \, (t \, -1)^2  \cdot \, h''^2 \,\, \,  \,\,
+4\,h' \cdot \, (t \cdot \, h' \, -h) \cdot \, \Bigl((t  -1) \cdot \, h' \, -h\Bigr)
\nonumber\\
\label{4okamoto}
&&\hspace{-.55in}
-\frac{1}{8} \cdot \, (N^2+3) \cdot \, h'^2 \,\,\, -\frac{1}{2^8} \cdot \, (N^2+3)^2 \cdot \, h'
\, \, \,
   -\frac{1}{2^{10}}\cdot \,  (N^2-1)^2
   \, \,  \, = \, \, \,  \,\, 0, 
\end{eqnarray}
where four functions  $\, h_j$ are solutions of (\ref{4okamoto}), 
and are related to $ \, t \, (t-1) \, df_j/dt\, $ by
(153)-(156) of~\cite{bmm}:
\begin{eqnarray}
\label{h1}
&&\hspace{-.4in}
h_1\, \, =\, \,\,  t\cdot   \, (t\, -1) \cdot \, \frac{d\ln f_1}{dt}
\, \,\,\,  -\frac{N^2\,+3}{16} \cdot \, t\, \,\, +\frac{N^2\, +3}{32}, 
\\
\label{h2}
&&\hspace{-.4in}
h_2\, \, =\, \,\,  t  \cdot \, (t\, -1) \cdot \,\frac{d\ln f_2}{dt}
\, \,\, \,-\frac{N^2\,-1}{16} \cdot \, t\, \, \,+\frac{N^2\,+3}{32}, 
\\
\label{h3}
&&\hspace{-.4in}
h_3\, \, =\, \,\,  t  \cdot \, (t\, -1)\cdot \,\frac{d\ln f_3}{dt}
\, \,\,\,  -\frac{N^2\,-1}{16} \cdot \, t\, \,\, +\frac{N^2\,-5}{32}, 
\\
\label{h4}
&&\hspace{-.4in}
h_4\, \, =\, \, \, t \cdot  \, (t\, -1)\cdot \,\frac{d\ln f_4}{dt}
\, \, \,-\frac{N^2\,-5}{16} \cdot \, t\, \,\, +\frac{N^2\,-5}{32}. 
\end{eqnarray}
From (\ref{ffactors}), (\ref{h1}), (\ref{h2}), (\ref{h3}), (\ref{h4}),
one gets:
\begin{eqnarray} 
 \label{onegets}
 &&\hspace{-.8in} 
\quad \quad \quad \sigma(0, \, N) \, \,\,  =
 \,  \, \,\, \, h_1 +h_2 +h_3 +h_4 \, \, \, \,  + 4 \cdot \, \Bigl( \frac{t}{16} \, \, + \frac{(N^2-1)}{32}  \Bigr). 
\end{eqnarray}  
The $\, \sigma_i$'s in the additive relation (\ref{sigmaaddfour}) such that they satisfy the
{\em same  non-linear differential equation},  are, thus, simply related to the previous $\, h_i$'s:
\begin{eqnarray} 
 \label{newsig}
 \quad 
\sigma_i(0,N;t) \, \, \, = \, \, \,\, \,  h_i  \, \, \,\,   +  \frac{t}{16} \, \, + \frac{(N^2-1)}{32}. 
\end{eqnarray} 
These $\, \sigma_i$'s are solutions of the {\em same}  non-linear differential equation 
obtained from (\ref{4okamoto}) by (\ref{newsig}), which reads:
\begin{eqnarray}
  &&\hspace{-.99in} \,\,
t^2 \cdot \, (t \, -1)^2  \cdot \, \sigma''^2 \,\, \,  \,\,
+4\, \sigma' \cdot \, (t \cdot \, \sigma' \, -\sigma) \cdot \,
\Bigl((t  -1) \cdot \, \sigma' \, -\sigma\Bigr)
\nonumber\\
&&\hspace{-.88in}
   +\frac{1}{4} \cdot \, \Bigl( (N^2+1) \cdot \, (t-1) \, -t^2 \Bigr)\cdot \, \sigma'^2 \,\,\,
   -\frac{1}{2^6} \cdot \,
   \Bigl( 16 \cdot \,  (N^2 \, +1 \, -2\, t ) \cdot \, \sigma  \, +N^2 \cdot \, t  \Bigr)\cdot \, \sigma'
   \nonumber\\
\label{4okamotoinsigma}
&&\hspace{-.55in} \quad  \quad 
   -\frac{1}{4} \cdot \, \sigma^2 \,  \, \,  \, +\frac{N^2}{2^6} \cdot \, \sigma \,  \,  \, \, 
 -{{ N^2 \cdot \, (N^2 \, -3) } \over { 2^{10}}} 
   \, \,  \, = \, \, \,  \,\, 0. 
\end{eqnarray}
This non-linear differential equation of the Painlev\'e type (\ref{4okamotoinsigma}) is of course of
the Cosgrove-Scoufis form  (\ref{cosgrove}), being reducible
to an Okamoto sigma-form of Painlev\'e VI equation (\ref{4okamoto})
up to a simple shift (see (\ref{newsig})).

\vskip .1cm

Do note that the four $\, h_i$'s are solutions of the {\em same}
Okamoto sigma-form of Painlev\'e VI  (\ref{4okamoto}).
The  boundary conditions for each $ \, h_j$  were not discussed in~\cite{bmm},
and must be properly chosen for additivity (\ref{sigmaaddfour})
to occur.

To do this we recall that in~\cite{bmm}, we found that there are in general four different
possible boundary conditions for the expansion of solutions analytic
at $ \, t= \, 0$ of any 
Okamoto sigma form of Painlev\'e VI equation  
\begin{eqnarray} \quad \quad \quad \quad 
h^{(i)}(t) \, \, = \, \, \,\,  \sum_{n=0} \,\,  a^{(i)}_n \,\,  t^n, 
\end{eqnarray}
where we denote by $ \, h^{(i)}\, $ the solution of (\ref{4okamoto}) with
boundary conditions of class $ \, (i)$ of appendix D of~\cite{bmm}.

The first few $ \, a^{(i)}_n\, $ were determined analytically in~\cite{bmm}.
For the present case with parameters (\ref{4factorparams}) we only 
need cases 1 and 4 of appendix D of~\cite{bmm} where we find
for case 1 that  
\begin{eqnarray}
  &&\hspace{-.9in} \,  \,   \quad 
  a_0^{(1)} \, = \, \, -\frac{N^2+4N-1}{32},~~~~a_1^{(1)}
  \, = \,  \,\frac{N -1}{16},~~~~a_2^{(1)} \,=  \,  \,
  \frac{N}{2^6},~~~~a_3^{(1)} \, = \,  \,\frac{N}{2^7}, 
\end{eqnarray}
with $ \, a^{(1)}_{(N+3)/2} \, $ arbitrary
and for case 4 that
\begin{eqnarray}
  &&\hspace{-.9in}   \quad  \quad 
 a_0^{(4)}  = \,  -\frac{N^2-4N-1}{32},~~~a_1^{(4)}  = \, -\frac{N+1}{16},
~~~a_2^{(4)} = \, -\frac{N}{2^6},~~~a_3^{(4)}  = \,  \-\frac{N}{2^7}, 
\end{eqnarray}
with $ \ \, a^{(4)}_{(N+1)/2}\, $ arbitrary. 

The four $\, h_i$'s solutions of (\ref{4okamoto})  can be written 
\begin{eqnarray}
  \label{fhi}
  \quad \quad \quad 
h_j \, \,  = \, \, \, t  \cdot \, (t-1)  \cdot \, \frac{d\ln {\cal H}_j(0, N; \, t)}{dt}, 
\end{eqnarray}
where the $\,  {\cal H}_j(0, N; \, t)$'s read\footnote[5]{Note that
the $\,  {\cal H}_j(0, N; \, t)$'s
  are Puiseux series :  $\, {\cal H}_1(0,5; \, t) \, \, = \, \, t^{1/8} \, + \cdots$,
  $\, {\cal H}_2(0,5; \, t) \, \, = \, \, t^{1/8} \, + \cdots$,
  $\, {\cal H}_3(0,5; \, t) \, \, = \, \, t^{11/8} \, + \cdots$,
  $\, {\cal H}_4(0,5; \, t) \, \, = \, \, t^{11/8} \, + \cdots$.
}
for instance for $\, N= \, 5$:
\begin{eqnarray}
  &&\hspace{-.98in}\, \, \quad 
     {\cal H}_1(0,5; \, t) \, \, = \, \, \,
     \frac{2}{3} \cdot \, (1-t)^{-7/8} \cdot \, t^{-7/8} \cdot \, 
     \left((2 t-1) \cdot  \, \tilde{E} \, \,  \, -(t-1)  \cdot \, \tilde{K}\right), 
  \\
  &&\hspace{-.98in}\, \, \quad 
  {\cal H}_2(0,5; \, t) \, \, = \, \, \, \frac{2}{3} \cdot \, (1-t)^{-5/8} \cdot \, t^{-7/8} \, 
     \left((t+1) \cdot \, \tilde{E} \,\,  \, +(t-1) \cdot \, \tilde{K}\right), 
  \\
  &&\hspace{-.98in}\, \, \quad 
     {\cal H}_3(0,5; \, t) \, \, = \, \, \,
     -\frac{8}{3}\cdot  \, (1 \, -t)^{-7/8} \cdot  \, t^{-5/8} \cdot  \, 
     \left((t-2) \cdot \, \tilde{E} \,\, \,  -2 \cdot \, (t-1) \cdot \, \tilde{K}\right), 
  \\
  &&\hspace{-.98in} \, \, \quad 
     {\cal H}_4(0,5; \, t) \, \, \, = 
\nonumber   \\
 &&\hspace{-.99in}\, \, \, \, \quad \quad \quad \, \, 
     -\frac{8}{3} \cdot \, (1-t)^{-5/8} \cdot \, t^{-5/8} \cdot \, 
     \left(3 \, \tilde{E}^2 \, \,  +2 \cdot \, (t-2) \cdot \, \tilde{E}\tilde{K}
     \, \, -(t-1)\cdot \, \tilde{K}^2\right). 
\end{eqnarray}
Let us display some expansions of  $\, \sigma_i$'s for example for  $ \, C(0,5)$:
\begin{eqnarray}
  &&\hspace{-.8in}
\sigma_1(0,5) \, \, = \,\,\,  \frac{5}{8} \, \, - \frac{5}{16} \, t  \, \, -\frac{5}{2^{6}}\,t^2 \,  \,  \,
 -\frac{5 \cdot    11}{2^{10}}\,t^3 \,   \,  \, -\frac{5}{2^7}\, t^4 \,\, \,
     -\frac{3^2\cdot 5\cdot 43}{2^{16}} \, t^5 \, \,\,
 -\frac{5\cdot 4817}{2^{20}}\, t^6
     \nonumber\\
\label{s051}
  &&\hspace{-.2in}
     -\frac{5\cdot 241\cdot 509}{2^{25}} \, t^{7} \, \, \, 
-\frac{5\cdot 397811}{2^{27}}\, t^8\,  \,  \,
-\frac{3\cdot 5\cdot 13\cdot 134401}{2^{31}}\, t^9 \,\,  \,
   +\cdots
 \end{eqnarray}
\begin{eqnarray}
&&\hspace{-.8in}
 \sigma_2(0,5) \, \, = \,\,\,  \frac{5}{8} \, \, - \frac{5}{16} \, t  \, \, 
     - \frac{5}{2^6} \,t^2   \,  \,\,-\frac{5^2}{2^{10}} \, t^3 \,   \, \,
-\frac{5}{2^9} \, t^4 \,  \,  \,
     -\frac{5\cdot 61}{2^{16}} \,t^5 \,  \, -\frac{5\cdot 23^2}{2^{20}} \,t^6
     \nonumber\\
 \label{s052}
  &&-\frac{5\cdot 10099}{2^{25}} \,t^7 \,   \, \,
     -\frac{5^2\cdot  71\cdot 73}{2^{27}} \, t^8 \,    \, \,
-\frac{5\cdot 281321}{2^{31}} \, t^9 \, \, \,  \,
   +\cdots
\end{eqnarray}
\begin{eqnarray}
  &&\hspace{-.9in}
 \sigma_3(0,5) \,  \, = \,\,\,  - \, \frac{5}{8} \,\, + \frac{5}{16} \, t  \,  \,
    + \frac{5}{2^{6}} \, t^2 \, \,   +\frac{5}{2^7} \,t^3 \,   \, \,
+\frac{3\cdot 5\cdot   13}{2^{13}} \, t^4 \,    \,\, +\frac{5\cdot 53}{2^{14}} \, t^5 \,  \,  \,
+\frac{5\cdot 11  \cdot 449}{2^{21}} \, t^6
     \nonumber\\
\label{s053}
  &&\hspace{-.2in}
     +\frac{5\cdot 19\cdot 397}{2^{22}} \, t^7 \,  \,   \,
+\frac{3\cdot 5\cdot 15907}{2^{25}} \, t^8 \,   \, \,
+\frac{5\cdot 77527}{2^{26}} \, t^9 \,   \, \, \,
   +\cdots
 \end{eqnarray}
\begin{eqnarray}
  &&\hspace{-.9in}
     \sigma_4(0,5) \, \,= \,\,\,  - \, \frac{5}{8} \,\, + \frac{5}{16} \, t  \, \,
   +  \frac{5}{2^6} \, t^2   \, \, \, +\frac{5}{2^7} \, t^3 \,   \, \,
 +\frac{5\cdot  41}{2^{13}} \, t^4 \,  \,  \,  +\frac{5\cdot 59}{2^{14}} \, t^5 \,   \, \,
 +\frac{5\cdot 5813}{2^{21}} \, t^6
 \nonumber\\
\label{s054}
  &&\hspace{-.2in}
     +\frac{5\cdot 47\cdot 199}{2^{22}} \, t^7 \,   \,  \,
+\frac{5\cdot 13\cdot 97\cdot 197}{2^{27}} \, t^8 \,   \, \,
+\frac{5\cdot 13\cdot 97\cdot 197}{2^{28}} \, t^9 \,  \,   \,  \,
+\cdots
\end{eqnarray}
to be compared\footnote[2]{One verifies easily that
  $\, \, \sigma_{+}(0,5) \, = \, \,  \sigma_1(0,5) \, + \sigma_3(0,5)$.}
with the expansion of $\, \sigma_{+}(0,5)\, $ solution of (\ref{nonlineareq}):
\begin{eqnarray}
  &&\hspace{-.8in}\,\,  
 \sigma_{+}(0,5) \,  \, = \,\,\,   - \, \frac{3 \cdot \, 5}{2^{10}} \cdot \, t^3 \,\,
     - \, \frac{5^3}{2^{13}} \cdot \, t^4 \,\,   - \, \frac{5^3 \cdot \, 7}{2^{16}} \cdot \, t^5 \,\,
     - \, \frac{3 \cdot \, 5^2 \cdot \, 313}{2^{21}} \cdot \, t^6 \,\, + \, \cdots 
\end{eqnarray}
The corresponding expansions for $ \, C(0,7) \, $ are given in \ref{appC}.

\subsection{Algebraic solutions of (\ref{4okamoto})  and  (\ref{4okamotoinsigma}) }
\label{lambdaext}

Let us now define
\begin{eqnarray}
  \hspace{-.6in} 
  A(t;N) \, \,  \, = \, \,  \, \,
  \frac{N}{2^3} \cdot \, \Bigl( 1 \, -{{t} \over {2}} \, -\sqrt{1-t}\Bigr)
  \,  \,\,  = \, \,  \, \,
  -\frac{N}{2^3} \cdot \, \sum_{n=2}\left(-\frac{1}{2}\right)_n \cdot \, \frac{t^n}{n!}, 
\end{eqnarray}
where $ \, (a)_n= \, a \, (a+1) \, \cdots \, (a+n-1) \, $ is the Pochhammer symbol.
We have the remarkable  (algebraic) result that 
(for case 1)
\begin{eqnarray}
  \label{exact1}
  \hspace{-.6in}  \quad  \, \,
  h^{(1)}_{A} \,\,  =  \,\,  \, A(t;N)  \,  \,  \, \, \,\,
  +\Bigl(\frac{N -1}{16}\Bigr) \cdot \, t \,  \, \,\, \, \,
  - \Bigl(\frac{N^2+4N -1}{32}\Bigr)
  \nonumber \\
 \hspace{-.4in}  \quad \quad  \, \, \,\,  =  \,\,  \,  \,  
   - {{N} \over {8}} \cdot \,  \sqrt{1-t} \, \, \, \, \, -\frac{t}{16} \, \,\,  - \frac{(N^2-1)}{32}, 
\end{eqnarray}
and (for case 4)
\begin{eqnarray}
  \label{exact4}
  \hspace{-.6in} \quad  
  h^{(4)}_{A} \,\,  = \,\,  \, -A(t;N) \, \, \, \,  \,
  -\Bigl(\frac{N+1}{16}\Bigr) \cdot \, t \, \, \,  \, \,
  -\Bigl(\frac{N^2-4N -1}{32}\Bigr)
  \nonumber \\
  \hspace{-.4in}  \quad \quad  \, \, \,\,  =  \,\,\,  \,  
  {{N} \over {8}} \cdot \,  \sqrt{1-t} \, \, \, \, \, -\frac{t}{16} \, \, \, - \frac{(N^2-1)}{32},
\end{eqnarray}
{\em are actually exact (algebraic) solutions of} (\ref{4okamoto}). Using (\ref{newsig})
these algebraic results correspond,
in fact, and more simply, to the fact that the corresponding algebraic sigma's
\begin{eqnarray}
  \label{sigmaA}
 \hspace{-.7in} \quad  \quad  
  \sigma^{(1)}_{A}(t; \, N) \,\,  = \,\, \,  
  \, \,  - \, {{N} \over {8}} \cdot \,  \sqrt{1-t},
  \quad  \quad \quad
  \sigma^{(4)}_{A}(t; \, N) \,\, =  \, \, \,  {{N} \over {8}} \cdot \,  \sqrt{1-t}, 
\end{eqnarray}
are, actually, (algebraic) solutions of (\ref{4okamotoinsigma}).

The introduction of  $\, A(t;N)$ corresponds to the
 {\em remarkable existence of  algebraic solutions} for the $\, g_j$'s.  Actually  $\, A(t;N)$  reads
\begin{eqnarray}
  \label{remarkable}
 \hspace{-.7in}
  A(t;N) \, \,  \, = \, \,  \, \,
  \frac{N}{2^3} \cdot \, \Bigl(1 \, -{{t} \over {2}}  \, -\sqrt{1-t}\Bigr)
    \,  \,\,  = \, \,  \, \, t \cdot \, (t\, -1) \cdot \, {{ d \ln({\cal A}(t))} \over {dt}},
\end{eqnarray}
where $\,  {\cal A}(t)$ is the
{\em algebraic function}\footnote[1]{A function  $\, {\cal A}(t)\, $ which is the exponential of the integral
  of an algebraic function  (here $\, A(t;N)/t/(t-1)$)  is called a Liouvillian function. Here we see that
$\, {\cal A}(t)\, $ is not only Liouvillian, it is algebraic.}:
\begin{eqnarray}
\label{remarkablealg}
  {\cal A}(t) \,  \, = \, \,  \,
  (1\, -t)^{N/16} \cdot \, \Bigl(   {{ 1 \, + (1-t)^{1/2} } \over {2}} \Bigr)^{-N/4}. 
\end{eqnarray}

\subsection{Lambda extension of the four factors of $ \, C(0,N)$ with $ \, N$ odd}
\label{lambdaext}

Using the {\em exact algebraic solutions} (\ref{exact1}) and (\ref{exact4}),
as well as the relations (\ref{newsig})
between the $\, \sigma_i$'s and the $\, h_i$'s, 
the recursive
expansions\footnote[5]{ By recursive we mean using the non-linear differential equation to get
  order by order a power series analytic at $\, t=0 \, $ solution of that  equation.}
of~\cite{bmm} can be extended to an arbitrary order
and generalized with the arbitrary boundary condition constant, to find the pattern of the
lambda extensions of the  $\, \sigma_i$'s.

For $ \, j= \, 1,2 \, $  (case 4) we found experimentally for $\, N= \, 5$
\begin{eqnarray}
&&\hspace{-.98in}
   \sigma_j(0,5;\lambda_j) \, \, = \, \, \,
    {{5} \over {8}} \cdot \,  \sqrt{1-t} \, \,  \,  \,  +\lambda_j \, t^3 \,   \, +\lambda_j \, t^4 \,  \, 
+\frac{ 163}{3\cdot2^{6}} \, \lambda_j \, t^5 \,   \,  \, 
   +\frac{1}{3} \cdot  \, \left(\lambda_j^2+\frac{67}{2^5} \, \lambda_j\right) \cdot \, t^6
   \nonumber\\
  &&\hspace{-.7in}
 \quad \quad \quad \quad 
     +\frac{1}{3} \cdot \, \left(\frac{5}{2} \,  \lambda_j^2 \, 
+\frac{5\cdot 11257}{2^{15}} \, \lambda_j\right) \cdot \, t^7 \,  \,   \, 
+\left(\frac{173}{2^7} \, \lambda_j^2 \,  \, 
     +\frac{7\cdot 29\cdot 229}{3\cdot 2^{15}} \, \lambda_j\right) \cdot \, t^8
     \nonumber\\
\label{nsigma12}
  &&\hspace{-.3in}
 +\left(\frac{1}{3^2} \, \lambda_j^3 \,  \, 
+\frac{7\cdot 199}{3\cdot 2^8} \, \lambda_j^2 \, 
     +\frac{7\cdot 347\cdot 1021}{3\cdot 2^{21}} \, \lambda_j\right) \cdot \, t^9
     \,  \,   \, \, \, + \, \cdots, 
\end{eqnarray}
where
\begin{eqnarray}
\label{L2minusL1}
  \quad  \quad   \quad  \quad  
\lambda_2 \, = \, \, \, -\lambda_1, 
\end{eqnarray}
with  $ \, \lambda_j$ for $ \, j= \, 1,2$,  the arbitrary constant 
$ \, a_{(N+1)/2}\, $ of class 4 solutions. 

For $ \, j=  \, 3,4\, $  (case 1) we found experimentally for $\, N= \, 5$
\begin{eqnarray}
 &&\hspace{-.98in}
\sigma_j(0,5;\lambda_j) \, \, = \, \,  \, \,  \,
 - \, {{5} \over {8}} \cdot \,  \sqrt{1-t} \,  \, \,  \,  \,+\lambda_j \, t^4 \,  \, \,  \,
+\frac{3}{2} \, \lambda_j \, t^5 \, \,  \,+\frac{19\cdot 23}{2^8} \, \lambda_j \, t^6 \, \,  \,  \,
 +\frac{5\cdot 181}{2^9} \, \lambda_j \, t^7
\nonumber\\
 \label{nsigma34}
&&\hspace{-.5in}
 +\left(\frac{1}{4}\lambda_j^2 \,\, 
  +\frac{7\cdot   8219}{2^{15}} \, \lambda_j\right) \cdot \, t^8 \, \,  \,
  +\left(\frac{7}{8}\lambda_j^2 \, \,
  +\frac{7\cdot 17\cdot 941}{2^{16}} \, \lambda_j \right) \cdot \, t^9
  \, \,\, \, \, + \, \dots
\end{eqnarray}
where 
\begin{eqnarray}
  \label{L4minusL3}
  \quad \quad \quad \quad 
\lambda_4 \, \, = \, \, \, -\lambda_3, 
\end{eqnarray}
with $ \, \lambda_j$ for $ \, j= \, 3,4 \, $ the arbitrary constant $ \, a_{(N+3)/2}\, $
for class 1 solutions.

When 
\begin{eqnarray}
  \label{lambda1}
   \quad \quad \quad \quad 
\lambda_2 \, \, = \, \, \, -\lambda_1 \, \, = \, \, \,\, \frac{3\cdot 5}{2^{10}}, 
\end{eqnarray}
the lambda expansions (\ref{nsigma12})  {\em actually reduce to} (\ref{s051}) and (\ref{s052}), 
and for
\begin{eqnarray}
  \label{lambda2}
  \quad \quad \quad \quad  \quad 
  \lambda_4 \,  \, = \, \, \, -\lambda_3 \, \, = \, \, \, \,\frac{5}{2^{13}},
\end{eqnarray}
the lambda expansions (\ref{nsigma34})  {\em actually reduce to} (\ref{s053}) and (\ref{s054}).

\vskip .1cm

More generally, for arbitrary $\, N$ for $ \, j= \, 1,2 \,$ (case 4), one has the
following general form for the lambda extension of the $\, \sigma_j(0,N)$'s
\begin{eqnarray}
  \label{sigma12}
  \hspace{-.8in} \quad 
\sigma_j(0,N;t;\lambda_j)\, \, 
  = \,\,  \,  \,   {{N} \over {8}} \cdot \,  \sqrt{1-t} \,  \,  \,  \,
  +\sum_{n=1}^{\infty} \, \Bigl(\lambda_j \cdot \, t^{(N+1)/2}\Bigr)^n \cdot \, B^{(4)}_n(0,N;t), 
\end{eqnarray}
and for $\, j= \, 3,4$  (case 1)
\begin{eqnarray}
\label{sigma34}
\hspace{-.8in} \quad 
\sigma_j(0,N;t;\lambda_j)  \,
=  \,  \,  \, \,  - {{N} \over {8}} \cdot \,  \sqrt{1-t} \,  \,  \,  \,  \,
+\sum_{n=1}^{\infty}  \, \Bigl(\lambda_j\cdot  \, t^{(N+3)/2}\Bigr)^n  \cdot \, B^{(1)}_n(0,N;t), 
\end{eqnarray}
where $\, B^{(4)}_n(0,N;t)$ and $\, B^{(1)}_n(0,N;t)$ are power
series\footnote[2]{They are in fact D-finite series (see \ref{appE}).},
in $\, t$, and where we have the following normalization for both $i= \, 1$ and $\, 4$ and all $ \, N$:    
\begin{eqnarray}
  \label{bnorm}
  \quad  \quad    
B^{(i)}_1(0,N;0) \, \, = \,\,  \, 1, \quad \quad \quad i \, = \, 1, \, 4.
\end{eqnarray}
In other words the two one-parameter solutions $\, \sigma_1(0,N;t;\lambda_1) \,$ and
$\, \sigma_2(0,N;t;\lambda_2)\, $
can be seen as a deformation
of the same algebraic function $\, N \cdot \,  \sqrt{1-t}/8$, when the other
two one-parameter solutions $\, \sigma_3(0,N;t;\lambda_3)\, $
and  $\, \sigma_4(0,N;t;\lambda_4) \,$
can be seen as a deformation
of the same algebraic function $\, - N \cdot \,  \sqrt{1-t}/8$.
We illustrate this for the series expansions of the lambda-extensions  
of $ \, \sigma_j(0,7) \, $ in \ref{appClambda}
(see also (\ref{nsigma12}) and (\ref{nsigma34}) for $\, N= \, 5$).

\vskip .1cm 

When $\, M\, = 0 \, $  with $ \, N$ odd, the additivity property (\ref{additive})
of the $\, \sigma$'s becomes the additivity property  (\ref{sigmaaddfour})
as a consequence of the additional factorization of the  $ \, g_{\pm}(0,N)\, $
in (\ref{twofactors}), yielding the factorization in four factors (\ref{fourfactors}).
Similarly the  lambda extension (\ref{lambdaadditive})
of the additivity property (\ref{additive})
becomes a lambda extension of the additivity property  (\ref{sigmaaddfour}).
One thus has
\begin{eqnarray}
\label{lambdasigmaaddtwo}
&&\hspace{-.98in} \quad  \quad \quad \quad \quad \quad
  \, \sigma(0, \, N; \, \lambda) \, \, = \, \,\,\,
\sigma_{+}(0, N;t;\lambda_{+})\, \,\,  +\sigma_{-}(0, N;t;\lambda_{-}), 
\end{eqnarray}
and also
\begin{eqnarray}
\label{lambdasigmaaddfour}
&&\hspace{-.99in} \quad \,  \quad   \quad 
\sigma(0, \, N; \, \lambda) \, \,\, \, = \, \,
  \\
&&\hspace{-.99in}  \quad \quad \quad \quad  \quad 
\sigma_1(0, \, N; \,  t; \, \lambda_1) \,  +\sigma_2(0, \, N; \, t; \,  \lambda_2) \, \,
  +\sigma_3(0, \, N; \, t; \, \lambda_3) \, \,  +\sigma_4(0, \, N; \, t; \,  \lambda_4),
   \nonumber  
\end{eqnarray}
 where $\, \lambda_{+} \, = \, -\lambda_{-}\, = \, \, \lambda \, $
 (see (\ref{lmbdapmlambda})) and where the well-suited $\, \lambda_i$'s
 and $\, \lambda_{\pm}$'s remain to be found.

\subsection{Constraints on the $\, \lambda_i$'s}
\label{Constraints}

One has the two following  relations: 
\begin{eqnarray}
\label{required1}
\hspace{-.4in} \quad  
\, \, \,  \, \, \, \,
\sigma_{+}(0, N;t;\lambda_{+})
  \, \, \,  = \, \, \, \, \, \sigma_1(0,N;t;\lambda_1) \,\, \,  +\sigma_3(0,N;t;\lambda_3), 
\end{eqnarray}
and
\begin{eqnarray}
\label{required2}
&&\hspace{-.4in} \quad 
\, \, \, \, \, \,\,
  \sigma_{-}(0, N;t;\lambda_{-})\, \, \,  = \, \, \,\,\, 
  \sigma_2(0, N;t; \lambda_2) \,\,  +\sigma_4(0,N;t;\lambda_4), 
\end{eqnarray}
where $\, \sigma_{\pm}(0,N;t;\lambda_{\pm}) \, $ are the sigma functions
(\ref{lambdasigmapm}) for the factors $\,\,  g_{\pm}(0,N;t)$
and where $\, \lambda_{+} \, = \, -\lambda_{-}\, = \, \, \lambda$
(see (\ref{lmbdapmlambda})).
These two relations (\ref{required1}) and (\ref{required2}) will only hold 
if there is a relation between
$\, \lambda_1\, $ and $\, \lambda_3$, as well as a similar relation between
$\, \lambda_2 \, $ and $\, \lambda_4$. 

\vskip .2cm

For different values of $\, N$ let us recall the form for the $\, \sigma_i$'s 
(see (\ref{sigma12}) and  (\ref{sigma34})) such that the $\,
\sigma_i$'s satisfy 
their respective non-linear differential ODE's.
Imposing that the  RHS of (\ref{lambdasigmaaddfour})
is solution the non-linear differential ODE (\ref{eqnmoddMzero})
for $\, \sigma(0, \, N; \, \lambda)$, one finds experimentally,
for different values of $\, N$, that
\begin{eqnarray}
\label{L2minusL1}
  \quad      
\lambda_2 \, = \, \, \, -\lambda_1, \quad  \quad  
  \quad  \quad \hbox{and:}   \quad \quad   \quad  \quad   
\lambda_4 \, = \, \, \, -\lambda_3.
\end{eqnarray}

Similarly imposing that $\, \sigma_{+}(0, N;t;\lambda_{+})\, $
and $\, \sigma_{-}(0, N;t;\lambda_{-})$, given respectively by 
(\ref{required1}) and (\ref{required2}), both verify the {\em same}
non-linear ODE (\ref{nonlineareq}), one finds that: 
\begin{eqnarray}
 \label{relL1L4first}
 \hspace{-.4in} \quad \quad \quad \quad  \quad \quad 
 \lambda_3\, \,  =\, \, \,   \frac{ \lambda_1}{4 \cdot \,  (N+1)},  
\end{eqnarray}

To determine this relation (\ref{relL1L4first}) it is, for instance, 
sufficient to consider the term $\,n=\,1\,$ in (\ref{sigma12}) and
(\ref{sigma34}). From (\ref{required1}) and (\ref{lambdasigmapm}) 
we obtain the condition
\begin{eqnarray}
  \label{l1l2cond}
  \hspace{-.4in}
  \quad 
  \lambda_1 \cdot \, B^{(4)}_1(t;N)\, \,\,
  +\lambda_3 \cdot \, t \cdot \, B^{(1)}_1(t;N)
   \, \,\, = \,\, \, \, \, \lambda_{+} \cdot \, B_1(0,N;t),  \quad 
\end{eqnarray}
where we recall from (\ref{htilde1}) that  $\, B_1\, $ reads:
\begin{eqnarray}
  \hspace{-.4in}
  \quad \quad  \quad  \quad 
B_1\, \, = \, \, \,
  {}_2F_1\Bigl([\frac{N}{2},\frac{N}{2}], \, [N+1], \, t\Bigr).
\end{eqnarray}

To proceed further, we require explicit forms for $ \, B^{(4)}_1(t;N)\, $ and
$ \, B^{(1)}_1(t;N)\,$ which are computed in  \ref{appE} as solutions (normalized to
unity at $ \, t= \, 0$) of linear differential equations:
\begin{eqnarray}
\label{b14}
  &&\hspace{-.98in}
     B^{(4)}_1(t;N) \, \, = \,  \,  \, 
     \frac{1}{2N} \cdot \,
     \Bigl( 2 \cdot \, t \cdot \,\sqrt{1-t} \cdot \, \frac{dB_1(t)}{dt} \, \, \,
     +N  \cdot  \, (1 \, +\sqrt{1-t}) \cdot \,  B_1(t)\Bigr),
 \\
 \label{b11}
&&\hspace{-.98in}
  B^{(1)}_1(t,N) \, \, = \, \, \,
 \nonumber   \\
&&\hspace{-.98in} \quad \quad 
 \, \, = \, \, \,
 -\frac{2  \, (N+1)  }{N \, t} \cdot \,
   \Bigl(2 \cdot \, t  \cdot \, \sqrt{1-t}  \cdot \, \frac{dB_1(t)}{dt}
   \, \,\,  -N \cdot \, (1 \, -\sqrt{1-t}) \cdot \, B_1(t)\Bigr)
   \\
&& \hspace{-.8in} \quad \quad \quad
 \, \,  = \, \, \,\,\,
 1 \, \, \,\,  +\Bigl({{N +1 } \over {4}} \Bigr) \cdot \, t
  \, \, \,\,
  +\Bigl( {{N^3 \, +8\,N^2 \, +20\,N \, +12 } \over {32 \cdot \, (N\, +3)}} \Bigr)  \cdot \, t^2
  \,  \, \, \, \, + \, \, \, \cdots
     \nonumber  
\end{eqnarray}
Thus we find:
\begin{eqnarray}
  &&\hspace{-.98in}   \quad \quad  \quad 
 \lambda_1 \cdot \, B^{(4)}_1(t;N) \, \, +\lambda_3 \cdot \, t \cdot \, B^{(1)}_1(t;N)
   \nonumber\\
  &&\hspace{-.86in} \, \, \, \,
     \quad  \quad  \quad  \, \, = \, \, \,  \,
\frac{t}{N}  \cdot \, \sqrt{1-t}  \cdot \,
\Big( \lambda_1 \,  \,
- 4  \cdot \, (N+1) \cdot \, \lambda_4 \Big)  \cdot \,  \frac{dB_1(t)}{dt}
  \\
  &&\hspace{-.76in} \, \, \, \,
     \quad  \quad    \quad \quad 
     +\frac{1}{2}  \cdot \, \Big( \lambda_1 \cdot \, (1 \, +\sqrt{1-t}) \, \,  \, 
     + \lambda_3 \cdot \, 4 \cdot \, (N+1) \cdot \,
     (1 \, -\sqrt{1-t})\Big)  \cdot \, B_1(t).
  \nonumber
\end{eqnarray}
By setting
\begin{eqnarray}
 \label{relL1L4}
  \quad  \quad  \quad \quad 
\lambda_3\, \,  =\, \, \,   \frac{ \lambda_1}{4 \cdot \,  (N+1)}, 
\end{eqnarray}
the coefficient of $ \,\, dB_1(t)/dt \,$ vanishes.  We find
\begin{eqnarray}
\,   \lambda_1 \cdot\, B^{(4)}_1(t;N) \, \,  \, +\lambda_3 \cdot \, t \cdot \, B^{(1)}_1(t,N)
    \,  \, \, = \, \, \,\,  \,  \lambda_1 \cdot \, B_1(t), 
\end{eqnarray}
which setting
\begin{eqnarray}
  \label{relL1Lplus}
  \quad  \quad  \quad 
\lambda\, \,  =\, \, \,   \lambda_{+} \, \,  =\, \, \,\lambda_1,
\end{eqnarray}
verifies (\ref{l1l2cond}) as desired.

\vskip .2cm 

{\bf To sum-up:}  There is a {\em one-parameter} family of  
$\sigma_j(0,N;t;\lambda)$'s solutions of Okamoto sigma form of Painlev\'e VI  for
which the lambda extension of the additive decomposition (\ref{lambdasigmaaddfour})
holds, namely, taken into account (\ref{relL1L4}) and (\ref{relL1Lplus}) 
\begin{eqnarray}
&&\hspace{-.98in}
\sigma(0,N;\lambda) \, \, = \, \, \,
\\
&&\hspace{-.9in}  \quad 
\sigma_1(0,N;\lambda) \, +\sigma_2(0,N;-\lambda)\, 
+\sigma_3\Bigl(0,N;\frac{ \lambda}{4 \cdot \,  (N+1)}\Bigr)
\, +\sigma_4\Bigl(0,N; -\, \frac{ \lambda}{4 \cdot \,  (N+1)}\Bigr).
\nonumber 
\end{eqnarray}

\vskip .2cm

{\bf Selected values of the $\, \lambda_i$'s:} The selected values
 of the $\, \lambda_i$'s, such that the $\, g_i$'s are homogeneous polynomial expressions
 of the elliptic integrals of the first and second kind
 $\, {\tilde K}$ and  $\, {\tilde E}$, read
\begin{eqnarray}
\label{selected}
 &&\hspace{-.98in} \quad  \quad  \quad   \quad  \quad  
  \lambda_2 \, = \, \, \, -\lambda_1
  \,\, = \, \, \, {{N+1} \over {2}} \cdot \, \alpha_{0,N}
  \, \,\, = \, \, \,\,
     {{ N! } \over {2^{2\, N\, +1} \cdot \, \Bigl(({{N-1} \over {2}})!\Bigr)^2 }}, 
\end{eqnarray}
and
\begin{eqnarray}
\label{selected2}
&&\hspace{-.98in}  \quad  \quad  \quad   \quad \quad
\lambda_4 \, = \, \, \, -\lambda_3
 \, \, = \, \, \, {{ \alpha_{0,N}} \over {8}}
  \, \,\, = \, \, \,\,
     {{ N! } \over {2^{2\, N\, +3} \cdot \, (N+1) \cdot \,
   \Bigl(({{N-1} \over {2}})!\Bigr)^2  }},  \quad
\end{eqnarray}
where $\, \alpha_{0,N} \, $ is given by (\ref{alphaMN}) for $\, M\, = \, 0$.  

\vskip .1cm

\section{Tracy-Widom viewpoint on the four $\, \sigma_i$'s}
\label{TracyWidom2}

Recalling section \ref{TracyWidom} one can try to see if, instead of the sums
(\ref{required1}) and (\ref{required2}), the difference\footnote[1]{Note that
the difference $\,\, \delta \, = \,  \, \sigma_2(0,N;t) \,\, \,  -\sigma_4(0,N;t) \, \,$
yields the same results
(\ref{orderthreeDelta}), (\ref{ordertwoDelta}),  (\ref{relation2}).
}
$\,\, \delta \, = \,  \, \sigma_1(0,N;t) \,\, \,  -\sigma_3(0,N;t) \, \,$
verifies a simple enough non-linear ODE. In that case the equivalent
of the Tracy-Widom-like relation (\ref{differencerelation})
becomes\footnote[5]{Similarly to (\ref{differencerelation})
 relation (\ref{TracyWidom3}) can also be simply obtained
 by guessing from the series expansion of $\,\, \delta$ and $\, \sigma$
 for different valus of $\, N$.}
\begin{eqnarray}
\label{TracyWidom3}
&&\hspace{-.96in} \quad  \quad \quad \quad \quad
   \, \delta^2 \,\, \, \, + 
   t  \cdot \, (t\, -1)  \cdot \, \sigma'  \,\, \,\,
   - {{t \, -1} \over {2}} \cdot \, \sigma \, \, \,\, \,
+ {{N^2} \over {16}} \cdot \, (t\, -1)
  \,  \,\, = \, \,\, \, \, 0,  
\end{eqnarray}
where $\, \sigma$ denotes, here, the sum (see (\ref{required1})) 
$\,\, \sigma_{+} \, = \,  \, \sigma_1(0,N;t) \,\, +\sigma_3(0,N;t)$.
In contrast with relation (\ref{differencerelation}),
relation (\ref{TracyWidom3}) depends on $\, N$.
Again, using Pantone's program one first finds
that this difference $\, \delta \,$ verifies an  order-three non-linear ODE
\begin{eqnarray}
\label{orderthreeDelta}
&&\hspace{-.98in}  \quad \quad \quad 
8 \cdot \, (t-1)^3 \cdot \, t^2 \cdot \, (t+1) \cdot \, \delta  \cdot \, \delta'''
\nonumber \\
&&\hspace{-.96in}  \quad  \quad  \quad \quad 
+8 \cdot \, t \cdot \, (t-1)^2 \cdot \,
\Bigl(2 \cdot \, (t^2+t-1) \cdot \, \delta \, \,
-t \cdot \, (t^2-1) \cdot \, \delta' \Bigr) \cdot \, \delta''
\nonumber \\
&&\hspace{-.96in}   \quad   \quad \quad   \quad \quad  
\, \,  -32 \cdot  \, (t^2 -1) \cdot \, \delta^3 \cdot \, \delta' \,\,\, \,\, 
   -2 \cdot  \, (t -1) \cdot \, (t +1)^2 \cdot \,(3\,t \, -1)  \cdot \, \delta \cdot \, \delta'
   \nonumber \\
&&\hspace{-.96in}   \quad  \quad  \quad \quad   \quad  \quad  \quad 
   \, \, +4 \cdot  \, t \cdot \, (t^2 -1)^2   \cdot \, \delta'^{\, 2}
   \, \, \, \,+32 \cdot \, t \cdot \, \delta^4
     \\
&&\hspace{-.96in}  \quad  \quad  \quad \quad \quad \quad  \quad \quad   \, \, \, \,  \, \, \,
     +\Bigl( 2\, t^2 \, +(N^2 \, +5) \cdot \, t \,
     -(N^2 \, -1) \Bigr)  \cdot \, (t-1)\cdot \, \delta^2 \, \, = \, \, \, 0,
      \nonumber 
\end{eqnarray}
which is similar to (\ref{orderthree}). In fact, with  more coefficients, one can
find\footnote[9]{Note two typos in the
  published version of this paper in J.Phys.A: $\, 16 \cdot \, t^3 \cdot \, (t-1)^4 \cdot \, \delta \cdot \,  \delta''^{\, 2}$
  must be changed into $\, 16 \cdot \, t^3 \cdot \, (t-1)^6 \cdot \,  \delta''^{\, 2}$ and the $\,(t^2\, -1)$
  factor in the $\, \delta'^{\, 2}$ term is a  $\,(t^2\, -1)^2$ factor. }
a second-order non-linear ODE like (\ref{ordertwoquadra}):
\begin{eqnarray}
  \label{ordertwoDelta}
&&\hspace{-.98in}  \quad  \quad  
   16 \cdot \, t^3 \cdot \, (t-1)^6  \cdot \,  \delta''^{\, 2} \, \, \, \, \,
   - 8 \cdot \, (t-1)^2  \cdot \, t^2 \cdot \, \Bigl(2   \cdot \, (t-1)^3   \cdot \, (t+1) \cdot \,  \delta'
   \nonumber \\
&&\hspace{-.98in}   \quad   \quad   \quad   \quad  \quad   \quad   \quad   \quad   \quad  
    + (t-1)^2\cdot \, (N^2 \, -2\,t-1)\cdot \, \delta  \,  -32 \cdot \, \delta^3 \Bigr)  \cdot \,  \delta''
   \nonumber \\
&&\hspace{-.96in}   \quad  \quad  \quad  \quad \,
   -4 \cdot \, t \cdot \, (t^2\, -1)^2 \cdot \,
   \Bigl( 16 \cdot \, \delta^2 \, -(1-t)^2 \Bigr) \cdot \, \delta'^{\, 2}
   \\
&&\hspace{-.96in}   \quad  \quad  \quad  \quad \,
+4 \cdot \, t \cdot \, (t^2\, -1) \cdot \,
\Bigl( 32 \cdot \, t  \cdot \, \delta^2 \, +(1 -t)^2 \cdot \,
\Bigl(N^2 \, -(2\, t \, +1) \Bigr) \Bigr) \cdot \, \delta  \cdot \, \delta'
\nonumber    \\
&&\hspace{-.98in} 
 +\Bigl( (t-1) \cdot \, (N^2 \, -t ) \, +16 \cdot \, \delta^2\Bigr) \cdot \,
\Bigl(  (t-1) \cdot \, (N^2\, t \, -(2\, t \, +1)^2 ) \, -16 \cdot \, \delta^2\Bigr)  \cdot \, \delta^2
\, \, = \, \, \, 0.
 \nonumber
\end{eqnarray} 
Note that equations (\ref{orderthreeDelta}) and  (\ref{ordertwoDelta}) are preserved by the Kramers-Wannier duality.
Denoting  $\, {\cal P}(t, \, \delta, \, \delta', \, \delta'', \, \delta''') \, $
and $\, P(t, \, \delta, \, \delta', \, \delta'')$
the LHS of (\ref{orderthreeDelta}) and  (\ref{ordertwoDelta}) one has the following covariance property:
\begin{eqnarray}
  \label{covP}
  &&\hspace{-.98in}  \quad   \quad   \quad   \quad
     P\Bigl( {{1} \over {t}}, \,  \, {{\delta} \over {t}}, \,  \, \delta \, - t\cdot \, \delta', \, \,   t^3 \cdot \, \delta'' \Bigr)
     \, \, = \, \, \, \, \, 
     {{1} \over {t^6}} \cdot \,  P(t,  \, \, \delta, \,  \, \delta', \,  \, \delta'').
\end{eqnarray} 
and
\begin{eqnarray}
 \label{covcallP}
&&\hspace{-.98in}  \quad   \quad   \quad  \quad  \quad   \quad 
{\cal P}\Bigl( {{1} \over {t}}, \, \,   \, {{\delta} \over {t}}, \, \,
 \, \delta \, - t \cdot \, \delta', \, \, \,    t^3 \cdot \, \delta'', \, \, \,   -t^5 \cdot \, \delta''' \, \,  - 3  \, t^4 \cdot \, \delta''\Bigr)
\nonumber \\
&&\hspace{-.98in}  \quad  \quad  \quad  \quad  \quad   \quad   \quad  \quad  \quad   \quad 
\, \, = \, \, \, \, \, 
 {{1} \over {t^4}} \cdot \,  {\cal P}(t, \, \delta, \, \delta', \, \delta'', \, \delta''').
\end{eqnarray} 
Again, similar to what has been achieved in section \ref{TracyWidom}, we can say that the Tracy-Widom-like
transformation (\ref{TracyWidom3})
will change the second order non-linear ODE (\ref{ordertwoDelta}) into a third-order  non-linear ODE in $\, \sigma$.
Using some differential algebra elimination, one can check that this last order-three non-linear ODE
is {\em actually compatible} with\footnote[5]{And we have seen that  (\ref{nonlineareq}) actually reduces to an Okamoto sigma-form of
Painlev\'e VI using a Landen change of variable (\ref{tx}) together with transformation (\ref{salah}).} (\ref{nonlineareq}).

Let us denote, again,  the LHS of the order-three non-linear ODE (\ref{orderthreeDelta}) by $\, {\cal R}_3$, 
and  the LHS of the order-two non-linear ODE (\ref{ordertwoDelta}) by $\, {\cal R}_2$,
we have the following relation similar to (\ref{relation}):
\begin{eqnarray}
\label{relation2}
&&\hspace{-.98in}   \quad  \quad   \quad  
\Bigl(4 \cdot \, t \cdot \, (t\, -1)^4 \cdot \,\delta'' \,  \, \,
 -2 \cdot \, (t \, +1) \cdot \, (t \, -1)^3  \cdot \, \delta'
\nonumber \\
&&\hspace{-.98in}  \quad   \quad  \quad \quad \, \, \, \, \, \,  \quad \quad
 -(t-1)^2\cdot \, (N^2 \, -2\, t-1) \cdot \, \delta   \, +32 \cdot \, \delta^3 \Bigr)
\cdot \, {\cal R}_3
\nonumber \\
&&\hspace{-.98in}  \quad \, \, \, \, \, \, \,  \quad   \quad
\, \, = \, \, \, \,
(t^2 \, -1) \cdot \, \delta  \cdot \, {{ d {\cal R}_2 } \over { dt}}
\, \, \, 
-2 \cdot \, \Bigl( (t^2 \, -1) \cdot \, \delta' \, +2 \cdot \,  \delta' \Bigr) \cdot \, {\cal R}_2.
\end{eqnarray}

Let us remark that for small  values of $\, N$, for instance $\, N=\, 9$, we get also another
second-order non-linear ODE in $\, \delta$. Combined with the previous second order
non-linear ODE (\ref{ordertwoDelta}) one eventually finds, eliminating $\,  \delta''$, 
a polynomial relation\footnote[1]{For fixed value of $\, t$ and for $\, N\, = \, \, 9$,
  the genus of the curve $\, P_N(x, \, y;  \, t) \, = \, \, 0 \, $  is zero (rational curve).}
$\,  P_N(\delta, \,\delta', \, t) \, \, \,\,  = \, \,\, \,  \, 0 \,\,  $ 
which, for $\, N=\, 9$, is of the form
\begin{eqnarray}
\label{polDeltaN9}
&&\hspace{-.98in}  \quad
   (16 \cdot \, \delta'  +35) \cdot \, (16 \cdot \, \delta' -15)  \cdot \,
   (16 \cdot \, \delta' +3)  \cdot \, (16 \cdot \, \delta' -63) \cdot \, t^8 \, \, \, \,  +  \, \cdots \,
\nonumber \\
&&\hspace{-.98in}  \quad\, \, \, 
   +(4  \cdot \,\delta -1) \cdot \, (4 \cdot \,\delta -9) \cdot \, (4 \cdot \,\delta -5)^2
   \cdot \, (4 \cdot \,\delta +7)^2 \cdot \, (4  \cdot \,\delta +3)^2  \, \, = \, \, \, 0,
\end{eqnarray}
this relation being {\em compatible} with the two previous second-order non-linear ODEs in $\, \delta$.
In fact we have the following situation.
Recalling the definition of the $\, \sigma_i$'s in terms of log-derivatives of factors of the $ \, C(0,N)$'s,
expressed in terms of the complete elliptic integrals $ \, {\tilde K}(k)$ and $ \, {\tilde E}(k)$, one
verifies easily that $\, \delta \, $ is a solution of (\ref{ordertwoDelta}) {\em as well as} (\ref{polDeltaN9}).
The power-series solutions of  second-order non-linear ODE (\ref{ordertwoDelta}) are actually 
{\em one-parameter families of solutions} of (\ref{ordertwoDelta}), which correspond
to {\em lambda extension of the previous} $\, \delta$
expressed in terms of  $ \, {\tilde K}$ and $ \, {\tilde E}$. By contrast, the power-series solutions
of relation (\ref{polDeltaN9}), valid for $\, N\, = \, 9$, correspond to  power series solutions of the form 
\begin{eqnarray}
\label{powerseriesN9}
&&\hspace{-.98in}  \quad  \quad  \quad  \quad \quad \quad \quad \quad \quad  \quad  \quad
     \delta \, \, = \, \, \, \sum_{n=0}^{\infty} \,  \delta_n \cdot \, t^n, 
\end{eqnarray}
where the first coefficient $\, \delta_0 \, $ can only take the following values $\, 1/4, \, -3/4$,
$5/4,\,  -7/4, \, 9/4$.
For  $\, \delta_0 \, = \, \,  9/4\, $ one finds easily, that the power series solution of (\ref{polDeltaN9})
is unique, and can be obtained order by order:
\begin{eqnarray}
\label{powerseriesN9}
&&\hspace{-.98in} \quad  \, \, \,   \quad  
   \delta \, \, = \, \, \, {{9} \over {4}} \, \, -{{9} \over {8 }}\cdot \, t \,  \,
   -{{9} \over {32}}\cdot \, t^2 \, \,  -{{9} \over {64}}\cdot \, t^3 \,
  \,  -{{45} \over {512}}\cdot \, t^4 \, \,  -{{16443} \over {262144}}\cdot \, t^5 \, \,
   \, \, +\, \cdots 
\end{eqnarray}
This series (\ref{powerseriesN9}) is nothing else but the expansion of the
$\, \delta$ expressed in terms of the complete elliptic integrals
$ \, {\tilde K}(k)$ and $ \, {\tilde E}(k)$
corresponding to the factors of $\, C(0,9)$.
In other words the polynomial relation (\ref{polDeltaN9}), which is compatible with the
order-two non-linear ODE  (\ref{ordertwoDelta}),
{\em actually selects in the one-parameter} (lambda-extension)
{\em family of solutions of}  (\ref{ordertwoDelta}), {\em the one corresponding to the ``physical''}
$\, C(0,9)$ (i.e. Toeplitz determinants and no longer Fredholm determinants). 

\vskip .1cm  

{\bf Remark 1:} Let us also note that the order-two non-linear ODE (\ref{ordertwoDelta}) can, also be obtained
performing some differential algebra eliminations using relations (\ref{TracyWidom3})
and (\ref{nonlineareq}) (for $\, M= \, 0$)  for  $ \,\,  \sigma_{+}\, \, = \, \sigma_1 \, +\sigma_3$.

\vskip .1cm 

{\bf Remark 2:}  Let us also note that performing some differential  algebra eliminations using relations
(\ref{TracyWidom3}), (\ref{polDeltaN9}), and  (\ref{nonlineareq}) for $\, M= \, 0$, 
{\em one also finds a polynomial relation} $ \, {\cal P}_N(\sigma, \, \sigma', \, t)\, = \, \, 0$,
for $\, \sigma = \, \sigma_{\pm}$, which, for $\, N=\, 9$, is of the form:
\begin{eqnarray}
\label{polDeltaN9sigma}
&&\hspace{-.98in}  \quad \quad
   (64 \cdot \, \sigma'  +225) \cdot \, (64 \cdot \,  \sigma' +3969)  \cdot \,
   (64 \cdot \,  \sigma' +1225)  \cdot \, (64 \cdot \,  \sigma' +9) \cdot \, t^8
   \, \, \, \,  \,  +  \, \cdots \,
\nonumber \\
&&\hspace{-.98in}  \quad \,  \quad  \quad \,  \quad\, 
   +2^{24} \cdot \, \sigma \cdot \, (\sigma -4)^2 \cdot \, (\sigma -7)^2 \cdot \, (\sigma -9)^2 \cdot \, (\sigma -10)
   \, \, \, = \, \,  \,\, 0.
\end{eqnarray}
The power-series solutions of relation (\ref{polDeltaN9sigma}), valid for $\, N\, = \, 9$,
correspond to  power series solutions of the form 
\begin{eqnarray}
\label{powerseriesN9bis}
&&\hspace{-.98in}  \quad  \quad  \quad  \quad \quad \quad \quad \quad \quad  \quad  \quad
     \sigma \, \, = \, \, \, \sum_{n=0}^{\infty} \,  \sigma_n \cdot \, t^n, 
\end{eqnarray}
where the first coefficient $\, \sigma_0 \, $ can only take the following values $\, 0, \, 4, \, 7, \, 9, \, 10$.
For $\, \sigma_0 \, = \, 0$  one finds easily, that the power series solution of (\ref{polDeltaN9sigma})
is unique, and can be obtained order by order:
\begin{eqnarray}
\label{powerseriesN9bis}
&&\hspace{-.98in}  \, 
   \sigma_{+} \,  = \, \, 
   {-{315} \over {262144}} \cdot \, t^5 \, \, -{{5103 } \over { 2097152 }}\cdot \, t^6 \,  \,
 -{{56133} \over {16777216}}\cdot \, t^7  \, \,  -{{1054053} \over {268435456}}\cdot \, t^8 \, 
   \, \, + \cdots 
\end{eqnarray}
This series (\ref{powerseriesN9bis}) is nothing else but the expansion of the $\, \sigma_{+} \, $
corresponding to the factors of $\, C(0,9)$, expressed in terms of the complete elliptic integrals
$ \, {\tilde K}$ and $ \, {\tilde E}$.

\vskip .1cm 

{\bf Remark 3:} Let us recall that
\begin{eqnarray}
\label{recall}
  &&\hspace{-.98in}  \,   \, \quad \quad \quad \quad  \quad \quad  
     \sigma_{+}\, \, = \, \, \sigma_1 \, +\sigma_3 \, \, \, = \, \, \, \,
     t \cdot \, (t-1) \cdot \, {{ d \ln(g_1\cdot \, g_3)} \over {dt}},
\end{eqnarray}
where the product $\, g_{13} \, = \, \, g_1\cdot \, g_3 \, \, $
is D-finite: it is solution of a linear differential equation of order {\em five}. Denoting $\, \Sigma \, $
the log-derivative of the product $\, g_{13}$,
one gets:
\begin{eqnarray}
\label{recall}
  &&\hspace{-.98in} \, \, 
  {{ g_{13}'} \over { g_{13}}}  \, = \, \, {{\sigma_{+}} \over {  t \cdot \, (t-1) }} \, = \, \, \Sigma,
 \quad  {{ g_{13}''} \over { g_{13}}} \, = \, \,     \Sigma' \, + \Sigma^2,
 \quad   {{ g_{13}'''} \over { g_{13}}} \, = \, \,    \Sigma'' \, + 3 \,  \Sigma \,  \Sigma' \, + \Sigma^3,
\, \, \, \cdots 
\end{eqnarray}
This order-{\em five} linear differential equation  can thus be rewritten in an order-four  Ricatti polynomial form: 
\begin{eqnarray}
\label{Ricatti}
  &&\hspace{-.98in}  \, \quad  \quad \quad  \quad  \quad   \quad  \quad  \quad  \quad   \quad 
 {\cal R}(\sigma, \, \sigma', \, \sigma'',\, \sigma^{(3)},\,\sigma^{(4)}; \, \, t)  \,\,  \,\,  = \, \,\,  \, \, 0.
\end{eqnarray}
We underline that the polynomial relation (\ref{polDeltaN9sigma}) can also be obtained performing
some differential algebra elimination between (\ref{nonlineareq}) and (\ref{Ricatti}).
This is a general result: the existence of polynomial relations\footnote[1]{Reminiscent of the ``invariants''
  one obtains for linear differential operators with selected differential Galois groups (see the relation
  $\, Q(f, \, f', \, f'') = \, Constant \,\,$ given in the
  introduction of~\cite{Weil}).},
like $ \, {\cal P}_N(\sigma, \, \sigma', \, t)\, = \, \, 0$,
which selects the D-finite (homogeneous polynomials of $ \, {\tilde K}$ and $ \, {\tilde E}$)
factors scenario in the one-parameter families
of solutions of a non-linear second order differential equation (like (\ref{nonlineareq})), is precisely
a consequence of this D-finite character (combined with the  non-linear second order differential equation).

\vskip .1cm 

{\bf Remark 4:} One should note that the differences of any of two $\, \sigma_i$'s
give similar results. Similarly, the sum of any of two $\, \sigma_i$'s give  results similar to
(\ref{nonlineareq}). However the sums of three among the four $\, \sigma_i$'s yield much more involved non-linear ODEs
as well as the linear combinations  $\, \sigma_i \, +\mu \cdot \sigma_j \, $ when $\, \mu$ is no longer equal to $\, \pm \, 1$.

\vskip .1cm

\section{The Determinants of Forrester-Witte} 
\label{detFW}

We began this paper  with examples of factorizations of Toeplitz determinants
(see (\ref{factoreven}),  (\ref{factorodd}))
and proceeded to show that this leads to a {\em one parameter family} of
sigma forms of Painlev{\'e} VI  which have {\em additive decompositions}.
To complete the discussion we need to determine the determinants of
the factors from the sigma functions which satisfy the Painlev{\'e}
equations.

 The $ \, \tilde{N}\times \tilde{N}$ Toeplitz determinants
of Forrester-Witte~\cite{fw} as given in~\cite{gil}  are
\begin{eqnarray}
 \label{fwdet}
  \hspace{-.6in}
  \quad  \quad  \quad  \quad 
D^{(p,p',\eta,\xi)}_{\tilde{N}}(t)
\,\, \,  = \, \,\,\, 
  {\rm det}\left[A^{(p,p',\eta,\xi)}_{j-k}(t)\right]_{j,k=0}^{\tilde{N}-1}, 
\end{eqnarray}
where
\begin{eqnarray}
  \label{Adef}
  \quad  \quad 
  A_m^{(p,p',\eta,\xi)}(t) \, \,\, = \, \,\, \,  \, A^{(1)}(t) \, \,\, +\xi \cdot \,  A^{(2)}(t), 
\end{eqnarray}
where in~\cite{bmm} $ \, A^{(1)}(t)$ and $ \, A^{(2)}(t)$ may be written as
\begin{eqnarray}
&&\hspace{-1in} \quad
   A^{(1)}_m(t) \, \, = \, \, \,
 \nonumber \\
\label{a1def2}
  &&\hspace{-1in} \quad \quad
     \frac{\Gamma(1+p') \cdot \, t^{(\eta-m)/2}}
 {\Gamma(1+\eta-m) \, \Gamma(1-\eta+m+p')} \cdot \,
 {}_2F_1[-p, \, -p'+\eta-m], \, [1+\eta-m], \, \, t), 
  \\
&&\hspace{-1in}  \quad
   A^{(2)}_m(t) \, \, = \, \, \,\,
   \nonumber \\
\label{a1def2}
  &&\hspace{-1in} \quad \quad
     \frac{ \Gamma(1+p) \cdot \, t^{(m-\eta)/2}}
{\Gamma(1-\eta+m) \, \Gamma(1+\eta-m+p)} \cdot \,
{}_2F_1([-p', \, -p-\eta+m], \, [1 -\eta+m], \, \, t), 
\end{eqnarray}
where each $ \, A^{(1)}_m$ and $ \, A^{(2)}_m$ separately gives a Toeplitz
matrix. For the purposes of this paper {\em it is sufficient to consider}
$ \, \xi= \, 0$ and 
see that after taking suitable limits (see eqs (128-130) of \cite{bmm})
\begin{eqnarray}
\label{Adef2}
A_m^{(p,p',\eta)}(t) \,  = \, \, \, \left\{
  \begin{array}{ll}
     A^{(1)}_m(t)   &  \hbox{for $m\, \le \, \eta$} \\
     A^{(2)}_m(t)   &  \hbox{for $m\, \ge \, \eta$}
    \end{array}
\right.
\end{eqnarray}
In~\cite{gil} it was shown that the sigma equations of these
determinants satisfy Painlev{\'e} VI sigma equations with
the Okamoto  parameters:
\begin{eqnarray}
 &&\hspace{-.3in}
     n_1 \, = \, \, (\tilde{N}+\eta+p-p')/2, \quad  \quad  \quad 
     n_2 \, = \, \, (\tilde{N}-\eta-p+p')/2,
     \nonumber\\
  \label{okamotofordet}
  &&\hspace{-.3in}
     n_3 \, = \, \, (\eta-\tilde{N}-p-p')/2, \quad  \quad \quad 
     n_4 \, = \, \, (\eta+\tilde{N}+p+p')/2.
\end{eqnarray}
In~\cite{bmm} we have computed the Okamoto parameters for the sigma
non-linear differential equations for
the two factors of $ \, C(M,N) \, $ for $\, \nu \, = \, -k \, $  with $ \, M+N$ odd,
$\, M \, \le \, N$, and for the four factors of
$ \, C(0,N)$ with $N$ odd. From (\ref{okamotofordet}) we see that the
parameters of the associated Forrester-Witte determinants are:
\begin{eqnarray}
\label{detparams}
\hspace{-1in}
\quad \quad 
{\tilde N}  = \, n_1+n_2, \quad\quad  \eta=\,n_3+n_4, \quad\quad
  p= \, -n_2-n_3, \quad \quad  p'= \, -n_1+n_4.
\end{eqnarray}
The non-linear differential equation is invariant under permutations of the parameters
$n_i$ and the change of sign of any pair. However, the parameters of
the determinants (\ref{detparams}) do not share this symmetry which
means that several different determinants have the same sigma equation. 
We must, of course select those choices for $n_i$ which make
$ \, \tilde N$, the size of the Toeplitz matrix, an integer. 

We must, therefore
consider all possible $ \, \tilde{N} \, \times \, \tilde{N}$ determinants which
can be obtained from a given set of Okamoto parameters. This is done
in \ref{appF} \ref{appG} and \ref{appH}, where we see that for both the two and  
four factor cases, the determinants can be grouped in sets of four, and it can be shown, by direct
computation that the ratios of these four determinants are up to constants, powers of 
 $ \, t \, $ and $ \, (1-t)$. 
These powers of $ \, t \, $ and $ \, (1-t) \, $ up to constants do not contribute to
the sigma equations and must be studied independenly to obtain the
associated factorization of the Forrester-Witte determinants.

\vskip .2cm 

\section{Conclusion}
\label{Conclusion}

The factorization of the low temperature correlation functions $\, C(M,N)\, $
for $\, \nu \, = \, -k\, $
with $\, M+N \, $ odd, $\, M \, \le \, N$ considered in this  paper, corresponds to a factorization
of Toeplitz determinants that has been seen in many papers, in particular
miscellaneous contexts (random matrices, see~\cite{tw,Probability,ForresterPerk}).
Here we address a much more rigid and strong property than a simple factorization
property. We try to understand how a sigma function, solution of
an Okamoto sigma form of Painlev\'e VI non-linear differential equation,
can actually be the  {\em sum} of several sigma functions
each being {\em also solution} of non-linear ODEs with a Painlev\'e property
that can be reduced to Okamoto sigma form of Painlev\'e VI.
This is some kind of {\em addition  formula of Painlev\'e transcendental functions}, 
similar to formulae of addition of elliptic functions\footnote[1]{There is, in fact,
a clear precedent for the phenomenon of multiplicative
identities for Painlevé $\tau$-functions and additive identities for the sigma-forms
observed here in the case of Painlev\'e VI, that was actually proven in the case
of Painlev\'e II and Painlev\'e III in as general setting as possible,
see equations (42), (43), (44) in Proposition 3 of~\cite{NSWitte}. Let us note
that paper~\cite{NSWitte} shows in that particular case that 
the factorisation and additive relations
directly follow from these (canonical/folding) transformations
(see Proposition 3 of~\cite{NSWitte}).}.
The case of the factorization of $\, C(0,N)$, for $\, \nu \, = \, -k \, $
with $\, N \, $ odd, in {\em four} factors corresponds to a quite remarkable
situation of four  sigma functions solutions of Okamoto sigma form of Painlev\'e VI, their sum
being also solution of an Okamoto sigma form of Painlev\'e VI. In that case
we do not have a change of the variable $\, t$. The more general case
of the factorization of the low-temperature $\, C(M,N)$, for $\, \nu \, = \, -k$
with $\, M+N \, $ odd, $\, M \, \le \, N$, in {\em two} factors is more illuminating, since it
actually introduces {\em selected} (Landen) {\em changes of variables} and functions,
enabling to understand what kind of mathematical structures have to be introduced. With
this last example, we provide the simplest example
of such kind of sum of two sigma Painlev\'e transcendental functions
being sigma Painlev\'e transcendental. This paper showed that these
factorization properties for the correlation functions, or the additivity properties on the
corresponding sigma's, {\em can actually be lambda-extended to one-parameter family of solutions}
of the corresponding Okamoto sigma form of Painlev\'e VI and non-linear ODE's reducible to
Okamoto sigma form of Painlev\'e VI.

The Painlev\'e transcendentals can be seen as {\em deformations of elliptic functions}
as very well illustrated in~\cite{Manin}. If one assumes that a non-linear ODE
with the Painlev\'e property, namely having fixed critical points, can be reduced
to  Painlev\'e VI, more precisely to an Okamoto sigma form of Painlev\'e VI equations,
these {\em algebraic} changes of variables {\em cannot be arbitrary}:  they have to be ``compatible''
with the underlying elliptic curve structure. These algebraic change of variables must not only
preserve the set\footnote[1]{Each of the critical points  $\, 0$, $\, 1$ and $\, \infty$
 do not have to be preserved individually.} of critical points $\, 0$, $\, 1$ and $\, \infty$, 
they must be compatible\footnote[5]{The new lattice of periods
must be included in the original lattice, or
conversely the original lattice of periods   must be included
in the new lattice of periods. } {\em with the lattice of periods}: in fact
these transformations are highly selected, they must be {\em isogenies},
{\em modular correspondences associated with modular curves}~\cite{Heegner,Modular,Hindawi}.
Other non-linear differential equations with the Painlev\'e property of having fixed critical
points,  have been found to reduce to Painlev\'e 
transcendentals\footnote[9]{In particular it  turns  out  that  there  are
  two  second-order  (but {\em fourth}-degree)
Painlev\'e-type equations, labelled as BP-IX and BP-X (see (1.10) with $\, m\, = 4$ in~\cite{Sakka}).
BP-IX and BP-X were solved in terms of elliptic functions
or the special case of the second Painlev\'e transcendent~\cite{Sakka}.} up to
change of variables and functions~\cite{cosgrove,Cosgrove,Clarkson}.
It would be interesting to see if more involved non-linear ODEs for other
Ising correlations $\, C(M,N)$
($\, \nu \, \ne \, -k$), can also be reduced to 
Painlev\'e VI transcendentals, possibly up to  changes of  variables
 corresponding to selected modular correspondences. 

\vskip .1cm

\vskip .3cm

{\bf Acknowledgments:} One of us (JMM) would like to thank  R. Conte and  I. Dornic 
for many discussions on Painlev\'e equations. 

\vskip .3cm

\appendix

\section{Examples of factorizations of $\, C(M,N)$, $\, M+N \, $ odd, $\, M \, \le \, N$}
\label{appA}

We give here examples of factors $\, g_{\pm}(M,N) \, $ of (\ref{twofactors}) 
for $\, C(M,N)$ with $\, k= \, -\nu \, $ and $\, M+N \, $ odd. We use the previous notation (\ref{elliptic})  
\begin{eqnarray}
 \hspace{-.3in} \,  \quad  \quad 
  {\tilde E}(k) \, = \, \, \,  {{2} \over {\pi}}  \cdot \,  E(k), \quad \quad  \quad 
  {\tilde K}(k) \, =\, \, \,  {{2} \over {\pi}} \cdot \, K(k), 
\end{eqnarray}
with $ \, K(k)$ and $ \, E(k)$ the complete elliptic integrals of the first
and second kind.

We note that the first $ \, N+4 \, $ terms in the term $ \pm\,  k^{N+1} = \, \pm \,  t^{(N+1)/2}\, $
in the form (\ref{expansion}) 
are fitted  by a constant times
\begin{eqnarray}
  \hspace{-1in} \quad \quad \quad \quad \quad 
{}_3F_2 \Big( [\frac{N+1}{2},\frac{N+M+2}{2},\frac{N-M+2}{2}],
[N+1,\frac{N+3}{2}], \, k^2 \Big). 
\end{eqnarray} 
This holds until large primes appear in the coefficients.

\subsection{Factors for $\, C(M,N) \, $ with $\, M+N$ odd and $\,  N$ even}
In all the examples shown the expansions are carried out to the point
that large primes appear in both the terms with even and odd powers of $\, k$.

\subsubsection{Factors for $ \, C(1,2)$\\}
The two factors of $ \, C(1,2)$ read:
\begin{eqnarray}
 && \hspace{-.9in}
g_{\pm}(1,2) \, \, = \, \, \,
    \frac{ (1 \, - k^2)^{-1/8}}{k} \cdot \,
    (1 \, \pm k)^{1/2}  \cdot \, \Bigl(\tilde{E} \, \,  -(1\, \mp k) \cdot \, \tilde{K}\Bigr)
 \nonumber\\
  && \hspace{-.8in}
    \, \,  = \, \,  \, 
    1  \,  \, \, \pm k^3  \cdot \, \Bigl(\frac{1}{2^4} \, \, +\frac{3}{2^6} \, k^2 \, \,  \,
+\frac{3 \cdot 5^2 }{2^{11}} \, k^4 \, \,  \,
+\frac{5 \cdot 7^2}{2^{13}} \, k^6 \, \,  \,
+\frac{3^3\cdot 5\cdot 7^2}{2^{18}} \, k^{8} 
\nonumber\\
  && \hspace{-.3in} \quad \quad 
     +\frac{3^3 \cdot 7 \cdot 11^2}{2^{20}} \, k^{10} \, \, \, 
+\frac{661\cdot 1949}{2^{26}} \, k^{12} \, \, \, \, +\cdots\Bigr) 
\\
  && \hspace{-.6in}
     +k^{8} \cdot \, \Bigl(\frac{3}{2^{14}} \, \, 
+\frac{3}{2^{13}} \, k^2 \, \, 
+\frac{3^3 \cdot 5}{2^{18}} \, k^4 \, \, 
+\frac{3 \cdot 5 \cdot 11}{2^{18}} \, k^6 \, \, 
+\frac{3 \cdot 5 \cdot 7 \cdot 7321}{2^{30}} \, k^8 \, \,  \, 
     +\cdots\Bigr).
     \nonumber
\end{eqnarray}

\subsubsection{Factors for $ \, C(1,4)$ \\}
The two factors of $ \, C(1,4)$ read:
\begin{eqnarray}
&& \hspace{-.8in} \quad 
g_{\pm}(1,4)\, \, = \,\, \,
 -\frac{4}{3} \cdot \, \frac{(1 \, - k^2)^{-1/8}}{k^4} \cdot \,  (1 \, \pm  k)^{1/2}  \cdot \,
     \Bigl((k^2 \,  \mp 3 k  \, +1)  \cdot \,  {\tilde  E}^2
\nonumber\\
&&\hspace{-.3in}   \quad  \quad   
\, +2 \cdot \, (1 \, \mp k)\cdot \, (k^2 \,  \pm k\, -1)) \cdot \,  {\tilde E}{\tilde K}
      \, \, \,  +(1\mp k) \, (1 \, -k^2) \cdot \,  {\tilde K}^2\Bigr) 
\nonumber\\
&& \hspace{-.6in}\quad 
     = \,  \, \,
  1 \,\,\, \,  \pm k^5 \cdot \, \{\frac{3}{2^8} \, +\frac{3\cdot 5}{2^{10}} \, k^2 \, \, 
+\frac{5\cdot 7^2}{2^{14}} \, k^4 \, \, 
+\frac{3^3\cdot 5\cdot 7}{2^{16}} \, k^{6} \, \, 
\nonumber\\
&& \hspace{-.6in} \, \, \quad  \quad \quad 
   +\frac{3^3\cdot 5 \cdot 7 \cdot 11^2}{2^{23}} \, k^{8} \, \, 
     +\frac{3 \cdot 7 \cdot 11^2 \cdot 13^2}{2^{25}} \, k^{10} \, \, 
+\frac{3^2 \cdot 5 \cdot 7 \cdot 11^2 \cdot 13^2}{2^{29}} \, k^{12} \, \, 
   \nonumber\\
&&\hspace{-.6in} \quad  \quad  \quad \quad 
+\frac{3^2 \cdot 5 \cdot 11 \cdot 13^2 \cdot 17^2}{2^{31}} \, k^{14}
\, \, +\frac{5 \cdot 1087 \cdot 267637}{2^{37}} \, k^{16} \, \, \,  \, +\cdots\Bigr) 
\nonumber\\
&&\hspace{-.5in} \quad 
 +k^{12} \cdot \, \Bigl( \frac{5}{2^{20}}\, \, 
+\frac{3\cdot 5}{2^{20}} \, k^{2} \, \, 
+\frac{3 \cdot 5 \cdot 7 \cdot 139}{2^{29}} \, k^{4} \, \, 
 +\frac{5 \cdot 7^2 \cdot 23}{2^{27}} \, k^{6}
\nonumber\\
&&\hspace{-.5in}\, \, \quad  \quad \quad  \quad 
+\frac{3^2\cdot 7 \cdot 19\cdot 827}{2^{34}} \, k^{8} \, 
     \, \,  \, +\cdots\Bigr).
\nonumber
\end{eqnarray}

\subsubsection{Factors for $ \, C(3,4)$ \\}
The two factors of $ \, C(3,4)$ read:
\begin{eqnarray}
&&\hspace{-.7in}
 g_{\pm}(3,4) \, = \, \, \,
 \frac{4}{45} \cdot  \,\frac{ (1 \, - k^2)^{-1/8}}{k^4} \cdot  \,  (1 \, \pm  k)^{1/2}  \cdot  \,
\Bigl(   (k^4 \, \pm 15k^3-16k^2 \, \pm 15k \, +1) \cdot \,  {\tilde E}^2
\nonumber\\
&&\hspace{-.6in} \quad \quad  \quad  \quad 
 -2 \cdot \, (1 \, \mp k) \cdot \, (3k^4 \, \pm  k^3 \, -2k^2 \, \pm 13 k \, +1) \cdot  \,{\tilde E}{\tilde K}
\nonumber\\
&&\hspace{-.6in} \quad \quad  \quad \quad  \quad  \quad 
  \, \,      -(1 \, -k^2)\cdot  \, (1 \, \mp k) \cdot  \, (3k^2 \, \mp 10 k \, -1) \cdot \,  {\tilde K}^2\Bigr)
\nonumber
\end{eqnarray}
\begin{eqnarray}
&&\hspace{-.6in}
 = \, \, \,
 1 \, \, \, \pm k^5 \cdot \, \Bigl( \frac{7}{2^8} \, \, + \frac{3^3}{2^{10}} \, k^2 \,  \, 
+ \frac{5 \cdot 7 \cdot 11}{2^{14}} \, k^4 \,  \, 
+ \frac{3 \cdot 5 \cdot 7 \cdot 13}{2^{16}} \, k^{6}
\nonumber\\
&&\hspace{-.6in} \quad \quad \quad  \quad 
     + \frac{3^4 \cdot  5^2 \cdot 7 \cdot 11}{2^{23}} \, k^{8} \,  \, 
+ \frac{3 \cdot 7 \cdot 11^2 \cdot 13\cdot 17}{2^{25}} \, k^{10} \,  \, 
   + \frac{3 \cdot 7 \cdot 11^2 \cdot 13^2 \cdot 19}{2^{29}} \, k^{12}
\nonumber\\
&&\hspace{-.6in} \quad  \quad \quad \quad \quad  \quad 
     +\frac{3^3 \cdot 5 \cdot 7 \cdot 11 \cdot 13^2 \cdot 17}{2^{31}} \, k^{14} \,  \, 
+ \frac{3^3\cdot 5 \cdot 7^3 \cdot 11 \cdot 3457}{2^{37}} \, k^{16} \,  \, 
+\cdots\Bigr)  
\nonumber\\
&&\hspace{-.6in} \quad  \quad \quad 
     +k^{12}  \cdot \, \Bigl( \frac{3 \cdot 7}{2^{20}} \,  \, 
+\frac{5 \cdot 11}{2^{20}} \, k^{2} \,  \, 
+\frac{3^4 \cdot 5\cdot 7 \cdot 17}{2^{29}} \, k^{4} \,  \, 
   +\frac{3 \cdot 5 \cdot 7 \cdot 163}{2^{27}} \, k^{6} \,
\nonumber\\
&&\hspace{-.6in} \quad  \quad  \quad \quad \quad \quad  \quad  \quad 
 +\frac{3^2 \cdot 7 \cdot 11 \cdot 4051}{2^{34}} \, k^{8} \,  \,  \,  \, +\cdots\Bigr). 
\end{eqnarray}

\subsection{Factors of $ \, C(M,N)$ with $ \, M+N$ odd,  $ \, N$ odd, $ \, M \,  \neq  \, 0$}
 
\subsubsection{Factors for $ \, C(2,3;t)$ \\} 
The two factors of $ \, C(2,3)$ read
\begin{eqnarray}
&&\hspace{-1in} \quad \quad \quad 
\label{g23}
g_{+}(2,3)\, \, = \, \, \,
  -\frac{2}{3} \cdot \, \frac{(1 \, - k^2)^{-1/8}}{k^2} \cdot \, (1-k^2)^{1/2}
\nonumber \\
&&\hspace{-1in}   \quad \quad \quad \quad \quad  \quad \quad 
\times \,
\Bigl( 3 \, \tilde{E}^2\, \, +(k^2-5) \cdot \, \tilde{E}\tilde{K} \, \, \, \,  -2 \cdot \, (k^2 -1)  \cdot \,  \tilde{K}^2\Bigr), 
\nonumber \\
&&\hspace{-1in}\quad \quad \quad 
g_{-}(2,3) \, \, = \,  \,\,
   \frac{2}{3} \cdot \, \frac{(1 \, - k^2)^{-1/8}}{k^2} \cdot \,
   \Bigl( (k^2+1) \cdot \, \tilde{E} \, \,  \,  +(k^2-1) \cdot \, \tilde{K}\Bigr),      
\end{eqnarray} 
which expand in the form (\ref{expansion}) as:
\begin{eqnarray}
\label{expg23}
&&\hspace{-.8in} \quad 
 g_{\pm}(2,3)\,\,  =\,\,\,   1 \,  \, \, +k^{10} \cdot \,  \Bigl( \frac{7}{2^{17}} \,\, 
 +\frac{3 \cdot 5^2  \cdot 7}{2^{22}}\, k^2\,\, +\frac{3^3 \cdot 5 \cdot 7^2}{2^{25}}\, k^4
\nonumber\\ 
 &&\hspace{-.8in} \quad \quad \quad \quad \quad \quad 
 +\frac{3 \cdot 5  \cdot 7 \cdot 11 \cdot 61}{2^{28}}\, k^6\, \,
 +\frac{5 \cdot 11 \cdot 107 \cdot 233}{2^{32}}\, k^8
 \, \, \, +\cdots \Bigr) 
\nonumber\\
 &&\hspace{-.8in} \quad  \quad 
 \pm k^4 \cdot \,  \Bigl( \frac{5}{2^7}\,  +\frac{5 \cdot 7}{2^{10}}\,  k^2\, \, 
+\frac{3^3 \cdot 5 \cdot 7}{2^{15}}\,  k^4 \, \, 
     +\frac{3 \cdot 7^2 \cdot 11}{2^{16}}\,  k^6\, \,
     +\frac{3^2 \cdot 5 \cdot 7 \cdot 11 \cdot 13}{2^{21}}\,  k^8 \, \, 
  \nonumber\\
 &&\hspace{-.8in} \quad \quad \quad  \quad \quad \quad \quad 
     +\frac{3^4 \cdot 5 \cdot 11^2 \cdot 13}{2^{25}}\,  k^{10}\, \, 
     +\frac{3 \cdot 5 \cdot 7 \cdot 11^2 \cdot 13^2 \cdot 17}{2^{31}}\,  k^{12}
 \nonumber\\
 &&\hspace{-.8in} \quad \quad \quad  \quad \quad  \quad \quad \quad \quad 
+\frac{5 \cdot 7^2 \cdot 257 \cdot 1049}{2^{32}}\, k^{14}\, \, \, 
+\cdots \Bigr).
\end{eqnarray}
{\bf Remark:} Recalling the previous variable $\, x$ (see (\ref{inverse}))
corresponding to the
square of the Landen modulus $\, k_L$ (see (\ref{inverse})),
one can actually also rewrite $ \, C(2,3;t)$ as the product of {\em two other factors}:
\begin{eqnarray}
\label{kL}
 &&\hspace{-.98in}
\quad   \quad \quad \, \, 
C(2,3;t) \, \, = \, \, \, \,  {{16 } \over {9 \, x^4 }} \cdot \,
 \Bigl( 4\cdot \, (x-1) \cdot \, (x-2) \, \,\, +(x^2 -8\, x +8) \cdot \, \sqrt{ 1 \, -x}  \Bigr)^{1/2}
 \nonumber\\
 &&\hspace{-.8in} \, \, \quad \quad \quad \quad \quad 
    \times \,
  \Bigl( (x \, -2) \cdot \, {\tilde E}_L \, \, \,  -2 \cdot \, (x \, -1)  \cdot \,  {\tilde K}_L \Bigr)
 \nonumber\\
&&\hspace{-.8in} \, \, \,  \quad \quad \quad \quad \quad \quad \quad 
 \times \,
 \Bigl(  3 \cdot \, {\tilde E}_L^2 \, \, \, \,  -(x \, -1)\cdot \, {\tilde K}_L^2
 \, \, \, \, +2 \cdot \, (x\, -2) \cdot \, {\tilde E}_L \, {\tilde K}_L  \Bigr)
\end{eqnarray}
where 
\begin{eqnarray}
&& \hspace{-.36in} \quad \quad \quad 
   {\tilde K}_L \, \, = \,  \,   \,\, 
   \frac{2}{\pi} \cdot \, K\Bigl( {{ 2 \, \sqrt{k} } \over {1 +k}} \Bigr) \, \, = \, \,\, \, 
   {}_2F_1\Bigl([\frac{1}{2},\frac{1}{2}], \, [1], \, x\Bigr),
   \quad
  \nonumber\\
\label{ellipticLanden}  
  &&\hspace{-.36in} \quad \quad \quad 
     {\tilde E}_L\,\,  = \,  \,  \,\, 
     \frac{2}{\pi} \cdot \,  E\Bigl( {{ 2 \, \sqrt{k} } \over {1 +k}} \Bigr)  \, \, = \,  \,\, \, 
  {}_2F_1\Bigl(\frac{1}{2},-\frac{1}{2}], \, [1],  \, x\Bigr), 
\end{eqnarray}
The corresponding $\, \sigma$'s for these two factors also verify Okamoto sigma forms
of the Painlev\'e VI equation. We have similar
results for all the (low-temperature) correlations $\, C(M, \, N)$ when $\, \nu = \, -k$.

\vskip .1cm 

\subsubsection{Factors for $ \, C(2,5)$ \\}
The two factors of $ \, C(2,5)$ read
\begin{eqnarray}
&&\hspace{-1in}
g_{+}(2,5)\, \, =\, \,\,
   -\frac{16}{45} \cdot \,  \frac{(1 \, - k^2)^{-1/8}}{k^6}  \cdot \,
 \Bigl( (7k^4-22k^2+7) \cdot \, {\tilde E}^3
     \,  \, -5 \cdot  \, (1 \, -k^2)^3 \cdot \,  {\tilde K}^3
 \nonumber\\
&&\hspace{-.9in}   \quad 
     -(11k^2-17) \cdot \, (1 \, -k^2)^2 \cdot \,  {\tilde E}{\tilde K}^2
     \, \,   -(1 \, -k^2) \cdot \, (2k^4 -33k^2+19) \cdot \, {\tilde E}^2{\tilde K}
     \Bigr),  
     \\
  &&\hspace{-1in}
     g_{-}(2,5)\, \, =\, \,\,
     -\frac{16}{45}\cdot \, \frac{(1 \, - k^2)^{-1/8}}{k^6} \cdot \, (1-k^2)^{1/2} \cdot \, 
   \Bigl( (2k^4+13 k^2+2)  \cdot \,  {\tilde E}^2
   \nonumber\\
  &&\hspace{-.2in}
     +(7 k^4-15 k^2-4) \cdot \,  {\tilde E} {\tilde K} \,\, \, \,
     +2 \cdot \, (1 \, +2 k^2) \cdot \, (1 -k^2) \cdot \, {\tilde K}^{2}\Bigr), 
\end{eqnarray}
which expand in the form (\ref{expansion}) as:
\begin{eqnarray}
  &&\hspace{-.98in} \quad \quad 
     g_{\pm}(2,5)\, \, = \, \, \,
     1 \, \, +k^{14}\cdot  \, \Bigl( \frac{3^2 \cdot 5}{2^{25}}\, \, 
+\frac{3^2 \cdot 7^2 \cdot 11}{2^{30}}\, k^2\, \, 
   +\frac{3 \cdot 5^2 \cdot 7^2 \cdot 11}{2^{32}}\, k^4
   \nonumber\\
  &&\hspace{-.4in} \quad 
     +\frac{3^2 \cdot 7 \cdot 11 \cdot 13 \cdot 239}{2^{37}}\, k^6\, \, 
     +\frac{3^2 \cdot 5 \cdot 7 \cdot 13 \cdot 12289}{2^{41}}\, k^8
     \, \, \,\,  +\cdots    \Bigr)
     \nonumber\\
  &&\hspace{-1in} \quad \quad \quad 
     \pm\,  k^6 \cdot \, \Bigl(\frac{7}{2^{10} }\, \,  +\frac{3^2 \cdot 5 \cdot 7}{2^{15}}\, k^2\, \, 
+\frac{3^2 \cdot 7 \cdot 11}{2^{16}}\, k^4\, \, 
   +\frac{3^2 \cdot 5 \cdot 7 \cdot 11 \cdot 13}{2^{22}}\, k^6
   \nonumber\\
  &&\hspace{-.6in} \quad \quad 
     +\frac{3^2 \cdot 5^2 \cdot 11^2 \cdot 13}{2^{25}}\,  k^8 \, \, 
    +\frac{3^2 \cdot 7 \cdot 11^2 \cdot 13^2 \cdot 17}{2^{31}}\,  k^{10}
   \nonumber\\
  &&\hspace{-.5in} \quad \quad 
     +\frac{5 \cdot 7 \cdot 11 \cdot 13^2 \cdot 17 \cdot 19}{2^{31}}\,  k^{12}\, \, 
+\frac{3^2 \cdot 7 \cdot 11 \cdot 13^2 \cdot 17^2 \cdot 19}{2^{36}}\,  k^{14}
     \nonumber\\
  &&\hspace{-.4in} \quad \quad 
 +\frac{3^2 \cdot 5 \cdot 7 \cdot 13 \cdot 17^2 \cdot 19^2 \cdot  23}{2^{40}}\, k^{16}\, \, 
+\frac{3 \cdot 5 \cdot 7^2 \cdot 11 \cdot 9284039}{2^{43}}\, k^{18}\, \, \, 
     +\cdots \Bigr). 
     \nonumber
\end{eqnarray}

\subsubsection{Factors for $ \, C(4,5)$ \\}
The two factors of $ \, C(4,5)$ read
\begin{eqnarray}
  &&\hspace{-.98in}   \quad\quad  \, \, \, \, \, 
  g_{+}(4,5) \, \, = \, \, \, \frac{16}{1575} \cdot \, \frac{(1 \, - k^2)^{-1/8}}{k^6} \cdot  \, 
   \Bigl( (2k^8+111k^6-34k^4+111k^2+2) \cdot  \, {\tilde E}^2
   \nonumber\\
  &&\hspace{-.4in} \quad \quad \quad 
     -(1 \, -k^2) \cdot  \, (43k^6-34k^4+179k^2+4)  \cdot \, {\tilde K}{\tilde E}
     \nonumber\\
  &&\hspace{-.4in} \quad \quad \quad \quad   \quad 
     -2 \cdot \, (1-k^2)^2 \cdot  \, (11k^4-34k^2-1) \cdot \, {\tilde K}^2\Bigr), 
\end{eqnarray}
\begin{eqnarray}
 &&\hspace{-.98in}\quad \quad 
     g_{-}(4,5)\, \, = \, \, \,\,
     \frac{16}{4725} \cdot \, \frac{(1 \, - k^2)^{3/8}}{k^6} \cdot \,
    \Bigl( (25k^6-825k^4-825k^2+25) \cdot \,  {\tilde E}^3
   \nonumber\\
  &&\hspace{-.4in} \quad  \quad  \quad 
  +3 \cdot \, (2k^8-219 k^6+121 k^4+631 k^2-23)  \cdot \,  {\tilde E}^2{\tilde K}
     \nonumber\\ \quad
  &&\hspace{-.4in} \quad \quad\quad  \quad \, \,\,
     +3 \cdot \,(1 \, -k^2) \cdot \, (47k^6-121k^4-459k^2+21)
    \cdot \, {\tilde E}{\tilde K}^2
   \nonumber\\ 
  &&\hspace{-.4in} \quad \quad \quad\quad  \quad \quad 
     +(1-k^2)^2 \cdot \, (69k^4+334k^2-19) \cdot \, {\tilde K}^3\Bigr), 
\end{eqnarray}
which expand in the form of (\ref{expansion}) as:
\begin{eqnarray}
  &&\hspace{-.8in} \quad 
  g_{\pm}(4,5)\, \,= \, \, \,
  1 \, \, +k^{14} \cdot \, \Bigl(\frac{3^3 \cdot 11}{2^{25}} \,
     +\frac{3^3 \cdot 7 \cdot 11 \cdot 13}{2^{30}} \, k^2
     \nonumber\\
  &&\hspace{-.3in} \quad  \quad 
     +\frac{7\cdot 11 \cdot 13 \cdot 197}{2^{32}} \,k^4 \,\,
+\frac{3^3 \cdot 7 \cdot 11 \cdot 13 \cdot 349}{2^{37}} \, k^6 \, \,
     +\cdots\Bigr) 
     \nonumber\\
  &&\hspace{-.6in}
     \pm k^6 \cdot \, \Bigl(\frac{3\cdot 7}{2^{10}} \,\,
+\frac{3^2 \cdot 7 \cdot 11}{2^{15}} \, k^2 \, \,
+\frac{3^2 \cdot 11 \cdot 13}{2^{16}} \, k^4 \, \,
     +\frac{3 \cdot 5^2 \cdot 7 \cdot 11 \cdot 13}{2^{22}} \, k^6
     \nonumber\\
  &&\hspace{-.3in} \, \,
     +\frac{3^2 \cdot 5^2 \cdot 11 \cdot 13 \cdot 17}{2^{25}} \, k^8 \,\,
+\frac{3^2 \cdot 7 \cdot 11^2 \cdot 13 \cdot 17 \cdot   19}{2^{31}} \, k^{10}
   \\
  &&\hspace{-.3in} \, \,
     +\frac{7^2 \cdot 11 \cdot 13^2 \cdot 17 \cdot 19}{2^{31}} \,k^{12} \,\,
+\frac{3^2 \cdot 7 \cdot 11 \cdot 13^2 \cdot 17 \cdot 19 \cdot  23}{2^{36}} \, k^{14}
     \nonumber\\
  &&\hspace{-.3in}
  +\frac{3^2 \cdot 5^3 \cdot 7 \cdot 13 \cdot 17^2 \cdot 19 \cdot  23}{2^{40}}\, k^{16}\,\,
+\frac{3 \cdot 11 \cdot 13 \cdot 29 \cdot 7757221}{2^{43}}\, k^{18} \,\,\,
     +\cdots  \Bigr).
     \nonumber
\end{eqnarray}

\vskip .1cm
\vskip .1cm 

\section{Sum decompositions of sigma functions}
\label{appB}

We give here examples of factors $\,\sigma_{\pm}(M,N;t) \, $ of (\ref{additive}). 

\subsection{Decomposition of $\,\sigma(M,N) \, $ with $\,M+N \, $ odd and $\, N$ even}

\subsubsection{Decomposition for $\,\sigma(1,2)$}

\begin{eqnarray}
  &&\hspace{-.5in}
  \sigma_{\pm}(1,2)\,\, = \,\,\,\,
   \,  \pm k^3 \cdot \, \Bigl( \frac{3}{2^5}\, \, +\frac{3}{2^7}\,k^2\,\, 
+\frac{3^2\cdot 5}{2^{12}}\,k^4\,\, 
     +\frac{3 \cdot 37}{2^{14}}\,k^6
     \nonumber\\
  &&\hspace{-.5in} \quad \quad \quad \quad 
     +\frac{3 \cdot 11 \cdot 19}{2^{17}}\,k^8 \,\, 
     +\frac{3^2 \cdot 5^2 \cdot 17}{2^{20}}\, k^{10} \,\,\,  +\cdots \Bigr) 
     \\
  &&\hspace{-.5in} \quad 
     +k^6 \cdot \, \Bigl(\frac{3}{2^9}\,\,  +\frac{3 \cdot 7}{2^{12}}\,k^2 \,\,  
+\frac{3^3 \cdot 5}{2^{15}}\,  k^4\,\,   +\frac{3\cdot 73}{2^{16}}\,k^6 \,\, 
     +\frac{3 \cdot 7741}{2^{23}}\, k^8 \,\,\,+\cdots\Bigr).
     \nonumber
\end{eqnarray}

\subsubsection{Decomposition for $ \,\sigma(1,4)$}

\begin{eqnarray}
  &&\hspace{-.5in}
 \sigma_{\pm}(1,4) \, \, = \, \, \,\,
 \, \,\, \pm k^5 \cdot \, \Bigl( \frac{3 \cdot 5}{2^9} \,\,
+\frac{3^2 \cdot 5}{2^{11}}\, k^2\, \, 
+\frac{3 \cdot 5^2 \cdot 7}{2^{15}}\, k^4 \, \, 
   +\frac{3^2 \cdot 5^2 \cdot 7}{2^{17}}\,  k^6
   \nonumber\\
  &&\hspace{-.3in} \quad  \quad 
     +\frac{3^4 \cdot 5^2 \cdot 7 \cdot 11}{2^{24}}\,  k^8\, \, 
     +\frac{3^2 \cdot 5 \cdot 23 \cdot 479}{2^{26}}\,  k^{10}
     \, \, \,\,   +\cdots \Bigr) 
  \nonumber\\
  &&\hspace{-.5in} \quad \quad 
     +k^{10} \cdot \, \Bigl(\frac{3^2 \cdot 5}{2^{17}}\, \, 
+\frac{3 \cdot 5 \cdot 23}{2^{19}}\, k^2 \, \, 
+\frac{3 \cdot 5^2 \cdot 7^2}{2^{22}}\, k^4 \,\, 
   +\frac{3^2 \cdot 5^2 \cdot 7 \cdot 43}{2^{26}}\, k^6
   \nonumber\\
  &&\hspace{-.2in} \quad \quad \quad 
     +\frac{3^3 \cdot 5^2 \cdot 7 \cdot 491}{2^{31}}\, k^8 \, \, 
     +\frac{3^3 \cdot 5 \cdot 11 \cdot 3209}{2^{32}}\, k^{10}
     \,\,\, +\cdots \Bigr). 
\end{eqnarray}

\subsubsection{Decomposition for $\,\sigma(3,4)$}

\begin{eqnarray}
  &&\hspace{-.6in}
 \sigma_{\pm}(3,4)\,\, = \,\,\,\,\,
  \, \, \pm k^5 \cdot \,
     \Bigl(\frac{5\cdot 7}{2^9}\,\,
+\frac{7^2}{2^{11}}\, k^2\, \, \,
+\frac{3^2 \cdot 7^2}{2^{15}}\, k^4 \, \, \,
   +\frac{3 \cdot 5 \cdot 7 \cdot 11}{2^{17}}\, k^6
   \nonumber\\
  &&\hspace{-.2in} \quad 
  +\frac{3 \cdot 5 \cdot 7^2 \cdot 11 \cdot 13}{2^{24}}\, k^8 \,\,
     +\frac{5 \cdot 7^2 \cdot 1301}{2^{26}}\, k^{10}
     \,\,\, +\cdots \Bigr) 
     \nonumber\\
  &&\hspace{-.5in} \quad 
     +k^{10} \cdot \, \Bigl(\frac{5 \cdot 7^2}{2^{17}} \,\,
+\frac{5^2 \cdot 7^2}{2^{19}}\, k^2 \, \,
+\frac{3^2 \cdot 7 \cdot 157}{2^{22}}\, k^4 \,  \,\,
     +\frac{3 \cdot 7 \cdot 7129}{2^{26}}\, k^6
 \nonumber\\
  &&\hspace{-.2in} \quad 
     +\frac{3 \cdot 5 \cdot 7^2 \cdot 11 \cdot 547}{2^{31}}\, k^8 \, \,
     +\frac{5 \cdot 7^2 \cdot 19 \cdot 37 \cdot 47}{2^{32}}\, k^{10}
     \,\, \, +\cdots \Bigr).
\end{eqnarray}

\subsection{Decomposition of $\, \sigma(M,N)$ with $\, M+N$ odd,$\, N$  odd, $\, M \,  \neq  \, 0$}

\subsubsection{Decomposition for $\, \sigma(2,3)$}

\begin{eqnarray}
\label{sigma23app}
  &&\hspace{-.4in} \,  
\sigma_{\pm}(2,3) \, \, = \,\,  \,
   \, \pm k^4 \cdot \, \Bigl(\frac{5}{2^6} \,  \,  \,
+\frac{5^2}{2^{10}}\, k^2 \, \,  \,
+\frac{3 \cdot 5 \cdot 7}{2^{13}} \, k^4 \,  \, \, 
   +\frac{3 \cdot 5^2 \cdot 7}{2^{16}}\, k^6
   \nonumber\\
  &&\hspace{-.2in} \quad \quad \quad 
     +\frac{5^2 \cdot 59}{2^{18}}\, k^8 \, \,  \, 
     +\frac{5 \cdot 28579}{2^{25}}\, k^{10}
     \,\,  \, \,  \,+\cdots \Bigr) 
   \nonumber\\
  &&\hspace{-.5in} \quad \quad \quad 
     +k^8 \cdot \, \Bigl(\frac{5^2}{2^{13}} \, \, \, 
+\frac{5\cdot 11}{2^{14}}\, k^2 \, \,  \, 
+\frac{3 \cdot 5 \cdot 7\cdot 31}{2^{20}}\, k^4 \,  \, \, 
     +\frac{3 \cdot 5^2 \cdot 7 \cdot 11}{2^{21}} \, k^6
     \nonumber\\
  &&\hspace{-.2in} \quad  \quad \quad \quad \quad \quad 
     +\frac{5^2 \cdot 17 \cdot1531}{2^{28} } \, k^8
     \, \, \,  \, \, +\cdots \Bigr). 
\end{eqnarray}

\subsubsection{Decomposition for $\,\sigma(2,5)$}

\begin{eqnarray}
  &&\hspace{-.6in}
\sigma_{\pm}(2,5) \,\,= \,\,\,\, 
   \,  \pm k^6 \cdot \, \Bigl( \frac{3 \cdot 7}{2^{10} } \, \, \, 
+\frac{3 \cdot 7^2}{2^{13}}\, k^2 \, \, 
+\frac{3^3 \cdot 5 \cdot 7}{2^{16}}\, k^4 \, \,  
     +\frac{3^2 \cdot 5\cdot 7^2 \cdot 11}{2^{21}}\, k^6
     \nonumber\\
  &&\hspace{-.5in} \quad 
     +\frac{3^2 \cdot 5 \cdot 7^2 \cdot 11\cdot 13}{2^{25}}\, k^8 \,  \, 
+\frac{3^3 \cdot 7^2 \cdot 11^2 \cdot 13}{2^{28}}\, k^{10} \,  \, 
     +\frac{3 \cdot 7^2 \cdot 94823}{2^{31}}\,k^{12}
     \,\, \, +\cdots \Bigr) 
     \nonumber\\
  &&\hspace{-.5in}\quad 
     +k^{12} \cdot \, \Bigl(\frac{3 \cdot 7^2}{2^{20}} \, \,  
+\frac{3 \cdot 7 \cdot 31}{2^{21}}\, k^2 \, \,  
+\frac{3^3 \cdot 5 \cdot 7 \cdot 131}{2^{28}}\, k^4 \,  \, 
     +\frac{3^3 \cdot 5 \cdot 7^2 \cdot 47}{2^{29}}\, k^6
     \nonumber\\  
  &&\hspace{-.3in} \quad \quad 
     +\frac{3^3 \cdot 5 \cdot 7^2 \cdot 11 \cdot 157}{2^{34}}\, k^8 \,  \, 
   +\frac{3^3 \cdot 7^2 \cdot 11 \cdot 1709}{2^{35}}\, k^{10}\, 
   \nonumber\\
  &&\hspace{-.3in} \quad \quad \quad \quad 
     +\frac{3\cdot 7^2 \cdot 131 \cdot 293 \cdot 1187}{2^{43}}\, k^{12}
     \, \, \,  +\cdots \Bigr). 
\end{eqnarray}`

\subsubsection{Decomposition for $\, \sigma(4,5)$}

\begin{eqnarray}
  &&\hspace{-.8in}
\sigma_{\pm}(4,5) \, \, = \,\,\,
\, \,\pm k^6 \cdot  \, \Bigl( \frac{3^2 \cdot 7}{2^{10} }\,  \, 
+\frac{3^3 \cdot 7}{2^{13}}\, k^2  \, \, 
+\frac{3^4 \cdot 11}{2^{16}}\, k^4  \, \, 
     +\frac{3^3 \cdot 5 \cdot 11 \cdot 13}{2^{21}}\, k^6
     \nonumber\\
  &&\hspace{-.8in} \quad 
     +\frac{3^2 \cdot 5^2 \cdot 7 \cdot 11 \cdot 13}{2^{25}} \, k^8  \, \, 
+\frac{3^4 \cdot 7 \cdot 11 \cdot 13 \cdot 17}{2^{28}}\, k^{10} \, \, 
     +\frac{3^3 \cdot 7 \cdot 23 \cdot 43 \cdot  47}{2^{31}}\, k^{12}
     \,\,\, +\cdots \Bigr) 
     \nonumber\\
  &&\hspace{-.6in}
     +k^{12} \cdot  \, \Bigl(\frac{3^3 \cdot 7^2}{2^{20}}\, \, 
+\frac{3^3 \cdot 7 \cdot 19}{2^{21}}\,k^2\, \, 
+\frac{3^4 \cdot 7 \cdot 11 \cdot 79}{2^{28}}\,k^4 \, \, 
     +\frac{3^2 \cdot 7 \cdot 11 \cdot 1409}{2^{29}}\, k^6
     \nonumber\\
  &&\hspace{.5in}
     +\frac{3^2 \cdot 5 \cdot 11 \cdot 13 \cdot 4651}{2^{34}}\, k^8 \,  \, 
     +\frac{ 3^4 \cdot 11 \cdot 13 \cdot 4871}{2^{35}}\, k^{10}
     \nonumber\\
  &&\hspace{.4in} \quad \quad  \quad 
     +\frac{3^3 \cdot 7 \cdot 37 \cdot 1932439}{2^{43}}\, k^{12}
\, \,  \,  \,  +\cdots\Bigr).  
\end{eqnarray}
Continuing in this fashion we obtain the form (\ref{lambdasigmapm}).

Note that the first $\,N+1$ terms of the  $\,\pm \,  k^{N+1}$ terms
are proportional to $\, {}_2F_1([\frac{N+M}{2},\frac{N-M}{2}],\, [N+1], \, k^2)$.

The series following $ \,\, \pm \,  k^{N+1} \, $ developes large primes at
order $\, k^{3(N+1)}$ and the series following $ \,\, k^{2\, (N+1)} \, $ developes
much larger primes at order $ \, k^{4 \, (N+1)}$. This is expected from the
result (\ref{final}) by making a recursive expansion of the sigma form of Painlev\'e VI
function $\, h(x)$ which satisfies (\ref{okamoto}) with parameters
(\ref{parameters}).

\vskip .1cm

\section{Reduction to Okamoto form: the $\, C(2, \, 3)$ example.}
\label{Reduc}

Let us illustrate the reduction of section  \ref{nonlin}  of the non-linear differential
equation (\ref{nonlineareq}) to the Okamoto sigma-form of Painlev\'e VI (\ref{okamotorewrit}),
using  the (Landen) substitution (\ref{tx}) together with transformation (\ref{salah}),
on a simple example associated with the two factors of the
low-temperature correlation function $\, C(2, \, 3)$. The two factors $\, g_{+}(2, \, 3) \, $
and $\, g_{-}(2, \, 3) \, $ have been given previously (see (\ref{g23})):
\begin{eqnarray}
\label{g23t}
 &&\hspace{-1in} \quad  \quad 
 g_{+}(2,3)\, \, = \, \, \,
  -\frac{2}{3} \cdot \, \frac{ (1\, -t)^{3/8} }{t} 
 \cdot  \,
    \Bigl( 3 \cdot \, \tilde{E}^2\, \, +(t \, -5) \cdot \, \tilde{E}\tilde{K}
    \, \, \,  -2 \cdot \, (t \,  -1)  \cdot \,  \tilde{K}^2\Bigr), 
 \nonumber \\
&&\hspace{-1in} \quad  \quad 
g_{-}(2,3) \, \, = \,  \,\,
   \frac{2}{3} \cdot \,
   \frac{ (1\, -t)^{-1/8}}{t} \cdot \, \Bigl( (t\, +1) \cdot \, \tilde{E} \, \,   +(t \, -1) \cdot \, \tilde{K}\Bigr),      
\end{eqnarray} 
It is straightforward to see that the corresponding   $\, \sigma_{\pm}(2, \, 3)$,
deduced from formula (\ref{loggpm}),
verify the non-linear equation (\ref{nonlineareq}) for $\, M\, = \, 2$
and $\, N\, =\, 3$. The expansion
of  $\, g_{+}(2, \, 3) \, $ and $\, \sigma_{\pm}(2, \, 3)$
are given previously (see (\ref{expg23}), (\ref{sigma23app})). 
One can rewrite  (\ref{salah}):
\begin{eqnarray}
\label{salah2}
\hspace{-.98in} \, \,  
  h(x) \, = \, \,  \,
  \frac{x}{2} \cdot \, \frac{1 \, +\sqrt{1-x}}{1 \, -\sqrt{1-x}} \cdot \,  \sigma(t) \, \, \,
  - \frac{3M^2-N^2+3}{16}
 \,  \,  \, +\frac{M^2-N^2 +1}{8 \cdot \, (1 \, -\sqrt{1 \, -x})} \cdot \,x.
\end{eqnarray}
The exact expressions of $\, \sigma_{\pm}(2, \, 3) \, $ are rational expressions
in terms of the complete elliptic integrals $\, {\tilde E} \, $ and  $\, {\tilde K} \,$
(see (\ref{elliptic})):
\begin{eqnarray}
 \label{elliptict}  
&& \hspace{-.96in} \quad \quad \quad \quad 
{\tilde K}(t) \, = \,  \,    {}_2F_1\Bigl([\frac{1}{2},\frac{1}{2}], \, [1], \,t\Bigr),
\quad \quad \quad 
{\tilde E}(t) \, = \,  \, 
  {}_2F_1\Bigl(\frac{1}{2},-\frac{1}{2}], \, [1],  \,t\Bigr).
\end{eqnarray}
Performing the (Landen) substitution (\ref{tx}) in these  exact expressions of
$\, \sigma_{\pm}(2, \, 3)$ and using  (\ref{salah2}) for $\, M\, = \, 2$ and $\, N\, =\, 3$, 
it is straightforward to verify (in Maple) that the corresponding function $\, h(x)$
{\em actually verifies the Okamoto sigma-form of Painlev\'e VI}
(\ref{okamotorewrit}) for $\, M\, = \, 2$ and $\, N\, =\, 3$.
The  expansions in $\, x$ of the corresponding  $\, h(x)$ read respectively
\begin{eqnarray}
 \label{exphpm}  
&& \hspace{-.98in}
h_{+} \, = \, 
-{{11 } \over { 8}} \,  +{{1 } \over {4}} \, x \, +{{5 } \over {64}}\, x^2
\, +{{5 } \over {128}}\, x^3\,
 +{{205 } \over {8192}}\, x^4  +{{295 } \over {16384}}\, x^5
 +{{29065 } \over {2097152}}\, x^6 \, + \, \cdots 
\end{eqnarray}
and:
\begin{eqnarray}
 \label{exphpm}  
&& \hspace{-.98in} 
h_{-}  = \,  
-{{11 } \over { 8}} \,  +{{1 } \over {4}} \, x \, +{{5 } \over {64}}\, x^2 \,
+{{5 } \over {128}}\, x^3\,
+{{195 } \over {8192}}\, x^4 +{{265 } \over {16384}}\, x^5
+{{24695 } \over {2097152}}\, x^6 \, +  \, \cdots 
\end{eqnarray}
to be compared with the expansion of the algebraic solution  (\ref{h0exact})
of (\ref{okamotorewrit}) for  $\, M\, = \, 2$ and $\, N\, =\, 3$:
\begin{eqnarray}
 \label{halgexp}  
  && \hspace{-.98in}
 h_{0} \, = \, \,   -{{11 } \over { 8}} \,  +{{1 } \over {4}} \, x \,
 +{{5 } \over {64}}\, x^2 \, +{{5 } \over {128}}\, x^3\,
 +{{25 } \over {1024}}\, x^4 \, +{{35 } \over {2048}}\, x^5 \,
 +{{105 } \over {8192}}\, x^6 \, \, \,  + \, \cdots 
\end{eqnarray}

\vskip .1cm

\section{Painlev\'e VI transcendentals as deformations of elliptic functions and the crucial role of modular correspondences}
\label{appManin}

Let us first recall (R. Fuchs~\cite{Fuchs}, 1907) that the Painlev\'e VI equation (\ref{gambier})
can be written (see (1.1) in~\cite{Manin}, here $\, X \, = \, y \, \, $ in (\ref{gambier})):
\begin{eqnarray}
\label{Fuchs}
&& \hspace{-.98in}  \quad \quad \quad \quad 
   t \cdot \,  (1\, -t) \cdot \, L_2 \cdot \,
   \int_{\infty}^{(X, \, Y)} \, {{ dx} \over { \sqrt{ x \cdot \, (x\, -1) \cdot \, (x\, -t)}}}   
 \nonumber   \\
&& \hspace{-.98in}    \quad \quad  \quad \quad  \quad \quad
   \, \, = \, \,  \, \,
   Y \cdot \,  \left(\alpha \, \, +\beta \cdot\,  \frac{t}{X^2} \,\,
   +\gamma \cdot \, \frac{t-1}{(X-1)^2} \, \,\, 
  +\delta \cdot\,  \frac{t \cdot \, (t-1)}{(X \, -t)^2}\right), 
\end{eqnarray}
where  $\,\, Y ^2 \, = \, \, X \cdot \, (X\, -1) \cdot \, (X\, -t)$, 
and where: 
\begin{eqnarray}
\label{Fuchswhere}
&& \hspace{-.98in} \quad \quad   \quad   \quad   \quad   \quad 
L_2 \, \,\, = \, \,  \,\,
t \cdot \, (1\, -t) \cdot \, {{d^2} \over { dt^2}}
\, \, +(1\, -2\, t) \cdot \, {{d} \over { dt}}  \, \, \, -{{1} \over {4}}.
\end{eqnarray}
Equation (\ref{Fuchs})  provides a clear illustration of the fact that the Painlev\'e VI transcendentals 
{\em can be seen as a deformation of elliptic functions}.  
The crucial role played by the second derivative
with respect to $\, \tau$ (the ratio of periods) displayed below
in the equations (\ref{Manin_sum}), (\ref{Manin_a0a1a0a1}), (\ref{Hitchin}) of \ref{subappManin},
is illustrated by the relation (see (1.18) in~\cite{Manin})
\begin{eqnarray}
\label{d2overdtau2}
&&\hspace{-.98in}  
 {{ \Pi_{i >j} \, (e_i \, -e_j)^2 } \over {9 \cdot \, (e_1\, e_2'  \, -e_2\, e_1')^2 }}
 \cdot \, (e_2 \, -e_1)^{-3/2} \cdot \,
  {{ d^2 } \over { d\tau^2}} \, \, = \, \, \,
  t \cdot \, (1 \, -t)   \cdot \, L_2 \cdot \, (e_2 \, -e_1)^{1/2},
\end{eqnarray}
where the order-two linear differential operator $\, L_2$ is given by (\ref{Fuchswhere}),  
and where the $\, e_i$'s read:
\begin{eqnarray}
\label{defei}
  &&\hspace{-.98in} \quad \quad
  e_i \, \, = \, \, \,  {\cal P}\Bigl({{T_i} \over {2}}, \, \tau\Bigr)
     \quad  \quad \hbox{ where:} \quad \quad
     (T_0, \, T_1, \, T_2, \, T_3) \, = \, \, (0, \, 1, \, \tau, \, 1\, +\tau), 
\end{eqnarray}
This means that, up to some dressing, {\em  the second derivative
with respect to $\, \tau$ is essentially the second order linear
differential  operator} $\,  L_2$
which annihilates the simplest elliptic function, namely the
complete elliptic integral of the first kind
$ \,\, _2F_1([1/2,1/2],\, [1], \, t)$.

\subsection{Modular correspondences and Painlev\'e VI transcendentals: the crucial role of the Landen transformations}
\label{subappManin}

Along a modular correspondence-line it is worth recalling Manin's idea~\cite{Manin}
that the Painlev\'e VI equation
for a particular choice of the four
Okamoto parameters, can be written extremely simply in terms of
 the ratio of periods $\, \tau$.
Let us denote $\,  {\cal P}(z, \, \tau)$  the  $\,  {\cal P}$-Weierstrass function
and
\begin{eqnarray}
  \label{defPz}
&&\hspace{-.58in} \quad \quad  \quad \quad   \quad  \quad   \quad  
   {\cal P}_z(z, \, \tau) \, = \, \,
{{ \partial {\cal P}(z, \, \tau) } \over { \partial z}}. 
\end{eqnarray}
The fundamental role of the {\em Landen transformation}~\cite{Heegner} is
illustrated by the following identity\footnote[1]{See the first equation
  without a number in section 1.6 of~\cite{Manin}.}
on the ${\cal P}$-Weierstrass function:
\begin{eqnarray}
  \label{P-Weier}
  {\cal P}_z\Bigl(z, \, {{\tau} \over {2}}\Bigr) \, \,= \, \,\,
         {\cal P}_z(z, \, \tau) \, \, + {\cal P}_z(z\, + {{\tau} \over {2}}, \, \tau). 
\end{eqnarray}
Manin's result means that the Painlev\'e VI equation can be written in a form
(see equation (1.16) in~\cite{Manin}): 
\begin{eqnarray}
  \label{Manin_sum}
 &&\hspace{-.58in} \quad \quad \quad \quad 
        {{ d^2 z(\tau)} \over { d\tau^2}} \, \,\, = \, \, \, \,
        \Bigl({{1} \over {2\, \pi \, i}}\Bigr)^2 \cdot \, \sum_{i=0}^{3}
\alpha_i  \cdot \, 
 {\cal P}_z(z\, + {{T_i} \over {2}}, \, \tau).
\end{eqnarray}
Switching from the $\, t$ variable to the $\, \tau$ variable,
which is a (differentially algebraic) transcendental
change of variable, changes the non-linear Painlev\'e VI equation
into another equation {\em superficially simpler but where
all the nonlinearity is encapsulated in the Weiertrass function}
$\, {\cal P}_z$.
Recalling~\cite{Manin}, we see that
if $\, z(\tau)$ is solution of the Painlev\'e VI equation
with parameters $ \, (\alpha_0, \, \alpha_1, \, \alpha_0, \, \alpha_1)$
one has
\begin{eqnarray}
  \label{Manin_a0a1a0a1}
 &&\hspace{-.58in} \quad \quad \quad \quad 
{{ d^2 z(\tau)} \over { d\tau^2}} \, \,\, = \, \, \, \,
\alpha_0  \cdot \Bigl(
 {\cal P}_z(z, \, \tau) \, + {\cal P}_z(z\, +{{\tau} \over {2}}, \,\tau) \Bigr)\, 
\nonumber \\
&&\hspace{-.58in} \quad \quad \quad \quad \quad  \quad \quad \quad 
\, \, +\alpha_1  \cdot \Bigl(
 {\cal P}_z(z \, + {{1} \over {2}}, \, \tau)\,
+  {\cal P}_z(z\, +{{1 \, +\tau} \over {2}}, \, \tau) \Bigr), 
\end{eqnarray}
which can be rewritten using the identity (\ref{P-Weier}), as:
\begin{eqnarray}
\label{Manin_a0a1a0a1bis}
\hspace{-.58in}  {{ d^2 z(\tau)} \over { d\tau^2}} \, \, = \, \, \,
   {{1} \over {4}} \cdot \, {{ d^2 z(\tau)} \over { d (\tau/2)^2}} \, \, = \, \, \,
\alpha_0  \cdot \Bigl(
  {\cal P}_z(z, \, {{\tau} \over {2}}) \Bigr)
\, \, +\alpha_1  \cdot \Bigl(
 {\cal P}_z(z \, + {{1} \over {2}},  \, {{\tau} \over {2}})  \Bigr), 
\end{eqnarray}
or
\begin{eqnarray}
\label{Manin_a0a1a0a1ter}
\hspace{-.58in} \quad \quad 
   {{ d^2 z(2\, \tau)} \over { d \tau^2}} \, \, = \, \, \,\, \,
4 \cdot \, \alpha_0  \cdot \Bigl(
  {\cal P}_z(z, \, \tau) \Bigr)
\, \, + 4 \cdot \, \alpha_1  \cdot \Bigl(
 {\cal P}_z(z \, + {{1} \over {2}},  \, \tau)  \Bigr), 
\end{eqnarray}
which means that $\, z(2\, \tau)$ { \em is also solution of the Painlev\'e VI equation
 but with parameters} $ \, (4\, \alpha_0, \, 4\, \alpha_1, \, 0, \, 0)$.
We thus see that we have a representation of the isogeny
$\, \tau \, \rightarrow \, 2 \, \tau$ on the Painlev\'e VI equations
with a price to pay, {\em namely that the parameters are changed}.
For $\, \alpha_1 \, = \, 0$  one gets the remarkable Hitchin's equation:
\begin{eqnarray}
  \label{Hitchin} \quad \quad  \quad \quad 
 {{ d^2 z} \over { d\tau^2}} \, \, = \, \, \,
    -{{1} \over {2 \, \pi^2}}  \cdot
 \, {{ \partial  {\cal P}(z, \, \tau) } \over { \partial z}}. 
\end{eqnarray}
In that simple heuristic case $ \, (\alpha_0, \, \alpha_1, \, \alpha_0, \, \alpha_1)$,
the Landen transformation
{\em preserves} the Gambier form (\ref{gambier}) of Painlev{\'e} VI or
the ``master Painlev\'e equation'' sigma-form  of
Painlev{\'e} VI: we are in the framework of the
so-called~\cite{TOS} ``folding transformations''.
In general  the Landen (or inverse Landen)  transformation matches
an Okamoto  sigma-form  of
Painlev{\'e} VI onto a second order non-linear ODE like (\ref{nonlineareq})
with the Painlev\'e property which is {\em not} of the Okamoto
sigma-form of  Painlev{\'e} VI .

\vskip .1cm

\subsection{More modular  correspondences  and Painlev\'e VI transcendentals. }
\label{deformManin}

Let us recall Mazzocco and Vidunas paper on cubic and quartic transformations on
Painlev\'e VI equation~\cite{CubicQuartic}
(and also Vidunas and Kitaev paper~\cite{VidunasKitev}). In\footnote[2]{Or
  in proposition 3.1 of ~\cite{CubicQuartic}.} equation (1.11) 
of~\cite{CubicQuartic}, one has the following transformation
(for the Tsuda, Okamoto, Sakai case~\cite{TOS}): 
\begin{eqnarray}
\label{cubicquarticTOS}
\hspace{-.98in} \quad 
 \tilde{t} \, \, = \, \, \,
 {{ (1\, + \sqrt{t})^2 } \over { 4 \, \sqrt{t} }},
 \quad \quad \hbox{or:}  \quad
       {{1} \over {\tilde{k} }} \, \, = \, \, \,
       {{2 \, \sqrt{k}} \over { 1\, + k}},
\quad \quad \hbox{where:}  \quad \, \, 
 \tilde{t} \,  = \, \,  \tilde{k}^2, \quad t \, = \, \, k^2, 
\end{eqnarray}
which makes crystal clear that this transformation is,
up to a Kramers-Wannier duality, again a {\em Landen transformation}.

In contrast, in the Picard's case, one has the algebraic transformation
(see proposition 1.4  in~\cite{CubicQuartic})
\begin{eqnarray}
\label{cubicquarticPicard}
\hspace{-.68in} \quad \quad  \quad \quad   \quad \quad \quad \quad  
 \tilde{t} \, \, \, = \, \, \, \,
 \Bigl( {{ t^{1/4} \, + \, 1 } \over {  t^{1/4} \, - \, 1 }} \Bigr)^4,
\end{eqnarray}
which can be rewritten in a more symmetric way
\begin{eqnarray}
\label{cubicquarticPicardsymmetric}
&&\hspace{-.98in} \quad \quad  \, \, \, 
t^4\cdot \, \tilde{t}^4 \, \, \, \, 
-4\cdot \, t^3\cdot \, \tilde{t}^3 \cdot \, (t+\tilde{t}) \, \, \, \, 
+2\cdot \, (t^2 \cdot \, \tilde{t}^2 \, +1) \cdot \,
(3 \cdot \, t^2 -376 \cdot \, t \cdot \, \tilde{t} +3 \cdot \, \tilde{t}^2)
\nonumber \\
&&\hspace{-.98in} \quad \quad \quad \quad
\, \, 
-4 \cdot \, (t \cdot \, \tilde{t}\, +1) \cdot \, (t +\tilde{t})
\cdot \, (t^2 +645 \cdot \, t \cdot \, \tilde{t} +\tilde{t}^2)
 \\
&&\hspace{-.98in} \quad \quad \, 
+t^4 +\tilde{t}^4 \, -752 \cdot \,t \cdot \, \tilde{t} \cdot \, (t^2 \, + \tilde{t}^2)
+13348\cdot \, t^2 \cdot \, \tilde{t}^2 \,
\, \, \, \, \, \, -4 \cdot \, (t +\tilde{t})
\, \, \,  \,  \,  +1 \, \, = \, \, \, \, 0.
\nonumber 
\end{eqnarray}
In order to have a relation between Hauptmoduls, let us perform the change of variables:
\begin{eqnarray}
\label{AB}
\hspace{-.98in} \quad \quad \quad \quad \quad \quad 
A \, \, = \, \, \,
{\frac {27 \cdot \, \,{t}^{2} \cdot \, (1 -t)^{2}}{4 \cdot \, ({t}^{2} -t +1)^{3}}},
\quad \quad \quad
B  \, \, = \, \, \,
{\frac {27 \cdot  \,{\tilde{t}}^{2} \cdot \, (1 \, -\tilde{t})^{2}}{
    4 \cdot \, \left( {\tilde{t}}^{2} -\tilde{t} +1 \right) ^{3}}}. 
\end{eqnarray}
The transformation (\ref{cubicquarticPicardsymmetric}) becomes the modular
equation\footnote[5]{It also corresponds to an isogeny of degree 3 of the underlying Legendre
  elliptic curve $\, w^2 \, = \, \, q\cdot \, (q-1) \cdot \, (q-t)$
  (see page 5 of~\cite{CubicQuartic}). } 
which corresponds to $\, \tau \, \rightarrow \, \, 4 \cdot \, \tau$,
or $\,  \tau \, \rightarrow \, \, \tau/4$
(see section 5.1.1 of~\cite{Modular}). This modular equation can be obtained
from the composition of the fundamental modular curve (corresponding
to the Landen transformation) with itself
(see section 5.1.1 of~\cite{Modular}).

\vskip .1cm 
\vskip .1cm

\section{Factorization of $ \, C(0,5)$ and $ \, C(0,7)$}
\label{appC}

\vskip .2cm

From~\cite {bmm} we can deduce the four factors $\, g_i(0,5)$  for $ \,C(0,5)$.
We display here the $\, \tilde{g}_i(0,N)$ related to the  $\, g_i(0,N)$ by 
\begin{eqnarray}
&&\hspace{-.98in}\, \,  \quad \quad \quad \quad 
  g_i(0,N) \, \, =  \, \,\,
   (1 \, -t)^{N/16} \cdot \, t^{-N/8} \cdot \,  \tilde{g}_i(0,N),
   \quad  \quad  \quad i \, = \, \, 1, 2,
   \nonumber \\
  &&\hspace{-.98in}\, \,  \quad \quad \quad \quad 
   g_i(0,N) \, \, =  \, \,\,
     (1 \, -t)^{-N/16} \cdot \, t^{N/8} \cdot \,  \tilde{g}_i(0,N), 
  \quad   \quad    \quad i \, = \, \, 3, 4.
\end{eqnarray}
These  $\, \tilde{g}_i(0,N)$'s read\footnote[1]{This normalization is chosen to have $\, \tilde{g}_i(0,N)$
series normalised as follows:  $\, \tilde{g}_i(0,N) \, = \, 1 \, + o(t^2)$.} for $\, N= \, 5$:
\begin{eqnarray}
  &&\hspace{-.98in}\, \,
     \tilde{g}_1(0,5) \, \, = \, \, \,
     \frac{2}{3} \cdot \, (1-t)^{-3/8} \cdot \, t^{-1} \cdot \, 
     \left((2t-1) \cdot  \, \tilde{E} \, \,  \, -(t-1)  \cdot \, \tilde{K}\right), 
  \\
  &&\hspace{-.98in}\, \,
  \tilde{g}_2(0,5) \, \, = \, \, \, \frac{2}{3} \cdot \, (1-t)^{-1/8} \cdot \, t^{-1}  \cdot \, 
     \left((t+1) \cdot \, \tilde{E} \,\,  \, +(t-1) \cdot \, \tilde{K}\right),
  \\
  &&\hspace{-.98in}\, \,
     \tilde{g}_3(0,5) \, \, = \, \, \,
     -\frac{8}{3}\cdot  \, (1 \, -t)^{1/4} \cdot  \, t^{-2} \cdot  \, 
     \left((t-2) \cdot \, \tilde{E} \,\, \,  -2 \cdot \, (t-1) \cdot \, \tilde{K}\right)
  \\
  &&\hspace{-.98in}\, \,
     \tilde{g}_4(0,5) \, \, = \, 
     -\frac{8}{3} \cdot \, (1-t)^{1/2} \cdot \, t^{-2} \cdot \, 
     \left(3 \, \tilde{E}^2 \,  +2 \cdot \, (t-2) \cdot \, \tilde{E}\tilde{K} \,
     -(t-1)\cdot \, \tilde{K}^2\right). 
\end{eqnarray}
For $\, N= \, 5$, these  $\, \tilde{g}_i(0,N)$'s  have the following expansions near $\, t= \, 0$:
\begin{eqnarray}
  &&\hspace{-1in}
     \tilde{g}_1(0,5) \, \, = \, \, \, \,
     1\, \, \, +\frac{5}{2^{7}} \, t^2 \, \,  \,
+\frac{3^2 \cdot 5}{2^{10}}\, t^3 \, \, \, +\frac{3\cdot 5^2\cdot 19}{2^{15}} \, t^4 \, \,  \,
     +\frac{5479}{2^{17}} \, t^5 \,\,  \, +\frac{5\cdot 11\cdot 3041}{2^{22}} \, t^6
     \nonumber\\
  &&\hspace{-.8in}
 +\frac{3^2 \cdot 5\cdot 7 \cdot 11 \cdot 23}{2^{21}} \, t^7 \,  \,
+\frac{3^4 \cdot 5 \cdot 7 \cdot 11^2 \cdot 227}{2^{31}} \, t^8 \,   \,
+\frac{5\cdot 7 \cdot 11^2 \cdot 17581}{2^{31}} \, t^9 \, \,  
+\cdots\\
  &&\hspace{-1in}
     \tilde{g}_2(0,5) \, \, = \, \, \,  \, 1 \,\,  +\frac{5}{2^7} \, t^2 \,\,  \,
     +\frac{5\cdot  7}{2^{10}} \, t^3 \, \, \, +\frac{3^3\cdot 5 \cdot 7}{2^{15}} \, t^4 \, \,  \,
+\frac{7\cdot 463}{2^{17}} \, t^5 \, \,  \,+\frac{3\cdot 5 \cdot 7\cdot   863}{2^{22}} \, t^6
     \nonumber\\
  &&\hspace{-.7in}
     +\frac{3^3 \cdot 5 \cdot 149}{2^{20}} \, t^7 \, \,  \,
+\frac{3^2 \cdot 5 \cdot 7 \cdot 11\cdot 19 \cdot 563}{2^{31}} \, t^8 \, \,  \,
+\frac{3\cdot 5 \cdot 17 \cdot 132199}{2^{31}} \, t^9 \,  \,  \,
+\cdots\\
  &&\hspace{-1in}
     \tilde{g}_3(0,5) \, \, = \, \, \,  \, 
     1 \, \, -\frac{5}{2^{7}} \, t^2 \, \, \, -\frac{5}{2^{7}} \, t^3 \, \,  \,
-\frac{5\cdot 113}{2^{14}}t^4-\frac{5\cdot 7^2}{2^{13}} \, t^5 \, \,  \,
     -\frac{5\cdot 7 \cdot 3119}{2^{22}} \, t^6
     \nonumber\\
&&\hspace{-.7in}
-\frac{5\cdot 19163}{2^{22}} \, t^7 \, \,  \,
-\frac{5\cdot 7\cdot 11\cdot 56443}{2^{30}} \, t^8 \, \,  \,
-\frac{5^2\cdot 11 \cdot 17657}{2^{23}} \, t^9 \, \, \,  \,
-\cdots\\ 
  &&\hspace{-1in}
     \tilde{g}_4(0,5) \, \, = \, \, \, \,
     1 \, -\frac{5}{2^7} \, t^2 \, \,  \,-\frac{5}{2^{7}} \, t^3 \, \,  \,
-\frac{3\cdot 5\cdot 19}{2^{13}} \, t^4 \, \,  \,-\frac{5^3}{2^{12}} \, t^5 \, \,  \,
   -\frac{5\cdot 22541}{2^{22}} \, t^6
   \nonumber\\
  &&\hspace{-.7in} \quad \quad 
     -\frac{3^4 \cdot 5\cdot 13\cdot 19}{2^{22}} \, t^7 \, \,  \,
-\frac{5^3 \cdot 47 \cdot 1951}{2^{29}} \, t^8 \, \,  \,
     -\frac{5\cdot 517129}{2^{27}} \, t^8
     \, \, \, \,   -\cdots
\end{eqnarray}
The four factors $\, g_i(0,7)$  for $\, C(0,7) \, $ read respectively 
\begin{eqnarray}
  &&\hspace{-1in}\, \, \,
    \tilde{g}_1(0,7) \, \, = \, \, \,
     \frac{8}{15t^2} \cdot \, (1-t)^{-1/4} \cdot \,
     \Bigl(2 \cdot  \, (t^2-t+1) \cdot  \, {\tilde E} \, \,  \,
     -(t-2) \cdot \, (t-1)  \cdot \, {\tilde K}\Bigr), 
\end{eqnarray}
\begin{eqnarray}  
  &&\hspace{-1in}\, \, \,
     \tilde{g}_2(0,7)\, \, 
   = \, \, \, \frac{8}{45t^2} \cdot \, (1-t)^{-1/2}\cdot  \, \Bigl( (4t^2+11t-11) \cdot \, {\tilde E}^2
   \nonumber\\
  &&\hspace{-.5in} \quad \quad  \quad  \quad  \quad \quad \,  \, 
     +8 \cdot \, (t-2) \cdot \, (t-1) \cdot \, {\tilde E}{\tilde K} \, \,
     - 5 \cdot \, (1-t)^2 \cdot \, {\tilde K}^2\Bigr), 
 \end{eqnarray}
 \begin{eqnarray}  
  &&\hspace{-1in}\, \, \,
     \tilde{g}_3(0,7)\, \, 
   = \, \, \, \frac{64}{45t^4} \cdot \, (1-t)^{5/8} \cdot \, \Bigl( (4t^2-19t+4) \cdot \, {\tilde E}^2
   \nonumber\\
  &&\hspace{-.5in} \quad \quad \quad \quad 
     -2 \cdot \, (8t^2-15t+4) \cdot \, {\tilde E}{\tilde K} \, \,  \, 
+(7t \, -4) \cdot \, (t \, -1) \cdot \, {\tilde K}^2\Bigr), 
  \\
  &&\hspace{-1in}\, \, \,
     \tilde{g}_4(0,7)\, \,      =\, \,  \,
     -\frac{64}{45t^4} \cdot \,  (1-t)^{3/8} \, \cdot\, \Bigl( (11t^2-11t-4) \cdot \, {\tilde E}^2
   \nonumber\\
  &&\hspace{-.5in} \quad \, 
     +2 \cdot \, (t-1) \cdot \, (3t^2-7t-4)  \cdot \, {\tilde E}{\tilde K}
         \, \, \,  -(t-1)^2 \cdot \, (3t+4) \cdot \, {\tilde K}^2\Bigr), 
\end{eqnarray}
 which have the following expansions at $ \, t= \, 0 $:
\begin{eqnarray}
  &&\hspace{-.6in}
   \tilde{g}_1(0,7) \,  \, = \,  \,  \, \,\,
     1 \, \,  \,+ \, \frac{7}{2^7} \, t^2 \,\, \, +\frac{7}{2^7} \, t^3 \, \, \,
+\frac{3^2\cdot 7 \cdot 13}{2^{14}} \, t^4 \,\, \, +\frac{7 \cdot 53}{2^{13}} \, t^5 \, \, \,
     +\frac{3^4\cdot 5 \cdot 7 \cdot 61}{2^{22}} \, t^6
     \nonumber\\
  &&\hspace{-.1in} \,\,
     +\frac{3 \cdot 7^2\cdot 13 \cdot 83}{2^{22}} \, t^7 \, \, \,
+\frac{3 \cdot 5 \cdot 7 \cdot 357293}{2^{30}} \, t^8 \, \, \,
   +\frac{5\cdot 7 \cdot 13 \cdot 19 \cdot 1009}{2^{28}} \, t^9
     \nonumber\\
\label{c071}
  &&\hspace{-.1in} \quad \quad \quad  \,\,
     +\frac{3^3 \cdot 7 \cdot 13 \cdot 17 \cdot 29 \cdot 3449}{2^{37}} \, t^{10} \, \,
 \,  \,  \, +\cdots
\end{eqnarray}
\begin{eqnarray}  
  &&\hspace{-.6in}
     \tilde{g}_2(0,7) \,  \, = \,  \,  \, \,\,
     1 \, \,  \,+\frac{7}{2^7} \, t^2 \, \,  \,+\frac{7}{2^{7}} \, t^3 \, \, \,
+\frac{7\cdot 61}{2^{13}} \, t^4 \,\, \, +\frac{7 \cdot 29}{2^{12}} \, t^5\,\, \,
     +\frac{5^2 \cdot 7 \cdot 11 \cdot 103}{2^{22}} \, t^6
     \nonumber\\
  &&\hspace{-.1in} \quad 
     +\frac{7\cdot 47\cdot 577}{2^{22}} \, t^7 \, \, \,
+\frac{7\cdot 13\cdot 41\cdot 6257}{2^{29}} \, t^8 \,\, \,
     +\frac{7\cdot 803461}{2^{27}} \, t^9
     \nonumber\\
  \label{c072}
  &&\hspace{-.1in} \quad \quad \quad  \quad 
     +\frac{7\cdot 23 \cdot 17281729}{2^{36}}\, t^{10}
     \, \, \,  \, + \, \, \, \cdots
 \end{eqnarray}
\begin{eqnarray}  
  &&\hspace{-.6in}
     \tilde{g}_3(0,7)\,\, = \,\,\, \,\,
     1\,\, \, -\frac{7}{2^7}\, t^2\,\,  \, -\frac{7}{2^{7}}\, t^3 \, \, \,
-\frac{3^2\cdot 5^2\cdot 7}{2^{15}}\, t^4 \, \, \, -\frac{7\cdot 1553}{2^{18}}\, t^5 \,\, \,
     \nonumber\\
  &&\hspace{-.5in} \quad
     -\frac{5\cdot 7\cdot 13\cdot 331}{2^{22}}\, t^6
     -\frac{3^2 \cdot 11\cdot 10631}{2^{25}}\, t^7 \, \, \,
-\frac{7\cdot 13\cdot 652831}{2^{31}}\, t^8 \,\, \,
       \nonumber\\
  \label{c074}
  &&\hspace{-.5in} \quad \quad 
     -\frac{5\cdot 7\cdot 11\cdot 13\cdot 42257}{2^{33}} \, t^9 \, \,\,
     -\frac{3^2\cdot 7\cdot 11\cdot 13\cdot 389\cdot 1733}{2^{38}}\,  t^{10} \,
\, \,\,-\cdots
\end{eqnarray}
\begin{eqnarray}  
  &&\hspace{-.6in}
     \tilde{g}_4(0,7)\, \, =\,\,\, \,\,
  1\,\, \,-\frac{7}{2^7}\,t^2 \,\,  \,  -\frac{7}{2^7}\, t^3 \,\,
-\frac{3^2\cdot 5^2\cdot 7}{2^{15}}\, t^4 \,\,
-\frac{3\cdot 7\cdot 11\cdot 47}{2^{18}}\, t^5 \,\,
       \nonumber\\
  &&\hspace{-.1in}
     -\frac{3\cdot 5\cdot 7\cdot 1429}{2^{22}}\, t^6
     -\frac{3^2 \cdot 116131}{2^{25}}\, t^7 \,\, \,
-\frac{3^2\cdot 7\cdot 11\cdot 89\cdot 953}{2^{31}}\, t^8 \,\, \,
       \nonumber\\
 \label{073}
  &&\hspace{-.1in}  \quad
     -\frac{5 \cdot 7 \cdot 11 \cdot 283 \cdot 1913}{2^{33}}\, t^9 \, \,\,
     -\frac{3^7 \cdot 7^2 \cdot 11 \cdot 13 \cdot 389}{2^{38}}\,t^{10}\,\, \,
     - \,\,\, \cdots
 \end{eqnarray}

Using the definition of sigma (\ref{sigmadef}) we find from 
(\ref{c071})-(\ref{c074}) that:
\begin{eqnarray}
  &&\hspace{-.98in}\, \, \,
 \sigma_1(0,7) \, \, = \,\,\,  \frac{7}{8} \, \, - \frac{7}{16} \, t  \, \, 
     -\frac{7}{2^6} \, t^2 \,\, \, -\frac{7}{2^7} \, t^3 \, \,\,
-\frac{5\cdot 7^2}{2^{13}} \, t^4 \, \, \,-\frac{7\cdot 41}{2^{14}} \, t^5 \, \,\,
   -\frac{7\cdot 3251}{2^{21}} \, t^6 \, \,\, -\frac{7\cdot 41 \cdot 103}{2^{22}} \, t^7
     \nonumber\\
  \label{s071final}
  &&\, \, \,
     -\frac{7\cdot 22853}{2^{25}} \, t^8 \,\, \, -\frac{7\cdot 32027}{2^{26}} \, t^9  \, \,\,
   -\frac{7\cdot 11848691}{2^{35}} \, t ^{10}
   \,  \, \,  \, +\cdots
  \\
  &&\hspace{-.98in}\, \, \,
     \sigma_2(0,7) \, \, =  \,\,\,  \frac{7}{8} \, \, - \frac{7}{16} \, t  \, \, 
     -\frac{7}{2^6} \, t^2 \, \,\, -\frac{7}{2^7} \, t^3 \, \,\,
-\frac{3^2\cdot 5\cdot 7}{2^{13}} \, t^4 \, \,\, -\frac{7\cdot 71}{2^{14}} \, t^5 \, \,\,
   -\frac{7\cdot 13 \cdot 577}{2^{21}} \, t^6
     \nonumber\\
  \label{s072final}
  &&\hspace{-.1in}\, \, \,
     -\frac{7\cdot 19\cdot 23 \cdot 29}{2^{22}} \, t^7 \, \,\,
-\frac{3\cdot 7 \cdot 115908}{2^{27}} \, t^8 \, \,\,
   -\frac{7\cdot 211 \cdot 2857}{2^{28}} \, t^9
   \nonumber\\
  &&\hspace{-.1in} \, \, \,\quad \quad 
     -\frac{7\cdot 13 \cdot 89\cdot  58321}{2^{35}} \, t^{10}\,\,
  \,    \,\, \, +\cdots
\end{eqnarray}
\begin{eqnarray}
  &&\hspace{-.78in}\, \, \,
     \sigma_3(0,7) \, \, =  \,\,\,  - \frac{7}{8} \, \, + \frac{7}{16} \, t  \, \, 
    +  \frac{7}{2^6} \, t^2 \, \,  \,+\frac{7}{2^7} \, t^3 \,  \, \,
+\frac{5 \cdot 7}{2^{10}} \, t^4 \,  \,  +\frac{7\cdot 17 \cdot 53}{2^{18}} \, t^5  \, \,
+\frac{7\cdot 11 \cdot 31}{2^{17}} \, t^6  \, \, 
     \nonumber\\
  \label{s074final}
  &&\hspace{-.6in} \quad  \quad \quad  \quad \, +\frac{7\cdot 34679}{2^{24}} \, t^7 \,  \, \,
     +\frac{7\cdot 28517}{2^{24}} \, t^8 \,  \,\,
     +\frac{7\cdot 673 \cdot 915}{2^{32}} \, t^9
     \nonumber\\
  &&\hspace{-.5in}\,\,  \,\,\quad \quad  \quad   \quad  \quad  \quad 
     +\frac{7\cdot 163 \cdot 520129}{2^{36}} \, t^{10}\,
  \,    \,  \,  \, +\cdots
\end{eqnarray}
\begin{eqnarray}
 \label{s073final}  
  &&\hspace{-.78in}\, \, \,
     \sigma_4(0,7) \, \, =  \,\,\,  - \frac{7}{8} \, \, + \frac{7}{16} \, t  \, \, 
     + \frac{7}{2^6}t^2 \,  \,  +\frac{7}{2^{7}}\, t^3\,\,\,
     +\frac{5\cdot  7}{2^{10}}\, t^4\, \, \, +\frac{3^4 \cdot 7\cdot 11}{2^{18}}\, t^5\, \,\,
+\frac{7 \cdot 331}{2^{17}}\, t^6\, \, \, 
     \nonumber\\
  &&\hspace{-.6in}\, \quad  \quad \quad  \quad  +\frac{5\cdot 7 \cdot 6381}{2^{24}}\, t^7\, \, \,
     +\frac{5 \cdot 7 \cdot 5279}{2^{24}}\, t^8\,\,\,
     +\frac{3\cdot 7\cdot 53 \cdot  83\cdot 421}{2^{32}}\,  t^9\, \,\,
   \nonumber\\
  &&\hspace{-.5in}\, \quad  \quad \quad  \quad  \quad  \quad 
     +\frac{7\cdot 61\cdot 311\cdot 3929}{2^{36}} \, t^{10}
    \,   \, \, +\cdots \quad 
 \end{eqnarray}

\subsection{Lambda extensions for $ \, C(0,7)$}
\label{appClambda}

The recursion procedure on the solution for the sigma function defined by
(\ref{sigmadef}) are expressed in the forms
(\ref{sigma12})-(\ref{sigma34}) for $ \, j= \, 1,2 \,$ as
\begin{eqnarray}
  &&\hspace{-.6in}
     \sigma_j(0,7;\lambda_j) \, = \, \, \,\,
     {{7} \over {8}} \cdot \,  \sqrt{1-t} \, \, \,  \,\, +\lambda_j \, t^4 \,  \,\,
+\frac{3}{2} \, t^5 \, \lambda_j \, \, \,
+\frac{5^2 \cdot 17}{2^{8}} \, t^6 \, \lambda_j \,  \,\,
     +\frac{5\cdot 13^2}{2^{9}} \, t^7 \, \lambda_j
     \nonumber\\
&&\hspace{-.1in}
   +\Bigl(\frac{197 \cdot 1301}{5\cdot 2^{15}} \, \lambda_j \,  \, \,
   +\frac{1}{4} \, \lambda_j^2\Bigr)  \cdot \, t^8 \, \,\,
   +\Bigl(\frac{7\cdot 73\cdot 929}{5\cdot 2^{16}} \, \lambda_j \,  \, \,
   +\frac{7}{8} \, \lambda_j^2\Bigr) \cdot  \, t^9
   \nonumber\\
  \label{s071recursion}
&&\hspace{-.1in} \quad \quad 
   +\Bigl(\frac{3\cdot 7\cdot 61\cdot 21713}{5\cdot 2^{22}} \, \lambda_j \, \,
   +\frac{  29 \cdot 41}{5 \cdot 2^7} \, \lambda_j^2\Bigr)  \cdot \, t^{10}
   \,  \, \,\, \, +\cdots 
\end{eqnarray}
and for $\, j= \, 3,4$
\begin{eqnarray}
  &&\hspace{-.6in}
     \sigma_j(0,7;\lambda_j) \,  \,\, = \, \, \, \,\,
     -\,  {{7} \over {8}} \cdot \,  \sqrt{1-t} \, \, \, \,
     +\lambda_j \, t^5 \, \, \, +2 \,t^6 \, \lambda_j \,  \, \,
+\frac{887}{5\cdot 2^6} \, t^7 \, \lambda_j \, \,  \,
     +\frac{1061}{5\cdot 2^6} \, t^8 \, \lambda_j
     \nonumber\\
  \label{s074recursion}
  &&\hspace{-.1in}
     +\frac{7\cdot 43049}{5\cdot 2^{14}} \, t^9 \, \lambda_j \,  \, \, \,
     +\Bigl(\frac{3\cdot 7\cdot 37\cdot 103}{5\cdot 2^{12}} \lambda_j \, \, \,
     +\frac{1}{5} \, \lambda_j^2\Bigr) \cdot \, t^{10}
     \, \, \, \,  \, + \, \, \, \cdots 
\end{eqnarray}
The algebraic functions $\, \sigma^{(4)}_A(t;7) \,$ and $\, \sigma^{(1)}_A(t;7)$ (see (\ref{sigmaA}))
\begin{eqnarray}
 &&\hspace{-.1in}
  \, \sigma^{(4)}_A(t;7)  \,\, = \,\, \frac{7}{8}\, \sqrt{1-t}, 
  \quad  \quad  \quad \sigma^{(1)}_A(t;7) \,\, = \,\, - \, \frac{7}{8}\, \sqrt{1-t}, 
\end{eqnarray}
are two solutions of the sigma form of Painlev\'e VI (\ref{4okamotoinsigma})
for $\, N\, = \, 7$.
Setting
\begin{eqnarray}
  \lambda_2 \, \, = \, \, \, -\lambda_1 \, \,   \, = \, \, \, \frac{5\cdot 7}{2^{13}},
 \quad \quad \quad \quad 
\lambda_4 \, \, = \, \, \, -\lambda_3 \, \,   \, = \, \, \, \frac{5\cdot 7}{2^{18}}, 
\end{eqnarray}
we see that (\ref{s071recursion}) and (\ref{s074recursion})   
reproduce (\ref{s071final})-(\ref{s074final}).

\vskip 1cm

\vskip .1cm 

\section{Computation of $ \, B_1^{(1)}(0,N;t) \, $ and $ \, B_1^{(4)}(0,N;t)$}
\label{appE}

To compute $ \, B_1^{(1)}(0,N;t)$ and $ \, B_1^{(4)}(0,N;t)$ in the expansion
(\ref{sigma12}) and (\ref{sigma34}) we put $ \, \sigma_{2}(0,N;t)$ and
$ \, \sigma_4(0,N;t)$ into the Okamoto equation (\ref{4okamoto}) using the
relations (\ref{newsig}) 
and set the
coefficient of each power of $ \,\lambda$ separately to zero. This gives
the following linear differential equations  for $ \, B_1^{(1)}(0,N;t)$ and $ \, B_1^{(4)}(0,N;t)$.
The D-finite function $ \, B_1^{(4)}(0,N;t) \, $ is solution of the second-order linear differential
operator:
\begin{eqnarray}
&&\hspace{-.7in}
L_2^{(+)} \, \, = \, \,   \, \,
4 \cdot \,t^2 \cdot \, (1 -t)^2 \cdot\, {{d^2} \over {dt^2}}
   \, \, \, \,   \,
   +2 \cdot \, t \cdot \, (t\, -1) \cdot \,
   \Bigl((2\, N\,  +1) \cdot \, t \, \,\, -2 \, N\Bigr)  \cdot  \, {{d} \over {dt}}
\nonumber\\
\label{b41eqn}
&&\hspace{-.5in} \quad \quad 
+N \cdot \, \Bigl( (t-1) \cdot \,  \Bigl((N-1) \cdot \, t \, +2\Bigr) \,\,
-(t -2) \cdot \, \sqrt{1-t} \Bigr).
\end{eqnarray}
The D-finite function $ \, B_1^{(1)}(0,N;t)$ is solution of the second-order linear differential
operator:
\begin{eqnarray}
&&\hspace{-.98in} \, \,   \, 
M_2^{(+)}  \, \, = \, \,   \, \,
4 \cdot \, t^2 \cdot  \, (1-t)^2 \cdot \,  {{d^2} \over {dt^2}}  \,  \, \,\,\,\, 
+2 \cdot \, t \cdot \, (t -1) \cdot \,
\Bigl( (2\, N \, +5)  \cdot \, t \,\,  -2\, N \, -4\Bigr)  \cdot \,  {{d} \over {dt}}
\nonumber\\
\label{b11eqn}
&&\hspace{-.98in}  \quad \quad \, \, 
+\Bigl((t-1) \cdot \, \Bigl((N+1) \cdot \, (N+2) \cdot \, t \,  \, -2 \, N\Bigr)
\, \, +N \cdot \, (t-2) \cdot \, \sqrt{1-t}\Bigr).
\end{eqnarray}
In order to get rid of the  $\, \sqrt{1-t}$ terms we do the following trick:  we introduce
the companion operators of $\, L_2^{(+)}$ (resp. of  $\, M_2^{(+)}$) which amount to changing
the sign of $ \, \sqrt{1-t}\,$. We denote   $\, L_2^{(-)}$ (resp. of  $\, M_2^{(-)}$) these
linear differential operators.
We calculate the LCLM (direct sum) of these two linear differential operators:
$\, L_4 \, = \, \, \, LCLM(L_2^{(+)} \, \oplus  L_2^{(-)}) \, \, = \, \,  L_2^{(+)} \, \oplus \,  L_2^{(-)}$,
which is an order-four linear differential operator
with polynomial coefficients (no square roots anymore).
Using the LCLM-DFactorisation of Maple we find another  LCLM (direct sum) for the order-four
linear differential operator $\, L_4$
\begin{eqnarray}
  \label{directsum}
  L_4 \, \,\,  = \, \, \, \, {\cal L}^{A}_2 \, \oplus \, {\cal L}^{B}_2
    \, \, = \, \, \,\,   LCLM( {\cal L}^{A}_2 , \, {\cal L}^{B}_2),  
\end{eqnarray}
where the two order-two linear differential operators $\, {\cal L}^{A}_2$ and $\, {\cal L}^{B}_2$
are, now,  linear differential operators with rational coefficients, reading respectively:
\begin{eqnarray}
  \label{f1eqn}
&&\hspace{-.28in} \,  \, 
{\cal L}^{A}_2  \, \, \, = \, \, \,\, \, 
 {{d^2} \over {dt^2}} \, \,\, +\frac{N+1}{t} \cdot \,  {{d} \over {dt}}
     \, \, \, \, \,   -\frac{N^2}{4 \cdot \, t \cdot \, (1-t)}, 
\\
\label{f2eqn}
&&\hspace{-.28in}\,  \,
{\cal L}^{B}_2  \, \,\,  = \, \, \,\, \, 
  {{d^2} \over {dt^2}}  \, \, \, +\frac{N+1}{t} \cdot \,  {{d} \over {dt}}
        \, \,\,  \,  \,  +\frac{N^2 \cdot \, (t-1) \,\, -(t\, -2)}{4 \cdot \, t \cdot \,(1-t)^2}. 
 \end{eqnarray}
 
The solution $\,  f_1(t)$ of (\ref{f1eqn}), which is analytic at $ \, t= \, 0$, reads:
\begin{eqnarray}
  \quad \quad \quad
  f_1(t) \, \, = \, \, \,
   {}_2F_1\Bigl([\frac{N}{2},\frac{N}{2}], \, [N+1], \, t\Bigr).
\end{eqnarray}
The solution $\,  f_2(t)$ of (\ref{f2eqn}), which is analytic at $ \, t= \, 0$, reads:
\begin{eqnarray}
  f_2 \, \, = \, \, \, \sqrt{1-t} \cdot \,
      {}_2F_1\Bigl([\frac{N}{2},\frac{N}{2}+1], \, [N+1], \, t\Bigr). 
\end{eqnarray}
The solution  of (\ref{b41eqn}) is  
a linear combination of $ \, f_1$ and $ \, f_2$  
\begin{eqnarray}
  \label{hf1f2}
  \quad  \quad  \quad 
 \, \,\,  \, c_1 \cdot \, f_1 \,\, \,   +c_2 \cdot \, f_2, 
\end{eqnarray}
which is determined using  (\ref{hf1f2})  in (\ref{b41eqn}).
This way we find that the solution $ \, B^{(4)}_1(0,N;t)$, which is normalized to
unity at $ \, t= \, 0$,  has $ \, c_1= \, c_2= \, 1/2 \, $ and  reads:
\begin{eqnarray}
\label{form1b41}
 &&\hspace{-.8in}
B^{(4)}_1(0,N;t) \, \, = \, \, \,
 \\
&&\hspace{-.7in} \,
   \, \, \, \, =  \, \,
 \frac{1}{2} \cdot  \, \Bigl( {}_2F_1\Bigl([\frac{N}{2},\frac{N}{2}], \, [N+1], \, t\Bigr) \, \,  \,
   +\sqrt{1-t} \cdot \, \, {}_2F_1\Bigl([\frac{N}{2},\frac{N}{2}+1], \, [N+1], \, t\Bigr)\Bigr)
\nonumber \\
&&\hspace{-.6in} \, \quad  \quad 
 \, = \, \, \, \, \,    1 \, \, \,\,  +  \Bigl({{N-1 } \over {4}}  \Bigr) \cdot \, t
\, \, \,\,  +  \Bigl({{N^3 \, +2\,N^2 \,-2\,N \, -2 } \over {32 \cdot \, (N\, +1)}} \Bigr)  \cdot \, t^2
\,  \, \, \, + \, \, \, \cdots
\nonumber
\end{eqnarray}
An alternative form for $ \, B^{(4)}_1(0,N;t) \, $ is obtained by use of the
identities (64) on page 64 and (38) on page 103 of~\cite{batemanv1} to 
write
\begin{eqnarray}
  &&\hspace{-.93in}
     {}_2F_1\Bigl([\frac{N}{2},\frac{N}{2}], \, [N+1], \, t\Bigr)
     \, \,   = \, \, \,
 ( 1 \, -t) \, \cdot \, {}_2F_1\Bigl([\frac{N}{2}+1,\frac{N}{2}+1], \, [N+1], \, t\Bigr)
   \\
 &&\hspace{-.83in}
 =  \,\,
  {}_2F_1\Bigl([\frac{N}{2},\frac{N}{2}+1], \, [N+1], \, t\Bigr) \, \, 
  -\frac{N\cdot \, t}{2\cdot \, (N+1)} \cdot \,
  {}_2F_1\Bigl([\frac{N}{2}+1,\frac{N}{2}+1], \, [N+2], \, t\Bigr),
 \nonumber
\end{eqnarray}
where using 
\begin{eqnarray}
\label{contiguous}
&&\hspace{-.3in}\quad 
{{d} \over {dt}} \, {}_2F_1\Bigl([\frac{N}{2},\frac{N}{2}], \, [N+1], \, t\Bigr)
\nonumber     \\
&&\hspace{-.23in} \quad \quad \quad = \, \, \,
\frac{N^2}{4 \cdot \, (N+1)} \cdot \,
     {}_2F_1\Bigl([\frac{N}{2}+1,\frac{N}{2}+1], \, [N+2], \, t\Bigr), 
\end{eqnarray}
we obtain the alternative expression (\ref{b14}):
\begin{eqnarray}
&& \hspace{-.8in} \quad \quad  \quad   \quad 
B^{(4)}_1(0,N;t)
\, \,  = \, \, \,\,\,
\frac{1}{2 \, N}  \cdot \,
\Bigl( 2 \cdot \, t \cdot \, \sqrt{1-t} \cdot \, \frac{d}{dt} \,  \,
{} _2F_1\Bigl([\frac{N}{2},\frac{N}{2}], \, [N+1], \, t\Bigr)
\nonumber     \\
&& \hspace{-.8in} \quad \quad \quad  \quad \quad \quad  \quad \,
    +N \cdot \, \, (1 \, +\sqrt{1 -t}) \cdot \,
    {}_2F_1\Bigl([\frac{N}{2},\frac{N}{2}], \, [N+1], \, t\Bigr)
    \Bigr).
     \nonumber 
\end{eqnarray}
This representation can be used to find a direct verification of (\ref{b41eqn}) 
which does not require the use of Maple.

\subsection{Computation of $ \, B_1^{(1)}(0,N;t) $ }
\label{appEB1}

The derivation of the exact expression (\ref{b11}) of $ \, B^{(1)}_1(0,N;t)\, $  from (\ref{b11eqn})
is done in a similar manner. The LCLM-DFactorisation
of the LCLM (direct-sum) of  $\, M_2^{(+)}$ and  $\, M_2^{(-)}$
gives two order-two linear differential operators $\, {\cal M}^{A}_2$ and $\, {\cal M}^{B}_2$
which are, now,  linear differential operators with rational coefficients, reading respectively:
\begin{eqnarray}
  \label{f1eqnbis}
&&\hspace{-.78in} \,  \, 
{\cal M}^{A}_2  \, \, = \, \, \,
{{d^2} \over {dt^2}} \, \,\, +\frac{N+3}{t} \cdot \,  {{d} \over {dt}} \, \, \, \,   \,
   +\frac{N^2 \cdot \, t \, \, + 4 \cdot \, (N +1) \cdot \, (t\, -1)}, 
\\
  \label{f2eqnbis}
&&\hspace{-.78in}\,  \,
{\cal M}^{B}_2  \, \, = \, \, \,
{{d^2} \over {dt^2}}  \, \, \, +\frac{N+3}{t} \cdot \,  {{d} \over {dt}}
\nonumber \\
&&\hspace{-.78in} \quad  \quad  \quad  \quad  \quad  \,  \, \, \,\,  \,
   +\frac{  1 \, \, + (t \, -1) \cdot \, (N^2 \, t \,  + (4\, N\, +3)\cdot \, (t-1) )
   }{4 \cdot \, t^2 \cdot \,(t \, -1)^2}. 
\end{eqnarray}
We find in analogy with  (\ref{form1b41})
\begin{eqnarray}
&& \hspace{-.8in} \quad \quad \quad \quad 
  B^{(1)}_1(0,N;t) \, \,
   = \, \, \,\, \frac{2 \cdot \, (N+1)}{t}   \cdot  \,
     \, \, {}_2F_1\Bigl([\frac{N}{2},\frac{N}{2}], \, [N+1], \, t\Bigr) \, \,
 \nonumber     \\
 && \hspace{-.8in} \quad \quad \quad \quad \quad \quad \quad 
    - \frac{2 \cdot \, (N+1)}{t}   \cdot  \,\sqrt{1-t} \cdot  \,
  {}_2F_1\Bigl([\frac{N}{2},\frac{N}{2}+1], \, [N+1], \, t\Bigr), 
\end{eqnarray}
and using (\ref{contiguous}) the alternative expression (\ref{b11}):
\begin{eqnarray}
  && \hspace{-.9in}      \quad  \, \, 
     B^{(1)}_1(0,N;t)\, \, 
  = \, \, \,
   -\frac{2 \cdot \, (N+1) }{N \, t}  \cdot \,
     \Bigl( 2 \cdot \, t \cdot \, \sqrt{1-t} \cdot \,
     \frac{d}{dt} \, \, {}_2F_1\Bigl([\frac{N}{2},\frac{N}{2}], \, [N+1], \, t\Bigr)
\nonumber     \\
 && \hspace{-.9in} \quad \quad  \, \quad \quad \quad  \quad   \quad  
   \, \, 
  -N \cdot  \, \Bigl(1 \,  -  \sqrt{1-t}\Bigr)  \cdot  \,
  {}_2F_1\Bigl([\frac{N}{2},\frac{N}{2}], \, [N+1], \, t\Bigr)\Bigr) 
\end{eqnarray}
\begin{eqnarray}
 && \hspace{-.8in} \quad \quad  
  \, \,  = \, \, \,\,\,
    2 \cdot \, (N \, +1) \cdot \, {{1\, -t} \over { t}}
    \cdot \, \,  {} _2F_1\Bigl([\frac{N}{2} \, +1,\frac{N}{2} \, +1], \, [N+1], \, t\Bigr)
    \nonumber     \\
 && \hspace{-.8in} \quad \quad \quad  \quad \quad \quad \quad \,
    -2 \cdot \, (N \, +1) \cdot \, {{ \sqrt{1-t} } \over {t}}
    \cdot \, \,  {} _2F_1\Bigl([\frac{N}{2},\frac{N}{2} \, +1], \, [N+1], \, t\Bigr)
\end{eqnarray}
\begin{eqnarray}
  && \hspace{-.8in} \quad \quad
  \, \,  = \, \, \,\,\, \,   1 \, \, \,\,  + \Bigl({{N+1 } \over {4}} \Bigr) \cdot \, t
  \, \, \,\,
  +\Bigl( {{N^3 \, +8\,N^2 \, +20\,N \, +12 } \over {32 \cdot \, (N\, +3)}} \Bigr)  \cdot \, t^2
     \, \,  \, \, \, + \, \, \, \cdots
     \nonumber  
\end{eqnarray}

\vskip .1cm

\section{Forrester-Witte determinants}
\label{appFW}

In this appendix we display the parameters of the Forrester-Witte
determinants for the {\em two}  factors of $ \, C(M,N)\, $ with $ \, M+N \, $ odd and the {\em four}
factors of $\, C(0,N)$ with $\, N$ odd, and give quite remarkable identities between Toeplitz
determinants related directly to the factorizations of $\, C(M,N)$ analysed in this paper. 

\subsection{Determinant parameters for $ \, C(M,N)$ with $ \, M+N$ odd}
\label{appF}
Let us first recall form eq. (125) of~\cite{bmm} that the low-temperature 
correlations $ \, C(M,N)$ with $M+N$ odd can also be written in 
terms of Forrester-Witte determinants with the Okamoto parameters
given in table G1 as
\begin{eqnarray}
\label{Cdef}
  \hspace{-.5in}
  C(M,N)  \, \, =\, \,\,
(1-k^2)^{((N-M)^2+1)/4} \cdot \, 
D\Big( N,0,\frac{M-N}{2},\frac{M-N}{2},k \Big),
\end{eqnarray}
where we use the notation $ \,D\Big( \tilde{N},\eta,p,p',k \Big) \, $ to mean
the Toeplitz determinant obtained from the $\,\tilde{N}\, \times \,\tilde{N} \, $
matrix with elements   $\,A_n^{(p,p',\eta)}$
(see eqs. (\ref{fwdet}) and (\ref{Adef2})).

One can verify that the other choice of parameters in table G1 
gives a similar expression since:
\begin{eqnarray}
&& \hspace{-.4in} \quad \quad  
  D\Big( N,0,\frac{M-N}{2},\frac{M-N}{2},k \Big) \,\,
  \nonumber \\ 
 && \hspace{-.4in} \quad \quad  \quad \quad \quad   
= \,\,\,(1-k^2)^{MN} \cdot \, 
D\Big( N,0,-\frac{M+N}{2},-\frac{M+N}{2},k \Big).
\end{eqnarray}

\begin{table}
\center
\caption{Determinant parameters  for $C(M,N)$} 
\begin{tabular}{|c|c|c|c|c|c|c|c|}\hline
$n_1$&$n_2$&$\tilde{N}$&$n_3$&$n_4$&$\eta$&$p$&$p'$\\ \hline
&&&&&&&\\
$N/2$&$N/2$&$N$&$-M/2$&$M/2$&$0$&$\frac{M-N}{2}$&$\frac{M-N}{2}$\\
$N/2$&$N/2$&$N$&$-M/2$&$M/2$&$0$&$-\frac{M+N}{2}$&$-\frac{M+N}{2}$\\
&&&&&&&\\
\hline
\end{tabular}
\end{table}

\subsection{Determinant parameters for the two factors of 
$\,C(M,N) \, $ with $\,M+N$ odd}

For the two factors of $\,C(M,N) \, $ with $\,M+N \,$ odd the Okamoto parameters are
chosen from the set in (\ref{parameters}). For $N$ even both factors
must have ${\tilde N}= \, N/2 \,$ and for $ \, N \,$ odd we must have 
$\,{\tilde N}\,=\,(N+1)/2 \, $ for one factor and $\,{\tilde N}\,= \,(N-1)/2 \,$ for the other.
There are many choices for $ \, n_i \,$ from the set (\ref{parameters}) which
are given in table G2 for $\,N\,$ even and  table G3 for $ \, N$ odd. 

\begin{table}
\center
\caption{Determinant parameters for $N$ even, $\tilde{N}= \, N/2$ and $ \, \eta \, > \, 0$} 
\begin{tabular}{|c|c|c|c|c|c|c|c|}\hline
$n_1$&$n_2$&$\tilde{N}$&$n_3$&$n_4$&$\eta$&$p$&$p'$\\ \hline
$\frac{M+N+1}{4}$&$-\frac{M-N+1}{4}$&$N/2$&$\frac{M+N-1}{4}$
&$-\frac{M-N-1}{4}$&$N/2$&$-\frac{N-1}{2}$&$-\frac{M}{2}$\\
$\frac{M+N+1}{4}$&$-\frac{M-N+1}{4}$&$N/2$&$-\frac{M-N-1}{4}$
&$\frac{M+N-1}{4}$&$N/2$&$\frac{M-N}{2}$&$-\frac{1}{2}$\\
$-\frac{M-N+1}{4}$&$\frac{M+N+1}{4}$&$N/2$&$\frac{M+N-1}{4}$
&$-\frac{M-N-1}{4}$&$N/2$&$-\frac{M+N}{2}$&$\frac{1}{2}$\\
$-\frac{M-N+1}{4}$&$\frac{M+N+1}{4}$&$N/2$&$-\frac{M-N-1}{4}$
&$\frac{M+N-1}{4}$&$N/2$&$-\frac{N+1}{2}$&$\frac{M}{2}$\\
\hline
$\frac{M+N-1}{4}$&$-\frac{M-N-1}{4}$&$N/2$&$\frac{M+N+1}{4}$
&$-\frac{M-N+1}{4}$&$N/2$&$-\frac{N+1}{2}$&$-\frac{M}{2}$\\
$\frac{M+N-1}{4}$&$-\frac{M-N-1}{4}$&$N/2$&$-\frac{M-N+1}{4}$
&$\frac{M+N+1}{4}$&$N/2$&$\frac{M-N}{2}$&$\frac{1}{2}$\\
$-\frac{M-N-1}{4}$&$\frac{M+N-1}{4}$&$N/2$&$\frac{M+N+1}{4}$
&$-\frac{M-N+1}{4}$&$N/2$&$-\frac{M+N}{2}$&$-\frac{1}{2}$\\
$-\frac{M-N-1}{4}$&$\frac{M+N-1}{4}$&$N/2$&$-\frac{M-N+1}{4}$
&$\frac{M+N+1}{4}$&$N/2$&$-\frac{N-1}{2}$&$\frac{M}{2}$\\
\hline
\end{tabular}
\end{table}
\begin{table}
\center
\caption{Determinant parameters for $N$ odd, $\tilde{N}=(N\pm 1)/2$ 
and $\eta>0$} 
\begin{tabular}{|c|c|c|c|c|c|c|c|}\hline
$n_1$&$n_2$&$\tilde{N}$&$n_3$&$n_4$&$\eta$&$p$&$p'$\\ \hline
$\frac{M+N+1}{4}$&$-\frac{M-N-1}{4}$&$(N+1)/2$&$\frac{M+N-1}{4}$
&$-\frac{M-N+1}{4}$&$(N-1)/2$&$-\frac{N}{2}$&$-\frac{M+1}{2}$\\
$\frac{M+N+1}{4}$&$-\frac{M-N-1}{4}$&$(N+1)/2$&$-\frac{M-N+1}{4}$
&$\frac{M+N-1}{4}$&$(N-1)/2$&$\frac{M-N}{2}$&$-\frac{1}{2}$\\
$-\frac{M-N-1}{4}$&$\frac{M+N+1}{4}$&$(N+1)/2$&$\frac{M+N-1}{4}$
&$-\frac{M-N+1}{4}$&$(N-1)/2$&$-\frac{M+N}{2}$&$-\frac{1}{2}$\\
$-\frac{M-N-1}{4}$&$\frac{M+N+1}{4}$&$(N+1)/2$&$-\frac{M-N+1}{4}$
&$\frac{M+N-1}{4}$&$(N-1)/2$&$-\frac{N}{2}$&$\frac{M-1}{2}$\\
\hline
$\frac{M+N-1}{4}$&$-\frac{M-N+1}{4}$&$(N-1)/2$&$\frac{M+N+1}{4}$
&$-\frac{M-N-1}{4}$&$(N+1)/2$&$-\frac{N}{2}$&$-\frac{M-1}{2}$\\
$\frac{M+N-1}{4}$&$-\frac{M-N+1}{4}$&$(N-1)/2$&$-\frac{M-N-1}{4}$
&$\frac{M+N+1}{4}$&$(N+1)/2$&$\frac{M-N}{2}$&$\frac{1}{2}$\\
$-\frac{M-N+1}{4}$&$\frac{M+N-1}{4}$&$(N-1)/2$&$\frac{M+N+1}{4}$
&$-\frac{M-N-1}{4}$&$(N+1)/2$&$-\frac{M+N}{2}$&$\frac{1}{2}$\\
$-\frac{M-N+1}{4}$&$\frac{M+N-1}{4}$&$(N-1)/2$&$-\frac{M-N-1}{4}$
&$\frac{M+N+1}{4}$&$(N+1)/2$&$-\frac{N}{2}$&$\frac{M+1}{2}$\\
\hline
\end{tabular}
\end{table}

However, one can make the following remarks
\begin{enumerate}

\item The FW-determinants associated to the first (resp. last) four rows of table G2,
when $\,N \, $ is even,
are all related. For example, one has
\begin{eqnarray}
&&\hspace{-.8in}
     D\Big( {{N} \over {2}}, {{N} \over {2}},  {{M -N} \over {2}}, -{{ 1} \over {2}},  \, k \Big) 
 \nonumber   \\
  &&\hspace{-.6in} \quad 
 = \, \, (-1)^{N/2}
\cdot \,  (1-k^2)^{N \; (M-1) /4} \,
\cdot \,
D\Big(  {{N} \over {2}},  {{N} \over {2}},  - {{M +N} \over {2}},  {{ 1} \over {2}}, \, k \Big),
\end{eqnarray}
and:
\begin{eqnarray}
  &&\hspace{-.8in}
  D\Big(  {{N} \over {2}},  {{N} \over {2}}, {{M -N} \over {2}}, {{ 1} \over {2}}, \, k \Big) 
  \nonumber  \\
  &&\hspace{-.8in} \quad  \quad 
 = \, \,   (-1)^{N/2}   \cdot \,  (1 -k^2)^{N \; (M+1) /4}
 \cdot \,
 D\Big( {{N} \over {2}},  {{N} \over {2}},  - {{M +N} \over {2}},  -{{ 1} \over {2}}, \, k \Big).
\end{eqnarray}

So we can use any one of them to represent the two factors 
appearing in $ \, C(M,N) \, $ when $ \, (M+N)\,$ is odd.
We have choosen in the following to use the Okamoto parameters
of row 2 and 6 in table G2.

\item Similarly, the FW-determinants associated to the first (resp. last) four rows of table G3,
when $ \, N$ is odd,
are also all related. For example, one has
\begin{eqnarray}
  &&\hspace{-.9in} \quad 
 D\Big( {{N-1} \over {2}}, {{N+1} \over {2}}, {{M -N} \over {2}}, {{ 1} \over {2}}, \,  k \Big)  
  \nonumber  \\
  &&\hspace{-.7in} \quad  \quad 
= \, \,  (1 \, -k^2)^{M \; (N-1) /4}   \,\cdot  \,
 D\Big(  {{N-1} \over {2}}, {{N+1} \over {2}}, -{{M +N} \over {2}}, {{ 1} \over {2}}, \,  k \Big),
 \end{eqnarray}
and 
\begin{eqnarray}
  &&\hspace{-.9in} \quad 
 D\Big( {{N+1} \over {2}},  {{N-1} \over {2}}, {{M -N} \over {2}}, -{{ 1} \over {2}},  \, k  \Big) 
 \nonumber  \\
  &&\hspace{-.7in} \quad  \quad 
 = \, \, (1-k^2)^{M \; (N+1) /4} \,\cdot  \,
     D\Big(  {{N+1} \over {2}}, {{N-1} \over {2}}, - {{M +N} \over {2}}, -{{ 1} \over {2}}, \, k \Big), 
\end{eqnarray}
So we can use any one of them to represent the two factors 
appearing in $ \, C(M,N)$ when $ \,(M+N) \, $ is odd, $\, M \, \le \, N$.
We have choosen in the following to use the Okamoto parameters
of row 2 and 6 in table G3.

\item We can now summarize the factorisations in two factors seen 
on $ \, C(M,N)$ when $ \, (M+N)$ is odd
by the following two identities on Toeplitz determinants
(we denote by $\,k_L\,=\,\, 2 \sqrt{k}/(1+k) \, $ the Landen tranform of $ \, k$)

\begin{itemize}

\item when $\,M$ is odd and $\,N$ is even
\begin{eqnarray}
\label{ff1}
  &&\hspace{-.98in}
     D\Big( N, 0,  {{M -N} \over {2}}, {{M -N} \over {2}},  \, k \Big)
     \nonumber \\
&&\hspace{-.9in}\quad 
   = \, \,\, (-1)^{N/2} \cdot \,  2^{N (N-2)/2} \cdot \,  k^{-N^2/4} \cdot \,  (1+k)^{N(2M-N)/2} 
  \\
&&\hspace{-.7in} \quad 
   \times \,  D\Big(  {{N} \over {2}}, {{N} \over {2}}, {{M -N} \over {2}}, {{1} \over {2}}, k_L \Big)
   \cdot \,  D\Big( {{N} \over {2}}, {{N} \over {2}},  {{M -N} \over {2}}, -{{1} \over {2}}, \,  k_L \Big).
   \nonumber
\end{eqnarray}

\item when $M$ is even and $N$ is odd
\begin{eqnarray}
\label{ff2}
&&\hspace{-.99in}\quad   
D\Big( N, 0,  {{M -N} \over {2}},  {{M -N} \over {2}}, k \Big)
\nonumber  \\
&&\hspace{-.6in} \quad   
   = \, \, \, \,  2^{(N-1)^2/2} \cdot \,  k^{-(N^2-1)/4} \cdot \,  (1 +k)^{(2MN-N^2-1)/2} 
\\
&&\hspace{-.99in} 
   \times \,
D\Big(  {{N-1} \over {2}},  {{N+1} \over {2}}, {{M -N} \over {2}}, {{ 1} \over {2}}, \,  k_L \Big)
 \cdot 
D\Big( {{N+1} \over {2}},  {{N-1} \over {2}}, {{M -N} \over {2}}, -{{ 1} \over {2}}, \, k_L \Big).
\nonumber
\end{eqnarray}

\end{itemize}

\end{enumerate}

Replacing these relations in eq. (\ref{Cdef}), we obtain the 
factorisation of $\, C(M,N)$, with $\, (M+N) \, $ even, in two factors.

These factorisation relations can be seen as a consequence of the symmetry of the
$\, N \times\,  N$
Toeplitz matrix associated to $ \, \eta= \, 0$ and $\, p=\, p'=\, (M-N)/2$
and Wilf relations
(\ref{factoreven}) when $\, N$ is even and (\ref{factorodd}) 
when $N$ is odd~\cite{wilf}.

\subsection{Expressions of $ \,  g_{\pm} (M,N,t) \, $ in $ \, C(M,N,t) \, $
      with $\, (M+N)$ odd, in terms of Toeplitz determinants}
\label{appG}

\begin{itemize}

\item when $ \, M$ is odd and $ \, N$ is even, the factors
  $ \,  g_{\pm} (M,N,t) \, $  of  $ \, C(M,N,t) \, $
 (see (\ref{twofactors}))  are given by: 
\begin{eqnarray}
 \hspace{-.8in} \quad 
 \, \, \, (-1)^{N/2} \cdot \,
  D\Big( {{N} \over {2}}, {{N} \over {2}},  {{M -N} \over {2}}, {{ 1} \over {2}}, \, k_L \Big)
\cdot \, \left( \frac{1+k}{1-k} \right)^{(N-M)/4} \cdot \,  S_1, 
\nonumber \\
  \hspace{-.8in} \quad 
   \, \,\,
  D\Big( {{N} \over {2}},  {{N} \over {2}}, {{M -N} \over {2}}, -{{ 1} \over {2}}, \, k_L \Big)
\cdot  \, \left( \frac{1+k}{1-k} \right)^{(M-N)/4} \cdot  \, S_1, 
\nonumber 
\end{eqnarray}
with
\begin{eqnarray}
  \hspace{-.98in}
S_1 \,\,  = \,\,\,
(-1)^{E((N+2)/4)} \cdot \,  2^{N(N-2)/4} \cdot \,  {{(1-k)^{(M-N)^2/8} } \over {k^{N^2/8} }}
\cdot  \, (1+k)^{(2MN+M^2-N^2)/8}, 
\nonumber 
\end{eqnarray}
where $\,E(x) \, $ denotes the integer part of $\, x$.

\item when $ \, M$ is even and $ \, N$ is odd, one has
\begin{eqnarray}
  \hspace{-.8in} \quad  \, \, 
 D\Big( {{N-1} \over {2}}, {{N+1} \over {2}},  {{M -N} \over {2}}, {{ 1} \over {2}}, \, k_L \Big)
\cdot \,  \left( \frac{1+k}{1-k} \right)^{(N-M)/4} \cdot \,  S_2, 
\nonumber \\
\hspace{-.8in} \quad   \,\,  
 D\Big({{N+1} \over {2}}, {{N-1} \over {2}},  {{M -N} \over {2}}, -{{ 1} \over {2}}, \,  k_L \Big)
\cdot  \, \left( \frac{1+k}{1-k} \right)^{(M-N)/4} \cdot  \, S_2, 
\nonumber 
\end{eqnarray}
with
\begin{eqnarray}
  \hspace{-.98in}
S_2 \, \, = \, \, \, 
(-1)^{E((N+2)/4)} \cdot \,  2^{(N-1)^2/4} \cdot \,  {{(1-k)^{(M-N)^2/8} } \over { k^{(N^2-1)/8} }}
\cdot \, (1 \, +k)^{(2MN+M^2-N^2-2)/8}.
\nonumber 
\end{eqnarray}

\end{itemize}
All these expressions are compatible with the series expansions 
of Appendix A.
From the above relations, we can also obtain closed expressions 
for $ \, f_1(t) \, $ and $ \, f_2(t) \, $ appearing in \ref{appA}.

\subsection{Determinant parameters for the four factors 
of $\, C(0,N)$ with $\, N$ odd}

For the four factors of $\, C(0,N) \, $ with $\, N$ odd the Okamoto parameters
are chosen from the set (\ref{4factorparams}).  The two cases must be
considered separately $\, N=\, 4\,  m \, \pm 1$. For $\, N=\, 4\, m\, +1$ the values of $\, \tilde N$
for the factors are
\begin{eqnarray}
  \hspace{-.1in}
  \quad  \quad \quad 
{\tilde N}\, = \, \, \,\, m, \, \, \, m, \,\, \,  \,m, \, \,\,  \,m+1, 
\end{eqnarray} 
and for $ \, n= \, 4 m\, -1$ the values of $\, \tilde N \, $ are:
\begin{eqnarray}
  \hspace{-.1in}
  \quad  \quad \quad 
{\tilde N} \, \, = \,\, \, \, m, \, \,\,  m, \,\, \, m, \,\, \, m-1.
\end{eqnarray}
The choices of $\, n_i$ which give integer $ \, {\tilde N} \, $ for $ \, N= \, 4 \,m \,+1 \, $ are
given in table G4 and for $\, N= \, 4\, m \, -1 \,\, $ in table G5. 

\begin{table}
\center
\caption{Determinant parameters for $\, C(0,N)$ with $ \, N= \, 4m+1$} 
\begin{tabular}{|c|c|c|c|c|c|c|c|}\hline
$n_1$&$n_2$&$\tilde{N}$&$n_3$&$n_4$&$\eta$&$p$&$p'$\\ \hline
$m+\frac{1}{2}$&$-\frac{1}{2}$&$\frac{N-1}{4}$&$m$&$0$&$\frac{N-1}{4}$
&$-m+\frac{1}{2}$&$-m-\frac{1}{2}$\\
$m+\frac{1}{2}$&$-\frac{1}{2}$&$\frac{N-1}{4}$&$0$&$m$&$\frac{N-1}{4}$
&$\frac{1}{2}$&$-\frac{1}{2}$\\
$-\frac{1}{2}$&$m+\frac{1}{2}$&$\frac{N-1}{4}$&$m$&$0$&$\frac{N-1}{4}$
&$-2m-\frac{1}{2}$&$\frac{1}{2}$\\
$-\frac{1}{2}$&$m+\frac{1}{2}$&$\frac{N-1}{4}$&$0$&$m$&$\frac{N-1}{4}$
&$-m-\frac{1}{2}$&$m+\frac{1}{2}$\\
\hline
$m+\frac{1}{2}$&$\frac{1}{2}$&$\frac{N+3}{4}$&$m$&$0$&$\frac{N-1}{4}$
&$-m-\frac{1}{2}$&$-m-\frac{1}{2}$\\
$m+\frac{1}{2}$&$\frac{1}{2}$&$\frac{N+3}{4}$&$0$&$m$&$\frac{N-1}{4}$
&$-\frac{1}{2}$&$-\frac{1}{2}$\\
$\frac{1}{2}$&$m+\frac{1}{2}$&$\frac{N+3}{4}$&$m$&$0$&$\frac{N-1}{4}$
&$-2m-\frac{1}{2}$&$-\frac{1}{2}$\\
$\frac{1}{2}$&$m+\frac{1}{2}$&$\frac{N+3}{4}$&$0$&$m$&$\frac{N-1}{4}$
&$-m-\frac{1}{2}$&$m-\frac{1}{2}$\\
\hline
$m$&$0$&$\frac{N-1}{4}$&$m+\frac{1}{2}$&$-\frac{1}{2}$&$\frac{N-1}{4}$
&$-m-\frac{1}{2}$&$-m-\frac{1}{2}$\\
$0$&$m$&$\frac{N-1}{4}$&$m+\frac{1}{2}$&$-\frac{1}{2}$&$\frac{N-1}{4}$&
$-2m-\frac{1}{2}$&$-\frac{1}{2}$\\
$m$&$0$&$\frac{N-1}{4}$&$-\frac{1}{2}$&$m+\frac{1}{2}$&$\frac{N-1}{4}$
&$\frac{1}{2}$&$\frac{1}{2}$\\
$0$&$m$&$\frac{N-1}{4}$&$-\frac{1}{2}$&$m+\frac{1}{2}$&$\frac{N-1}{4}$
&$-m+\frac{1}{2}$&$m-\frac{1}{2}$\\
\hline
$m$&$0$&$\frac{N-1}{4}$&$m+\frac{1}{2}$&$\frac{1}{2}$&$\frac{N+3}{4}$
&$-m-\frac{1}{2}$&$-m+\frac{1}{2}$\\
$0$&$m$&$\frac{N-1}{4}$&$m+\frac{1}{2}$&$\frac{1}{2}$&$\frac{N+3}{4}$
&$-2m-\frac{1}{2}$&$\frac{1}{2}$\\
$m$&$0$&$\frac{N-1}{4}$&$\frac{1}{2}$&$m+\frac{1}{2}$&$\frac{N+3}{4}$
&$-\frac{1}{2}$&$\frac{1}{2}$\\
$0$&$m$&$\frac{N-1}{4}$&$\frac{1}{2}$&$m+\frac{1}{2}$&$\frac{N+3}{4}$
&$-m-\frac{1}{2}$&$m+\frac{1}{2}$\\
\hline
\end{tabular}
\end{table}

\begin{table}
\center
\caption{Determinant parameters for $C(0,N)$ with $N=4m-1$} 
\begin{tabular}{|c|c|c|c|c|c|c|c|}\hline
$n_1$&$n_2$&$\tilde{N}$&$n_3$&$n_4$&$\eta$&$p$&$p'$\\ \hline
$m-\frac{1}{2}$&$\frac{1}{2}$&$\frac{N+1}{4}$&$m$&$0$&$\frac{N+1}{4}$
&$-m-\frac{1}{2}$&$-m+\frac{1}{2}$\\
$m-\frac{1}{2}$&$\frac{1}{2}$&$\frac{N+1}{4}$&$0$&$m$&$\frac{N+1}{4}$
&$-\frac{1}{2}$&$\frac{1}{2}$\\
$\frac{1}{2}$&$m-\frac{1}{2}$&$\frac{N+1}{4}$&$m$&$0$&$\frac{N+1}{4}$
&$-2m-\frac{1}{2}$&$-\frac{1}{2}$\\
$\frac{1}{2}$&$m-\frac{1}{2}$&$\frac{N+1}{4}$&$0$&$m$&$\frac{N+1}{4}$
&$-m+\frac{1}{2}$&$-\frac{1}{2}+m$\\
\hline
$m-\frac{1}{2}$&$-\frac{1}{2}$&$\frac{N-3}{4}$&$m$&$0$&$\frac{N+1}{4}$
&$-m+\frac{1}{2}$&$-m+\frac{1}{2}$\\
$m-\frac{1}{2}$&$-\frac{1}{2}$&$\frac{N-3}{4}$&$0$&$m$&$\frac{N+1}{4}$
&$\frac{1}{2}$&$\frac{1}{2}$\\
$-\frac{1}{2}$&$m-\frac{1}{2}$&$\frac{N-3}{4}$&$m$&$0$&$\frac{N+1}{4}$
&$-2m+\frac{1}{2}$&$\frac{1}{2}$\\
$-\frac{1}{2}$&$m-\frac{1}{2}$&$\frac{N-3}{4}$&$0$&$m$&$\frac{N+1}{4}$
&$-m+\frac{1}{2}$&$m+\frac{1}{2}$\\
\hline
$m$&$0$&$\frac{N+1}{4}$&$m-\frac{1}{2}$&$\frac{1}{2}$&$\frac{N+1}{4}$
&$-m+\frac{1}{2}$&$-m+\frac{1}{2}$\\
$0$&$m$&$\frac{N+1}{4}$&$m-\frac{1}{2}$&$\frac{1}{2}$&$\frac{N+1}{4}$
&$-2m+\frac{1}{2}$&$\frac{1}{2}$\\
$m$&$0$&$\frac{N+1}{4}$&$\frac{1}{2}$&$m-\frac{1}{2}$&$\frac{N+1}{4}$
&$-\frac{1}{2}$&$-\frac{1}{2}$\\
$0$&$m$&$\frac{N+1}{4}$&$\frac{1}{2}$&$m-\frac{1}{2}$&$\frac{N+1}{4}$
&$-m-\frac{1}{2}$&$m-\frac{1}{2}$\\
\hline
$m$&$0$&$\frac{N+1}{4}$&$m-\frac{1}{2}$&$-\frac{1}{2}$&$\frac{N-3}{4}$
&$-m+\frac{1}{2}$&$-m-\frac{1}{2}$\\
$0$&$m$&$\frac{N+1}{4}$&$m-\frac{1}{2}$&$-\frac{1}{2}$&$\frac{N-3}{4}$
&$-2m+\frac{1}{2}$&$-\frac{1}{2}$\\
$m$&$0$&$\frac{N+1}{4}$&$-\frac{1}{2}$&$m-\frac{1}{2}$&$\frac{N-3}{4}$
&$\frac{1}{2}$&$-\frac{1}{2}$\\
$0$&$m$&$\frac{N+1}{4}$&$-\frac{1}{2}$&$m-\frac{1}{2}$&$\frac{N-3}{4}$
&$-m+\frac{1}{2}$&$m-\frac{1}{2}$\\
\hline
\end{tabular}
\end{table}

One can make similar remarks as in the previous section:
\begin{enumerate}

\item The Okamoto parameters of the FW-determinants are displayed
in tables G4 and G5 in four groups of four rows. 
We can use any row in each group 
to represent the four factors appearing in $ \, C(0,N)$ when $\, N$ is odd.

\item As in the previous section, 
we can summarize the factorisations in four factors seen 
on $ \, C(0,N)$ when $ \, N$ is odd
by the following identities on Toeplitz determinants
\begin{itemize}

\item when $\, N =\,  1 \mod\,  4$, i.e. $\, N= \,  1, 5, 9, 13, \, \cdots $
\begin{eqnarray}
  &&\hspace{-.98in}  \quad 
     D\Big( {{N-1} \over {2}}, {{N+1} \over {2}}, -{{N} \over {2}}, {{ 1} \over {2}}, \,  k_L \Big)
 \nonumber   \\
&&\hspace{-.6in} \quad 
= \, \, \, \, 
(-1)^{(N-1)/4} \cdot \, 
\; \left( \frac{1+k} {1-k} \right)^{(N-1)^2/8} 
   \\
&&\hspace{-.98in} \quad  \quad  \quad 
   \times \,
D\Big( {{N-1} \over {4}}, {{N-1} \over {4}},  {{ 1} \over {2}},  -{{ 1} \over {2}}, \,  k \Big) \cdot
   \,  D\Big( {{N-1} \over {4}}, {{N+3} \over {4}}, - {{ 1} \over {2}},  {{ 1} \over {2}}, \,  k \Big), 
   \nonumber 
\end{eqnarray}
and:
\begin{eqnarray}
  &&\hspace{-.98in} \quad 
  D\Big( {{N+1} \over {2}}, {{N-1} \over {2}}, -{{N} \over {2}}, -{{ 1} \over {2}}, \, k_L \Big)
\nonumber   \\
&&\hspace{-.6in} \quad 
=  \, \, \, 
(1 \, +k) \cdot  \, \left( \frac{1+k}{1-k} \right)^{(N-1)(N+3)/8} 
\cdot  \\
&&\hspace{-.98in} \quad  \quad  \quad 
\times  \,
D\Big( {{N-1} \over {4}}, {{N-1} \over {4}}, {{ 1} \over {2}}, {{ 1} \over {2}},  \, k \Big) \cdot
\, D\Big( {{N+3} \over {4}}, {{N-1} \over {4}}, -{{ 1} \over {2}}, -{{ 1} \over {2}}, \, k \Big).
\nonumber 
\end{eqnarray}

\item when $\, N = \, 3 \, \mod \,  4$, i.e. $\, N=  \, 3, 7, 11, 15, \, \cdots $
\begin{eqnarray}
&&\hspace{-.98in} \quad 
D\Big(  {{N+1} \over {2}},  {{N-1} \over {2}}, -{{N} \over {2}}, -{{ 1} \over {2}}, \,  k_L \Big)
\nonumber \\
&&\hspace{-.6in} \quad 
= \, \, \,  \, 
(-1)^{(N+1)/4} 
\cdot \,  \left( \frac{1+k}{1-k} \right)^{(N+1)^2/8} \, 
 \\
&&\hspace{-.98in} \quad  \quad  \quad 
 \times \,
 D\Big(  {{N+1} \over {4}},  {{N+1} \over {4}},  -{{ 1} \over {2}}, {{ 1} \over {2}}, \, k \Big) \cdot
\, D\Big( {{N+1} \over {4}}, {{N-3} \over {4}},  {{ 1} \over {2}}, - {{ 1} \over {2}},\,  k \Big), 
\nonumber 
\end{eqnarray}
and:
\begin{eqnarray}
  &&\hspace{-.98in} \quad 
 D\Big(  {{N-1} \over {2}}, {{N+1} \over {2}}, -{{N} \over {2}}, {{ 1} \over {2}}, \,  k_L \Big)
  \nonumber \\
&&\hspace{-.6in} \quad 
= \, \, \, 
(1+k) \cdot  \, \left( \frac{1+k}{1-k} \right)^{(N+1)(N-3)/8} 
  \\
&&\hspace{-.98in} \quad  \quad  \quad 
  \times \,
D\Big(  {{N+1} \over {4}},  {{N+1} \over {4}}, -{{ 1} \over {2}},-{{ 1} \over {2}}, \, k \Big) \cdot
 \,   D\Big( {{N-3} \over {4}},  {{N+1} \over {4}}, {{ 1} \over {2}}, {{ 1} \over {2}}, \,  k \Big).
\nonumber 
\end{eqnarray}

\end{itemize}

\end{enumerate}

Replacing these relations in eqs. (\ref{Cdef}) and using (\ref{ff1}) and (\ref{ff2}), 
we obtain the factorisation of $ \, C(0,N)$, with $ \, N$ even, in four factors.

\subsection{Expressions of $\, g_{i} (M,N,t) \, $ in $\, C(0,N,t)\, $ with $\, N$ odd,
  in terms of Toeplitz determinants}
\label{appH}

If we denote $\, S=\, (-1)^{E((N+4)/8)}$, then the factors appearing in eq. (\ref{fourfactors}),
solutions of the nonlinear equation (\ref{4okamotoinsigma}),
with the coefficient of their leading term normalised to one, are given by

\begin{itemize}

\item for $ \, N =  \, 1 \mod \,  4$, i.e. $ \, N= \, 1, 5, 9, 13, \,  \, \cdots $
\begin{eqnarray}
\hspace{-1.1in}
&& \quad 
D \Big( {{N-1} \over {4}}, {{N-1} \over {4}}, {{ 1} \over {2}}, -{{ 1} \over {2}}, \, k \Big)
\cdot  \, {{ (1-k^2)^{-1/16} } \over {k^{ (N+1)^2/16 } }} \cdot \,  2^{(N-1)(N-3)/8}  \cdot  \, S,
\nonumber \\
\hspace{-1.1in}
  && \quad 
(-1)^{(N-1)/4} \cdot \,
D \Big( {{N-1} \over {4}}, {{N-1} \over {4}}, {{ 1} \over {2}}, {{ 1} \over {2}}, \, k \Big)
        \cdot  \, {{(1-k^2)^{3/16} } \over {k^{ (N+1)^2/16 } }}  \cdot \,  2^{(N-1)(N-3)/8} \cdot \,  S ,
\nonumber \\
\hspace{-1.1in}
&& \quad 
D \Big( {{N-1} \over {4}}, {{N+3} \over {4}}, -{{ 1} \over {2}}, {{ 1} \over {2}},  \, k \Big)
\cdot \, {{ (1-k^2)^{-1/16} } \over { k^{ (N+1)(N-3)/16 }}} \cdot  \, 2^{(N^2-1)/8} \cdot \,  S ,
\nonumber \\
\hspace{-1.1in}
&& \quad 
D \Big( {{N+3} \over {4}}, {{N-1} \over {4}}, -{{ 1} \over {2}}, -{{ 1} \over {2}}, \, k \Big)
\cdot \,  {{ (1-k^2)^{3/16} } \over { k^{ (N+1)(N-3)/16 }  }}  \cdot \,  2^{(N^2-1)/8} \cdot  \,  S .
\nonumber 
\end{eqnarray}
\vskip .1cm

\item for $ \, N = \,  3 \mod  \, 4$, i.e. $N= \,  3, 7, 11, 15,  \,  \, \cdots $
\begin{eqnarray}
\hspace{-1.1in}
  &&  \quad
     D \Big( {{N-3} \over {4}},  {{N+1} \over {4}}, {{ 1} \over {2}}, {{ 1} \over {2}}, \, k \Big)
     \cdot \, {{ (1-k^2)^{3/16} } \over {k^{ (N-1)(N+3)/16 }  }}  \cdot  \, 2^{(N-1)(N-3)/8} \cdot \,  S.
\nonumber \\
\hspace{-1.1in}
  && \quad 
     D \Big( {{N+1} \over {4}}, {{N-3} \over {4}}, {{ 1} \over {2}}, -{{ 1} \over {2}},  \, k \Big)
     \cdot \,  {{(1-k^2)^{-1/16} } \over { k^{ (N-1)(N+3)/16 } }}  \cdot \,  2^{(N-1)(N-3)/8} \cdot \,  S,
\nonumber \\
\hspace{-1.1in}
  &&  \quad
(-1)^{(N+1)/4} \cdot \,
D \Big( {{N+1} \over {4}}, {{N+1} \over {4}}, -{{ 1} \over {2}}, -{{ 1} \over {2}}, \, k \Big)
    \cdot \, {{ (1-k^2)^{3/16} } \over {k^{ (N-1)^2/16 } }}  \cdot \,  2^{(N^2-1)/8} \cdot \,  S ,
\nonumber \\
\hspace{-1.1in}
&&  \quad 
   D \Big( {{N+1} \over {4}},  {{N+1} \over {4}}, -{{ 1} \over {2}}, {{ 1} \over {2}}, \,  k \Big)
   \cdot \,  {{ (1-k^2)^{-1/16} } \over {k^{ (N-1)^2/16 }  }} \cdot  \, 2^{(N^2-1)/8} \cdot \, S,
\nonumber
\end{eqnarray}

\end{itemize}
All these expressions are compatible with the series expansions of \ref{appC}.

\vspace{.2in}

\vskip .2cm 


\end{document}